\documentclass[aps,prd,superscriptaddress,floatfix,nofootinbib,reprint,amsmath,amssymb]{revtex4-2}

\usepackage[normalem]{ulem}
\usepackage{bm}
\usepackage{graphicx}
\usepackage{dcolumn}
\usepackage{color}
\usepackage[dvipsnames]{xcolor}
\usepackage[colorlinks]{hyperref}
\hypersetup{
    colorlinks=true,
    linkcolor=blue,
    filecolor=blue,      
    urlcolor=blue,
    citecolor=blue,
    pdfpagemode=FullScreen
} 

\newcommand{\dd}{{\rm d}}
\newcommand{\ii}{{\rm i}}

\newcommand{\bvec}[1]{\mathbf{#1}}
\newcommand{\orcid}[1]{\hspace{0.5mm}\href{https://orcid.org/#1}{\includegraphics[height=0.3cm,keepaspectratio]{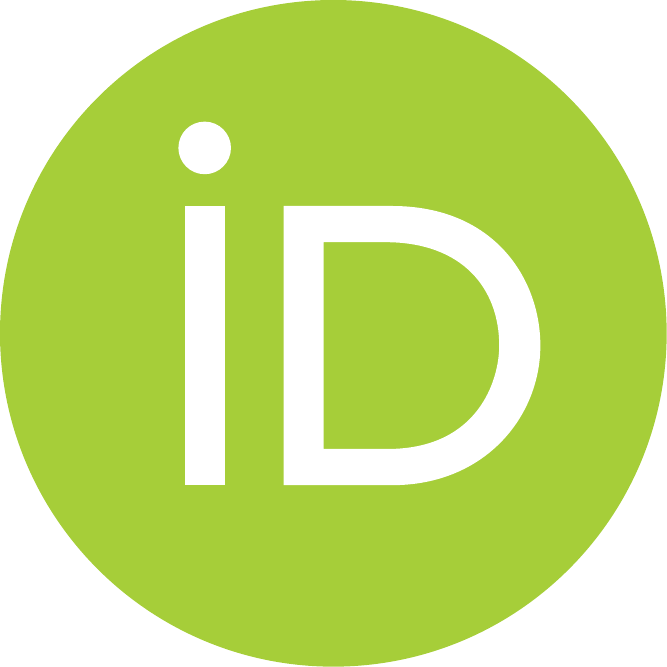}}}

\usepackage{aas_macros}

\begin{document}

\title{Novel aspects of cosmic ray diffusion in synthetic magnetic turbulence}

\author{Andrej Dundovic\orcid{0000-0001-9009-8379}}
\email{andrej.dundovic@gssi.it}
\author{Oreste Pezzi\orcid{0000-0002-7638-1706}}
\email{oreste.pezzi@gssi.it}
\author{Pasquale Blasi\orcid{0000-0003-2480-599X}}
\author{Carmelo Evoli\orcid{0000-0002-6023-5253}}
\affiliation{Gran Sasso Science Institute (GSSI), Viale F. Crispi 7, 67100 L’Aquila, Italy}
\affiliation{INFN, Laboratori Nazionali del Gran Sasso (LNGS), 67100 Assergi, L'Aquila, Italy}

\author{William H. Matthaeus\orcid{0000-0001-7224-6024}}
\affiliation{Bartol Research Institute and Department of Physics and Astronomy, University of Delaware, Newark, Deleware 19716, USA}

\begin{abstract}
The diffusive motion of charged particles in synthetic magnetic turbulence with different properties is investigated by using numerical simulations with unprecedented dynamical range, which allow us to ensure that both the inertial range and the long wavelength part of the turbulent spectrum are properly described. This is of particular importance in evaluating previous suggestions that parallel and perpendicular diffusion coefficients differ in their energy dependence, an assertion at odds with the many claims of universality of the $D_{\perp}$ and $D_{\parallel}$ as functions of particle energy. Cases with and without an ordered magnetic field are discussed. Results of the numerical simulations are compared with available theoretical models, for slab, slab/2D and isotropic turbulence. We find widespread evidence that universality is broken, and that the ratio $D_{\perp}/D_{\parallel}$ is not independent of energy. The implications of this finding for the physics of cosmic ray transport are discussed in depth. 
\end{abstract}

\maketitle

\section{Introduction}

A proper understanding of the interaction of charged particles with magnetic fluctuations in a plasma is an essential ingredient of the description of cosmic ray (CR) transport in astrophysical environments, as well as in the solar wind. Despite much progress in the field, there are still fundamental questions that are left unanswered, both in terms of the nature of turbulence and in terms of particle transport parallel and perpendicular to an ordered magnetic field on which turbulence is superposed. 

From the theoretical point of view, the problem of particle transport in an ensemble of Alfv\'en waves with amplitude $\delta B$ propagating along an ordered magnetic field \(\bvec{B_{0}}\), with $\delta B\ll B_{0}$, has been solved long ago using a perturbative approach, the so-called quasilinear theory (QLT)~\cite{jokipii1966cosmicray}. The theory has been very successful in predicting that the particle motion along the direction of \(\bvec{B_{0}}\) is diffusive. The same theory connects the diffusion coefficient parallel to \(\bvec{B_{0}}\) with the properties of the turbulence, notably its  amplitude and spectrum. On the other hand, the theory has also shown some weak points: 1) it predicts that the pitch angle diffusion coefficient at $90^{o}$ vanishes, so as to create the problem of crossing such a singular point in velocity space; 2) it is limited to small amplitude magnetic fluctuations; and 3) it fails in describing the motion of particles perpendicular to magnetic field lines. While the first two problems have been properly addressed in weakly non linear versions of QLT (for instance in the second order version of QLT~\cite{shalchi2005second}), the third problem appears to be more subtle and remains only partially addressed, as we discuss below. 

From the physics point of view, the problem of particle transport in turbulent magnetic fields is crucial to understand the origin of CRs and to make sense of the wealth of information that is becoming available with CR observations from space (see \cite{blasi2013origin,2019NCimR..42..549B} for recent reviews). The escape of CRs from the Galaxy is usually modelled as diffusive with an effective diffusion coefficient that depends on particle momentum and sometimes position (although at low energies advection might play a role). This effective treatment is prolific in producing results that compare well with observations~\cite{2019PhRvD..99j3023E,2020PhRvD.101b3013E}.
However it remains an effective treatment in which the diffusion coefficient is interpreted as some sort of spatial average (over a sufficiently large volume compared with the turbulence coherence length, perhaps the Galaxy) of the diffusion parallel and perpendicular to the ordered field. The latter is reasonably well known in the disc, but poorly known, if at all, in the Galactic halo. The question remains: are CRs escaping the Galaxy mainly along magnetic field lines or perpendicular to the field lines? And connected to the first question: do the diffusion coefficients parallel and perpendicular to the ordered magnetic field share the same energy dependence? This last question was first raised, as far as we know, in Ref.~\cite{de2007numerical}, where numerical simulations of CR transport in synthetic turbulence were carried out and some hints of a steeper energy dependence of $D_{\perp}$ was found compared with $D_{\parallel}$. In other words, the authors found that the ratio $D_{\perp}/D_{\parallel}$ is a growing function of energy at low energy, where parallel scattering is still the result of resonances with turbulence in the inertial range. On the other hand, based on the limited dynamical range of such simulations, the authors could only claim a hint of such a trend. Moreover, previous simulations~\cite{casse2001transport} did not seem to find such an effect. This tentative claim~\cite{de2007numerical} was later used as a basis to introduce anisotropic diffusion, though in a phenomenological way, in models of CR transport in the Galaxy~\cite{2012PhRvL.108u1102E,2017JCAP...10..019C}. 

More recently, thanks to better computational resources, it became possible to reassess this issue and several new hints appeared that the ratio $D_{\perp}/D_{\parallel}$ might be increasing with energy.
One such case is that of isotropic turbulence superposed on an ordered magnetic field with $\delta B/B_{0}\lesssim 1$, and at low enough energies that the gyro radius of the particles is much smaller than the energy containing scale of the turbulence \cite{2018JCAP...07..051G}. This result is consistent with \cite{de2007numerical}.

The issue of describing particle transport in realistic turbulence is even more complex than described so far, in that there is now overwhelming evidence that Alfv\'enic turbulent cascades develop in anisotropic way~\cite{ShebalinEA83,goldreich1995toward,OughtonEA15}, with most of the energy channelled into the perpendicular wave numbers, for $k$ in the inertial range. This implies that there is less power available in the modes that may potentially resonate with the gyromotion of particles and lead to particle diffusion \cite{BieberEA94}. Magnetosonic modes should be less affected by this process and play an important role in CR transport~\cite{2004ApJ...614..757Y}. The damping of these modes and the implications for CR transport were recently discussed in~\cite{2014ApJ...782...36E,2016ApJ...826..166X}. Much work done in evaluating scattering theories is accomplished by numerically representing turbulence at magnetohydrodynamics (MHD) scales propagating particles in such turbulence~\cite{2011ApJ...728...60B,2016A&A...588A..73C}.

A major limitation and challenge in this approach is that the limited dynamical range imposed by the available computational resources does not allow extension of these results in a straightforward way to particle energies of relevance for CR physics. 

In carrying out numerical studies of test particle trajectories, an important and somewhat delicate issue is the construction of realizations of turbulence based on various approximations and theoretical models of turbulent fluctuations. 
The distribution of power in wave vector is a central issue, and both the scale dependence and rotational symmetry have leading order influence on scattering. 
There are several models that are of historical, technical and physical relevance. 
Some popular models are the traditional but oversimplified one-dimensional ``slab'' model \cite{jokipii1966cosmicray} that appears in many plasma physics textbooks, often associated with parallel propagating waves.  
Here the importance of this model is that it concisely captures parallel resonances and therefore the physics of pitch angle scattering, an effect of primary importance in energetic particle transport. Even a small amount of power in parallel wave numbers or slab turbulence can dominate pitch angle scattering rates and control spatial diffusion \cite{bieber1994proton}.
Another standard limit is the isotropic model, familiar in hydrodynamics and expected in plasma turbulence when fluctuations are strong and no preferred direction is of significance~\cite{subedi2017charged}.
The two dimensional (2D) model admits variations of the correlation functions only in the two directions perpendicular to the mean magnetic field, providing an idealized representation of the tendency for turbulence in 
magnetized plasma to preferentially produce gradients transverse to the magnetic field direction~\cite{RobinsonRusbridge71, shebalin1983anisotropy}. 

The composite slab + 2D model~\cite{bieber1994proton} is a useful but {\it ad hoc} parametrization that accommodates both the dynamical tendency towards two dimensionality as well as the numerous possibilities for producing an admixture of parallel gradients due to wave particle interactions, shear instability, and several other effects.  
Other specialized models include reduced MHD (RMHD), popular in laboratory plasma and coronal studies, that describes low frequency fluctuations with a very strong magnetic field~\cite{OughtonEA17-rmhd} and the ``critical balance'' model \cite{goldreich1995toward,OughtonMatthaeus20} that also permits weak variations along the magnetic field direction but lacks 
(slablike) resonant power at higher wave numbers. 
Since parallel scattering is almost always an important ingredient in transport, in the following sections we will model fields of the slab, composite (slab + 2D) and isotropic types.  

In addition to the anisotropic cascades of MHD turbulence, it has been known for long time that CRs can also generate their own scattering centres through streaming instability, either resonant \cite{1969ApJ...156..445K,1975MNRAS.170..251H} or nonresonant \cite{2004MNRAS.353..550B}. The role of streaming instability for CR Galactic transport and its implications for observations has been recently investigated in detail (see \cite{2018AdSpR..62.2731A,2019Galax...7...64B} for recent reviews). The role of self-excited nonresonant streaming instability in the escape of CRs from the Galaxy has been recently discussed in \cite{2019PhRvL.122e1101B}.

Progress in this field has proceeded by continuously seeking a compromise between retaining as much physics as possible and extending the dynamical range of simulations so as to being able to apply results to CR transport. With this spirit, in this article we developed and used numerical simulations with the largest dynamical range ever achieved to simulate test particle transport in synthetic turbulence with an assigned spectrum and different topological properties. We focus on the cases of slab/2D turbulence and isotropic turbulence, both with and without an ordered field. In order to check the validity of the simulations we also test them versus known cases, such as pitch angle diffusion for slab turbulence and compare the numerical results with QLT and its second order extension. 

While there have been claims of universality of the $D_{\perp}/D_{\parallel}$ ratio (independence of the ratio on energy) \cite{hussein2015influence}, we find that this universality is evidently broken in the energy range of relevance for CR physics. In particular, for slab/2D turbulence we show that, for $\delta B_{\rm slab}/B_{0}\lesssim 1$, the ratio becomes constant only at very low energies, typically too low to be important for CR physics. The ratio $D_{\perp}/D_{\parallel}$ in this situation drops with energy in the inertial range of the turbulence responsible for particle scattering. Moreover we find that the nonlinear guiding center (NLGC) theory is a poor description of the $D_{\perp}$ obtained from our simulations, for the cases in which energy is roughly equally shared between slab and 2D. On the other hand it provides a sufficiently good description of the results when most of the energy is in the 2D modes. 
The numerical simulations we perform in this case have unprecedented dynamical range and provide strong confirmation that in the range of energy that is accessible at the present time, the ratio $D_{\perp}/D_{\parallel}$ is not constant in energy. Contrary to the case of slab/2D turbulence, for isotropic turbulence the ratio is found to be a growing function of energy, which is in agreement with the preliminary results obtained in Refs. \cite{de2007numerical,2018JCAP...07..051G}. More specifically, the parallel diffusion coefficient has a behaviour which appears to be consistent with the picture in which particles 
at low energy
scatter resonantly on perturbations that travel parallel to the local magnetic field, namely its energy dependence is $\propto r_{g}^{1/3}$ ( $\propto r_{g}^{1/2}$) for Kolmogorov (Kraichnan) turbulence for $r_{g}<l_{c}$ where $l_{c}$ is the coherence scale of the turbulence. For high energies $r_{g}>l_{c}$, $D_{\parallel}\propto r_{g}^{2}$ irrespective of the turbulent spectrum. As expected, in the limit of strong turbulence, $\delta B/B_{0}\gg 1$, one recovers the condition that $D_{\perp}\simeq D_{\parallel}$  \cite{subedi2017charged,2016MNRAS.457.3975S}. 

The difference in slope, at low energies, between parallel and perpendicular diffusion coefficient is $\sim 0.2$. For instance, for the case of Kolmogorov turbulence, $D_{\parallel}\propto E^{1/3}$ and $D_{\perp}\propto E^{0.5}$ at low energies. This example shows how the effective diffusion coefficient used in calculations of CR transport in the Galaxy might resemble a $D(E)\propto E^{0.5}$ even for the case of Kolmogorov turbulence if the effective coefficient were dominated by perpendicular transport. This would reflect, for instance, in the slope of the boron-to-carbon ratio. Similar considerations would apply to other spectra of turbulence. 

The article is organised as follows: in \S \ref{sec:methods} we discuss in detail all definitions used (spectra of slab, 2D and isotropic turbulence, diffusion coefficients) and methods adopted in the numerical simulations of particle transport. In \S \ref{sec:theory} we briefly summarize the theoretical models that are typically used to compare the results of numerical simulations with. In \S \ref{sec:results} we illustrate our main results for the different types of turbulence investigated here. We conclude in \S \ref{sec:discuss}.

\section{Definitions and Methods}
\label{sec:methods}

The motion of a charged particle in an assigned electromagnetic field is described by the Netwon-Lorentz equation:
\begin{equation}
 \label{eq:lorentz}
  \frac{\dd}{\dd t}\bvec{p}(\bvec{r}, t) = q\left[\bvec{E}(\bvec{r}, t) + \frac{\bvec{v}}{c}\times \bvec{B}(\bvec{r}, t)\right]\,
\end{equation}
where \(q\), \(\bvec{v}\) and \(\bvec{p}=m\gamma\bvec{v}\) are, respectively, the particle's charge, velocity and momentum, \(\bvec{E}\) and \(\bvec{B}\) represent the electric and magnetic field,  respectively, and $\gamma=\sqrt{1+\bvec{p}^2/m^2c^2}$ is the Lorentz factor, $m$ being the particle mass and $c$ the speed of light. 

Here, we focus on the test-particle regime, in which particles are independent on each other and do not affect the electromagnetic field. Furthermore, since we are particularly interested in relativistic particles moving in a nonrelativistic environment, the evolution of the underlying magnetofluid is neglected and the electric component of the field can be neglected, \(\bvec{E}=0\). For the same reasons, we restrict to the magnetostatic case, i.e.~$\partial \bvec{B}/\partial t=0$. The magnetic field can be split into a regular and fluctuating part, \(\bvec{B}(\bvec{r}) = \bvec{B}_0 + \delta\bvec{B}(\bvec{r})\), with the following properties: \(\bvec{B}_0 = B_0 \bvec{\hat{z}}\), \(\langle\delta\bvec{B}\rangle = 0\), and \(\delta B^2 \equiv \langle\delta\bvec{B}^2\rangle\), where $\langle \dots \rangle$ denotes the ensemble average.

In Nature, the turbulent magnetic fluctuations, $\delta\bvec{B}(\bvec{r})$, are typically the result of perturbations that evolve due to a complex interplay of mode couplings and result in spectra of the turbulent fluctuations. While this phenomenon is typically described by using MHD numerical simulations, here we focus on the transport of charged particles in synthetic turbulence, which allows us to address the specific issues discussed in the Introduction that are still matter of debate in the community. We are fully aware of the fact that an active turbulent cascade towards small scales may heavily affect the transport of CRs, both through the magnetic fields and through electric field 
fluctuations.
The results of such an
investigation will be presented 
separately in a forthcoming article, where we discuss the propagation of test particle CRs in a snapshot of a MHD simulation where turbulence is fully developed~\citep{cohet2016cosmic}. In this approach, all the features of MHD turbulence, such as intermittency and spectral anisotropy, are described self-consistently. However, the dynamical range is limited by the finite space resolution of the numerical simulation and performing high-resolution numerical simulation requires a huge numerical effort. In the case of synthetic turbulence, some of the main features of MHD turbulence are mimicked only through selection of an appropriate magnetic field two-point 
correlation tensor (see, e.g.~\citep{mertsch2019test} for a recent review on these methods). Higher correlations as well as the information contained in possible 
phase
correlations of magnetic fluctuations are discarded. These methods in general have a minor computational cost with respect to direct simulations. In the current work, we opt for synthetic models of turbulence, since 
our aims are 
to compare numerical results with theoretical considerations and to provide some hints for further theoretical developments.

\subsection{Models of homogeneous turbulence}

As anticipated above, magnetic fluctuations in synthetic models of MHD turbulence are described through the two-point correlation tensor:
\begin{equation}
    R_{\ell m}(\bvec{x}, \bvec{x}') = \left\langle \delta B_\ell(\bvec{x}) \delta B_m(\bvec{x}') \right\rangle, \quad \ell,m=x,y,z .
    \label{eq:Rlm}
\end{equation}
If the turbulence is homogeneous then the correlation depends only on the spatial {\it lag}, $\bvec{r} = \bvec{x}-\bvec{x}'$.
Calculating the trace gives the correlation function $R(\bvec{r}) \equiv R_{\ell\ell}(\bvec{r})$.
The Fourier transform of \(R_{\ell m}(\bvec{r})\)
(in a homogeneous infinite volume)  
provides a definition of the magnetic spectral  tensor:
\begin{equation}
    S_{\ell m}(\bvec{k} ) = \frac{1}{(2\pi)^3}
    \int \dd^3 r\,  R_{\ell m}(\bvec{r}) \, e^{-\ii\bvec{r}\cdot \bvec{k} }. 
\label{spectensor}    
\end{equation}
This transform is well defined whenever the correlation functions
(elements of the correlation tensor) fall off rapidly enough at infinity. 
It is also possible to provide a more singular 
relationship that can be used to define the spectrum in the 
homogeneous case,
namely  
\begin{equation}
    \delta(\bvec{k} + \bvec{k}') S_{\ell m}(\bvec{k} ) = \left\langle \delta \tilde{B}_\ell(\bvec{k}) \delta \tilde{B}_m(\bvec{k}') \right\rangle \, ,
    \label{specsingular}
\end{equation}
where $\delta(\bvec{k}+\bvec{k}')$
is the Dirac delta function and the the Fourier transform of the magnetic field fluctuations \(\delta \tilde{\bvec{B}}(\bvec{k})\) is formally defined as:
\begin{equation}
  \delta\tilde{\bvec{B}}(\bvec{k}) = \frac{1}{(2\pi)^3}\int \dd^3 r\, \delta\bvec{B}(\bvec{r}) \, e^{-\ii\bvec{r}\cdot \bvec{k} } \ .
\label{fcB}
\end{equation}
Note that both the $\delta$-function and $\delta\bvec{B}$ are singular objects and require careful limiting procedures to 
properly define them. However, since the analogues of Eqs. (\ref{specsingular}--\ref{fcB})
are well defined in a periodic domain of 
arbitrarily large size, this becomes the basis of generating synthetic fields in the treatment below. 

The magnetic correlation tensor 
(i) has to be consistent 
with the $\mathbf{\nabla\cdot B}=0$ condition, namely \(\mathbf{k}\cdot \delta\tilde{\bvec{B}}(\bvec{k}) = 0\), and (ii) may 
be determined by selecting an appropriate model for the energy spectrum. 
The first requirement, in the case of isotropic turbulence \cite{batchelor1982}, leads to
\begin{align}
    S_{\ell m}(\bvec{k} ) = \frac{S(k)}{2} & \left[ \left(\delta_{\ell m} - \frac{k_\ell k_m}{k^2}\right) \right. \nonumber \\
    &\quad \left. + \ii \sigma(\bvec{k}) \sum_n \epsilon_{\ell m n} \frac{k_n}{k}\right],
    \label{eq:corr_tensor}
\end{align}
where \(S(k)\) is a real function that reflects the geometry of the turbulence and its energy spectrum, \(\delta_{lm}\) is the Kronecker delta, \(\epsilon_{\ell m n}\) is the Levi-Civita symbol, and \(\sigma(\bvec{k}) \to \sigma(|\bvec{k}|)\) 
is the magnetic helicity which will be assumed to vanish in this work.
For axisymmetric turbulence \cite{montgomery1981anisotropic, matthaeus1981structure,oughton1997general} about a direction $\hat z$, the spectral form is slightly more complicated, with two scalar functions of cylindrical coordinates $(k_\parallel = \bvec{k}\cdot \hat z,k_\perp = |\bvec{k} \times \hat z|)$ instead of just one, defining the symmetric part of the tensor.

The omnidirectional energy spectrum of turbulence, usually denoted as \(E(k)\), is defined in terms of \(S_{\ell \ell}(\bvec{k})\) \cite{monin2013statistical, davidson2015turbulence}:
\begin{equation}
        E(k) \equiv \frac{1}{2}
        \int S_{\ell \ell}(\bvec{q}) 
        \delta(|{\bf q}| - k) d^3q \ .
    \label{eq:E_spectrum}
\end{equation}
The Einstein summation convention is implied for \(S_{\ell \ell}\). Due to the energy cascade, for a large system one postulates the existence of an inertial range in which \(E(k)\) becomes self-similar and, in some sense, universal. That is, for \(k\) large 
compared to $1/L$, the energy residing near scale $L$, and \(k\) small compared to the reciprocal of the dissipation scale, the spectrum satisfies the scaling \(E(k) \propto k^{-s}\) where \(s\) is the slope of the spectrum in that range. For example, $s=5/3$ for the Kolmogorov theory of turbulence and $s=3/2$ for Kraichnan turbulence. Moreover, by the solenoidal constraint, the property of homogeneity, and Cram\'er's theorem, the energy spectrum obeys the scaling \(E(\bvec{k})\propto k^q\) when \(k \rightarrow 0\) \cite{batchelor1982}, where \(q>0\) depends on the turbulence dimensionality \citep{matthaeus2007spectral}. The total energy density is normalized to the rms magnetic field strength squared:
\begin{equation}
    \int_0^\infty E(k) \dd k = \frac{1}{2}\delta B^2 \ ,
    \label{eq:norm_req}
\end{equation}
or, alternatively, \(\delta B^2 = R_{\ell \ell}(0) = \int \dd \bvec{k} S_{\ell \ell}(\bvec{k}) \).

Another important quantity that characterizes a turbulent field is the correlation length \(l_c\), defined as [cf. Eq. (\ref{eq:Rlm}) 
et seq.]
\begin{equation}
    l_c \equiv \frac{1}{R(0)} \int_0^\infty \dd r R(r) \ ,
    \label{eq:corr_length_definition}
\end{equation}
where we recall that $R(r)$ is the trace of the correlation matrix, 
Eq. (\ref{eq:Rlm}), and we make use of homogeneity and isotropy to write the argument as the scalar $r=|\bvec{r}|$. 

Note that, since we are focusing on the magnetostatic case, the time dependence of the dynamical correlation function, in general contained in the $k$-space correlation tensor, is not retained here. 

\subsubsection{Synthetic spectrum models}
The above properties of homogeneous turbulence 
provide a framework for development
of explicit realizations of turbulence  with prescribed properties,
for use in particle trajectory calculations. Generally speaking the purpose will be twofold -- on the one hand to make contact with scattering and transport theories that are applied to explanation of cosmic ray observations, and, on the other hand, to provide magnetic field realizations for particles trajectory calculations that are implemented in finite domains, often periodic domains that are nominally much larger than all relevant physical scales.

According to the aforementioned prescription, we model all spectra with a smooth function following the form of spectrum equations in \cite{sonsrettee2015magnetic, shalchi2009random}:
\begin{equation}
    E(q, s, l, \delta B, k) = 2 C(q, s) \delta B^2 l  \frac{\left(k l\right)^q}{\left[1 + (k l)^2 \right]^{(s+q)/2}},
    \label{eq:smooth_spectra}
\end{equation}
\(k\) being the wave number, and \(l\) the bend-over scale. The normalization constant $C(q,s)$, chosen to fulfill Eq. (\ref{eq:norm_req}) (see next subsection), is:
\begin{equation}
    C(q, s) = \frac{\Gamma\left(\frac{s+q}{2}\right)}{2 \Gamma\left(\frac{s-1}{2}\right)\Gamma\left(\frac{q+1}{2}\right)},
    \label{eq:C_norm}
\end{equation}
where $\Gamma$ is the Euler gamma function. 

In the following subsections we describe specific correlation tensors in three different geometries which are adopted in this work. In order to control properties when the models are applied to infinite homogeneous media, the models are described in terms of Fourier transforms and unbounded domains. 
The appropriate steps for conversion to Fourier series representation are discussed in Sec. \ref{sec:numerical_methods}.

\subsubsection{Slab model}

The first turbulence model considered is the slab model, that is also one of the first models historically introduced \citep{jokipii1966cosmicray, bieber1993longterm}. The slab turbulence model is one-dimensional and the wave vector is parallel to the imposed
background magnetic field \({\bf B}_0\). It resembles the propagation of Alfv\'en waves along the ordered magnetic field. Thus, by requiring \(\delta B_z = 0\) and \(\delta B_i(\bvec{r}) = \delta B_i(z)\), the axisymmetric tensor analogous to Eq. (\ref{eq:corr_tensor}) gives the slab spectral tensor \cite{oughton1997general}:
\begin{equation}
   S^{\rm slab}_{\ell m} \left({\bf k}\right) = \frac{S^{\rm slab}(k_\parallel, k_\perp)}{2} \delta_{\ell m} \, ,
\label{eq:Plm_slab}
\end{equation}
for \(\ell,m=x,y\), while other components are zero due to \(\delta B_z = 0\). The term \(k_\ell k_m/k^2\) can be omitted, since \(\delta B_i(\bvec{r})=\delta B_i(z)\), that implies the presence of \(\delta(k_x) \delta(k_y)\). Wave-vectors follow the axial symmetry of \(\bvec{B}_0\): \(k_\perp^2=k_x^2+k_y^2\) and \(k_{\parallel}=k_z\); while the slab spectral function is:
\begin{align}
S^{\rm slab}(k_\parallel, k_\perp) &= E^{\rm slab}(k_\parallel)
\frac{\delta(k_\perp)}{2\pi k_{\perp}}  \nonumber \\
&= 2 \frac{C^{\rm slab}(s) \delta B_{\rm slab}^2 l_{\rm slab}}{ \left[1 + (k_\parallel l_{\rm slab})^2 \right]^{s/2}}\frac{\delta(k_\perp)}{2\pi k_{\perp}}  ,
\label{eq:S_slab}
\end{align}
in which \(E^{\rm slab}(k_\parallel) \equiv E(q=0, s, l_{\rm slab},\delta B_{\rm slab}, k_\parallel) \), $l_{\rm slab}$ is the slab bend-over scale; and $\delta B_{\rm slab}$ is the (rms) strength of magnetic field fluctuations. The slope of the spectrum in the inertial range is controlled by specifying the parameter \(s\). 
The normalization constant reads:
\begin{equation}
C^{\rm slab}(s) \equiv C(q=0, s) = \frac{1}{2\sqrt{\pi}} \frac{\Gamma\left(\frac{s}{2}\right)}{\Gamma\left(\frac{s-1}{2}\right)} \ .
    \label{eq:C_slab}
\end{equation}
The correlation length for the slab model equals $l_{c, {\rm slab}}=2\pi C^{\rm slab}(s) l_{\rm slab}$. Note that, Eq. (\ref{eq:S_slab}), which contains the factor \(\delta(k_\perp) / 2\pi k_\perp = \delta(k_x) \delta(k_y) \) is a spectrum correctly normalized in a 3D \(k\)-space (e.g.,~Ref.~\cite{bieber1994proton}).

\subsubsection{2D model}
The presence of a background magnetic field ${\bf B}_0$ favors the turbulent cascade in the direction transverse to ${\bf B}_0$ \citep{shebalin1983anisotropy, OughtonEA15,goldreich1995toward}. The 2D model, proposed to take into account this perpendicular {\it complexity} \citep{matthaeus1990evidence}, is characterized by perpendicular magnetic field fluctuations ($\delta B_z = 0$), that depend only on the perpendicular coordinates, i.e., \(S^{\rm 
    2D}(k_\parallel, k_\perp) \propto \delta(k_\parallel)\).
 
As the slab above, the 2D spectral tensor follows also from the axisymmetric correlation tensor \cite{oughton1997general}:
\begin{equation}
   S^{\rm 2D}_{\ell m} \left({\bf k}\right) = \frac{S^{\rm 
    2D}(k_\parallel, k_\perp)}{2} \left(\delta_{\ell m} - \frac{k_{\ell} k_{m}}{k^2}\right) 
\label{eq:Plm_2D}
\end{equation}
for $\ell,m=x,y$ while $S_{\ell z}= S_{z m}=S_{zz}=0$, again, due to \(\delta B_z = 0\). In detail the 2D spectrum function reads:
\begin{align}
   S^{\rm 2D}(k_\parallel, k_\perp) &= E(q, s, l_{\rm 2D},\delta B_{\rm 2D}, k_\perp)\frac{2 \delta(k_\parallel) }{\pi k_\perp} \\
   &= \frac{4 C(q,s)}{\pi k_\perp} \frac{\delta B_{\rm 2D}^2 l_{\rm 2D} \left(k_\perp l_{\rm 2D}\right)^q}{\left(1 + k_\perp^2 l_{\rm 2D}^2 \right)^{(s+q)/2}}  \delta(k_\parallel)
\label{eq:S_2D}
\end{align}
where $l_{2D}$ is the 2D bend-over scale; $\delta B_{\rm 2D}$ is the magnetic field fluctuations strength (rms); $s$ is the slope of the spectrum in the inertial range; and $q$ is the spectral slope in the energy containing range. To satisfy homogeneity, we set \(q=3\) \citep{matthaeus2007spectral}. The 2D correlation length is $l_{c, {\rm 2D}}=\frac{4\sqrt{\pi}}{s+1} C(q, s) l_{\rm 2D}$, which for \(q=3\) and \(s=5/3\) reads \(l_{c, {\rm 2D}} \simeq 0.59 l_{\rm 2D}\). It is worth clarifying that by setting $q>0$ the correlation length of the 2D model is always finite (see also \cite{shalchi2009random,hussein2015influence}), opposite to some previous definitions of the 2D spectral form (e.g., \citep{shalchi2009}). These properties of the 2D case are discussed in detail in Ref. \citep{matthaeus2007spectral}.

\subsubsection{Slab/2D (composite) model}

In the composite slab/2D model, the slab component is combined with the 2D component \citep{bieber1994proton, bieber1996dominant, matthaeus2007spectral}. The composite correlation tensor is defined by:
\begin{equation}
   S^{\rm comp}_{\ell m} \left({\bf k}\right) = S^{\rm slab}_{\ell m} \left({\bf k}\right)  + S^{\rm 2D}_{\ell m} \left(\bvec{k}\right)
\label{eq:Plm_composite}
\end{equation}
where $S^{\rm slab}_{\ell m} \left({\bf k}\right)$ and $S^{\rm 2D}_{\ell m} \left({\bf k}\right)$ are given by Eqs. (\ref{eq:Plm_slab}) and (\ref{eq:Plm_2D}), respectively. In such composite model, one needs to specify the ratios of magnetic field fluctuation amplitudes $\delta B_{\rm 2D}^2/\delta B_{\rm slab}^2$ and of bend-over scales $l_{\rm 2D}/l_{\rm slab}$ in order to properly balance the two different components. Fiducial values, estimated by comparing the synthetic model with {\it in situ} observation of the solar wind \citep{bieber1996dominant} and numerical simulations \citep{hossain1995phenomenology}, are $\delta B_{\rm 2D}^2/\delta B_{\rm slab}^2=4$ ($80\%/20\%$) and $l_{\rm 2D}=0.1\, l_{\rm slab}$.

\subsubsection{Isotropic model}

The isotropic model is a fully three-dimensional (3D) model, in which all the three components of the magnetic field are present and the correlations and spectra 
depend only on the magnitude of spatial lag and $k=|\bvec{k}|$, 
respectively. 
The spectral tensor is defined as:
\begin{align}
    S^{\rm iso}_{\ell m}(\bvec{k} ) &= \frac{S^{\rm iso}(k)}{2} \left( \delta_{\ell m} - \frac{k_\ell k_m}{k^2} \right)
    \label{eq:Plm_iso}
\end{align}
with the spectrum function as in \citep{sonsrettee2015magnetic}:
\begin{align}
    S^{\rm iso}(k) &= E(q=4, s, k, l_{\rm iso},\delta B_{\rm iso} )\frac{1}{2\pi k^2} \nonumber \\
   &= \frac{C(q=4,s)}{\pi k^2} \delta B_{\rm iso}^2 l_{\rm iso} \frac{\left(k l_{\rm iso}\right)^4}{\left(1 + k^2 l_{\rm iso}^2 \right)^{s/2+2}} \, ,
    \label{eq:S_iso}
\end{align}
for $\ell,m=x,y,z$, $l_{\rm iso}$ being the bend-over scale; and $\delta B_{\rm iso}$ the rms strength of the magnetic field fluctuations.

The correlation length for the isotropic case reads:
\begin{align}
 \l_{c,{\rm iso}} &= \frac{4\pi}{\delta B_{\rm iso}^2} \int_0^\infty \dd r \int_0^\infty \frac{\sin(k r)}{kr} k^2 S^{\rm iso}(k) \dd k
 \nonumber \\
 &= \frac{4\pi}{s(s+2)} C(q=4,s) l_{\rm iso} \ .
 \label{eq:corr_length_isotropic}
\end{align}
which for \(s=5/3\) gives \(l_{c,{\rm iso}} \simeq 0.498 l_{\rm iso}\).

The spectra presented above comply with the turbulence homogeneity requirement, contrary to those used in \citep{hussein2015influence}, where the authors adopted similar shapes as in Eq. (\ref{eq:smooth_spectra}), but with \(q=3\) for the isotropic case (instead of, at least, 4), and \(q=2\) for the 2D case (instead of, at least 3). We remind the reader that this requirement arises from the fact that the diagonal elements of the correlation tensor must be even functions of \(\bvec{k}\) to satisfy \(S_{\ell m}(\bvec{k}) = S_{m \ell}(-\bvec{k})\) \citep{batchelor1982,matthaeus2007spectral}.

\subsection{Numerical methods}
\label{sec:numerical_methods}

Learning from technical difficulties of previous numerical studies \cite{shalchi2004nonlinear,de2007numerical}, we decided to use two independent numerical implementations. This redundancy enabled detailed comparisons of all results and minimized chances for subtle errors in codes that could affect results. One implementation uses and extends CRPropa \citep{alvesbatista2016CRPROPA}, a publicly available framework to study the propagation of cosmic rays, while the other one is based on a novel, locally developed code specifically tailored for the purposes of this work. Both codes perform two main duties: the first is to generate turbulent magnetic fields having
assigned properties and the second is to actually simulate the propagation of particles in those fields in order to determine 
diffusion coefficients as functions of energy. 

Magnetic fluctuations are generated on a \(k\)-space grid according to the models described in previous section: the amplitude of Fourier coefficients are chosen to reproduce the spectral tensors, i.e., the power spectrum, as described above, while phases are chosen randomly. Realizations are constructed in 
a large periodic simulation box. In order to satisfy the solenoidality condition, we construct the field through the magnetic potential in case of the 2D model \citep{matthaeus2007spectral,shalchi2009}, while the procedure described in Ref.~\citep{sonsrettee2015magnetic} is adopted for the 3D isotropic model. The solenoidal condition for the slab model of turbulence is fulfilled automatically. The wave number range is determined by the grid size and the box size and spans from $k_{\rm min}=k_0=2\pi/L_{\rm box}$ to the Nyquist wave number $k_{\rm max}= k_0 (N/2-1)$, where $L_{\rm box}$ is the box size along one axis and $N$ is the number of grid points along the same axis. The bend-over scale \(l\) is always chosen to be several times smaller than the box size in the isotropic case, or many times smaller in the slab and 2D cases, in order to approximate the 
property of homogeneity. 

Besides the obvious requirement to capture most of the turbulent energy within the box, larger values of the ratio \(L_{\rm box}/l\) ensure the presence of more correlation lengths inside the box, thus 
mitigating potential concerns due to periodic boundaries within the simulations. However, as the grid size and, consequently, the wave number range are limited by available computer memory (RAM), increasing the \(L_{\rm box}/l\) ratio shrinks the bandwidth of the simulated inertial range of turbulence. We recall that the inertial range is of prime interest here as for most astrophysical applications particle transport is dominated by resonant scattering off perturbations in such regime. Moreover well known standard trends of the parallel diffusion coefficient typically refer to the inertial range. In the case of isotropic turbulence, \(L_{\rm box}/l_{\rm iso} = 8\)  was found to be a good compromise. The used grid size of \(2048^3\) guarantees that particle scattering is properly described in at least 1 -- 1.5 decades in energy (or equivalently in wave number in the inertial range). For the 2D model and especially the slab model the ratio can be much larger,
effectively retaining several decades of the inertial range due to weaker memory constraints. In the slab case, the grid size is limited up to \(2^{30}\) and the ratio is chosen to be at most \(L_{\rm box}/l_{\rm slab} = 10^6\), while for the 2D case the grid size is \(2^{14}\) in each direction and \(L_{\rm box}/l_{\rm 2D} = 10^3\). The specific parameters for each set of performed simulations are given in the next section.

As the final step, the magnetic field in physical space is reconstructed by using fast Fourier transform. Within CRPropa, that is accomplished with FFTW \cite{FFTW05} while in the second code we use the procedure based on Ref.~\cite{press1992numerical}. CRPropa-based vector grids use single-precision floating point while the second implementation adopts double precision. This setting did not introduce discrepancies in the numerical results of particles diffusion. We remark that the box is 1D (along the $z$ axis) in the slab case, 2D (along $x$ and $y$) in the 2D case and 3D in both the slab/2D and the 3D-isotropic cases.

In addition, we employ interpolation methods to evaluate the magnetic field between the equidistant grid points. Results presented in this work have been obtained with a trilinear interpolation method \citep{birdsall1991}. We made sure that the results presented here are not affected by the usage of a more accurate yet slower 3D-cubic spline method (not shown here).

An alternative technique of synthesizing the required turbulent field 
is the so-called wavelet technique in which the field is generated directly, \textit{and only as a function of time}, at the particle position through a superposition of plane waves, thus eliminating the need for interpolations and in terms of performance, trading memory space for CPU time \citep{juneja1994synthetic,giacalone1999transport,casse2001transport,malara2016fast,pucci2016energetic}. A fast method that reduces the computational time and significantly extends the dynamical range of simulations has been recently proposed in Ref.~\citep{malara2016fast}. These methods usually do not set the wave number binning and the number of modes, and that may affect features of resonance interaction, that needs at least few wave numbers around the resonant one \citep{mace2012velocity,mertsch2019test}. Considering that, here we focus on the above mentioned grid method and leave for future work the comparison with a wavelet-based method.

Test-particle simulations are performed by numerically integrating Eq. (\ref{eq:lorentz}) adopting the symplectic Boris method \citep{HockneyEastwood,QinEA13-boris,webb2014symplectic,ripperda2018comprehensive}.
CRPropa also supports the Runge-Kutta method with the Cash-Karp coefficients, and some key results have been repeated with this integrator: no differences between the two have been found. For one set of parameters, the integration is repeated for an ensemble of $N_p$ particles injected homogeneously throughout the box with the initial directions of particles uniformly sampled on the unit 3D sphere. 
Additionally, in the case of CRPropa simulations, every simulation experiment is repeated and averaged over at least five different realizations of the turbulent field. Results obtained within the two numerical methods are consistent. The total number of particles injected in one run, $N_p$, depends on the specific context. However, at least $N_p = 2000$ have been used to ensure statistical convergence and, for several cases, we have injected $N_p=5000--10000$. We emphasize that, according to several tests we have carried out (not shown), injecting many particles ($\sim 8000$) in a single field realization is equivalent to injecting fewer particles ($\sim2000$) in several field realizations, provided that the single field realization contains many correlation lengths, i.e., $L_{\rm box}/l_c$ is large ($\gtrsim10$).  In other words, a statistically homogeneous sample is achieved in both ways, making them equivalent for the purposes of the present study.

\subsection{Calculation of diffusion coefficients}

The spatial diffusion coefficient, defined as
 \begin{equation}
     D_{xx} \equiv \lim_{t\rightarrow \infty} \frac{\langle (\Delta x (t))^2 \rangle}{2t}\, ,
 \end{equation}
is estimated 
numerically  through the practical formulation of the time-dependent running diffusion coefficient:
 \begin{equation}
     D_{xx}(t) = \frac{\langle (\Delta x(t))^2 \rangle}{2t}\, ,
     \label{eq:D_runn1}
 \end{equation}
 or in the derivative form:
 \begin{equation}
     d_{xx}(t) = \frac{1}{2}\frac{\dd}{\dd t}\langle (\Delta x(t))^2 \rangle\, .
  \label{eq:D_runn2}
 \end{equation}
Here $\Delta x(t)$ 
is the displacement of a particle during a time interval $t$.
A familiar approach to  
empirical determination of
the diffusion coefficient is to 
continuously compute 
$D_{xx}(t)$ or $d_{xx}(t)$
until a stable value is attained. 
Analogous definitions hold for the pitch angle diffusion coefficient $D_{\mu \mu}$. Eq.~(\ref{eq:D_runn1}) and Eq.~(\ref{eq:D_runn2}) are equivalent at sufficiently late times, when the saturation of the running diffusion coefficients is achieved. However, we emphasize that the time needed to reach such saturation may be different with the two definitions. In particular Eq.~(\ref{eq:D_runn1}), although easier to implement and less computationally demanding, takes longer time to reach the running diffusion coefficient's saturation with respect to Eq.~(\ref{eq:D_runn2}). Particles propagation is stopped once the diffusion plateau is reached.

\section{Theories of Diffusive Particle Transport}
\label{sec:theory}

In this section we revisit existing theoretical results on the diffusion of charged particles parallel and perpendicular to the regular (background) magnetic field and a limiting case of vanishing regular field, all of which are compared with our numerical results in the next section. For the sake of clarity, these cases are discussed in separate subsections.

\subsection{Physics of parallel scattering}

The basic picture of particle transport parallel to an ordered magnetic field, in the presence of a turbulent magnetic field can be described in a simple way: the unperturbed motion of a charged particle in the ordered magnetic field \(\bvec{B_0}\) simply consists of a circular orbit in the plane perpendicular to \(\bvec{B_0}\) and a uniform motion along \(\bvec{B_0}\), with velocity \(v_\parallel=v\mu\), where \(\mu\) is the cosine of the pitch angle, the angle between the regular field \(\bvec{B_0}\) and the particle velocity vector \(\bvec{v}\). The presence of turbulence dramatically changes this simple picture, in that if perturbations exist with wavelength comparable with the particle's gyroradius \(r_g\), the pitch angle resonantly changes. This phenomenon leads to pitch angle diffusion and, consequently, parallel diffusion in physical space.

This idea was successfully implemented for the first time in the mid-sixties, in the QLT \citep{jokipii1966cosmicray}. The basic assumption of the QLT is that the particle's perpendicular components of velocity and position can be approximated as the unperturbed values, while turbulence only affects the parallel velocity vector, provided the level of turbulence is very small compared with the original field \(\bvec{B_0}\). This assumption makes QLT increasingly more inaccurate for larger amplitudes of the fluctuations and for later times when the discrepancy between the unperturbed and real orbits is accumulating. Moreover, for magnetostatic turbulence, the resonance function is assumed to be Dirac's delta with the gyroresonance condition being $v_\parallel/\Omega_g = \mu r_g = 1/k_{\parallel}$, where $\Omega_g=qB/mc\gamma$ is the gyration frequency. The
wave number $k_\parallel= (\mu r_g)^{-1}$ is called 
the resonant wave number. 

From the Fokker-Planck equation and using the assumptions mentioned above, QLT returns the pitch angle diffusion coefficient \(D_{\mu\mu}\) in the following form:
\begin{equation}
    D_{\mu\mu} = \frac{\pi \Omega_g (1-\mu^2)}{2 B_0^2 v \mu} E^{\rm slab}\left(k_\parallel = \frac{\Omega_g}{v\mu}\right) \ .
    \label{eq:Dmumu_SLAB}
\end{equation}

By inserting the slab model spectrum from Eq.~(\ref{eq:S_slab}), it is straightforward to obtain:
\begin{equation}
    D_{\mu\mu} = \frac{\pi C^{\rm slab}(s) v \,\delta\tilde{B}_{\rm slab}^2 }{l_{\rm slab}} \frac{\left(1-\mu^2\right) \mu^{s-1}}{\left(1+\mu^2\tilde{r}_g^2\right)^{s/2}}\tilde{r}_g^{s-2},
    \label{eq:Dmumu_SLAB2}
\end{equation}
where $\delta\tilde{B}_{\rm slab}=\delta B_{\rm slab}/B_0$ and $\tilde{r}_g=r_g/l_{\rm slab}$.
The pitch angle diffusion coefficient controls parallel spatial diffusion \(D_\parallel\):
\begin{equation}
    \lambda_\parallel = \frac{3 D_\parallel}{v} =\frac{3 v}{8} \int_{-1}^1 d\mu \frac{(1-\mu^2)^2}{D_{\mu\mu}},
    \label{eq:lambda_par}
\end{equation}
where \(\lambda_\parallel\) is known as the parallel mean free path\citep{jokipii1966cosmicray,HasselmannWibberenz70}.

Finally, by using Eq. (\ref{eq:Dmumu_SLAB2}), the parallel mean free path reads:
\begin{equation}
\begin{split}
    \lambda_\parallel = & \frac{6 l_{\rm slab}}{8 \pi C^{\rm slab}(s) \, \delta\tilde{B}_{\rm slab}^2} \tilde{r}_g^{2- s} \\ 
    & \left[ \frac{1}{2-s} {}_2F_1\left(1-\frac{s}{2},-\frac{s}{2},2-\frac{s}{2}, -\tilde{r}_g^{2}\right) \right. \\
    & \left. - \frac{1}{4-s} {}_2F_1\left(2-\frac{s}{2},-\frac{s}{2},3-\frac{s}{2}, -\tilde{r}_g^{2}\right) \right] ,
    \label{eq:lambda_par_SLAB}
\end{split}
\end{equation}
where \({}_2F_1\) are the ordinary hypergeometric functions.

For \(s=5/3\), it is easy to recover the known scaling $\lambda_\parallel\sim \tilde{r}_g^{\frac{1}{3}}$ and $\lambda_\parallel\sim \tilde{r}_g^{2}$, in the limit of small and large rigidities, respectively.

QLT is characterised by several limitations: as described above, it does not describe well diffusion as the strength of the turbulent fluctuations increases. Furthermore, as it is based on a sharp resonance function, it leads to the well known $90^\circ$ ($\mu\sim0$) problem, namely a divergent parallel mean free path at $\mu=0$, for any reasonable spectral slope of the turbulence spectrum~\citep{bieber1994proton}. Besides that, for the perpendicular transport, which is the topic of the next subsection, QLT fails to reproduce numerical results. In order to cope with these issues, nonlinear theories were developed through the years. 

Here we provide a brief overview of one weakly nonlinear theory of parallel transport proposed in \cite{shalchi2005second} and named as second order QLTs (SO-QLTs). This formulation adopts the second order correction of the unperturbed orbit. The resonance function is broadened and the broadening width is taken as \(\sigma_z \sim \delta B_{\rm slab}/B_0\). The pitch angle diffusion coefficient then reads:
\begin{align}
    D_{\mu\mu} =& \frac{\Omega_g^2 (1-\mu^2)}{2 B_0^2} \int_0^{\infty} d k_{\parallel} E^{\rm slab}(k_\parallel) \int_0^{\infty} dt \nonumber \\
   &\left\{ \cos[(k_\parallel v \mu + \Omega_g)t] + \right. \nonumber \\
   & \left. \quad \cos[(k_\parallel v \mu - \Omega_g)t] \right\} \exp{\left[-\frac{\sigma_z^2(t)k_\parallel^2}{2}\right]}  \ .
\label{eq:Dmumu_SLAB_SOQLT}
\end{align}
A closure for the broadening width, that in general depends on both time and pitch angle, has to be chosen to evaluate the pitch angle scattering. Here, we follow the so-called $90^\circ$-late-time approximation~\citep{shalchi2009}, namely:
\begin{equation}
  \sigma_z^2(t, \mu)\sim \frac{v^2 \delta\tilde{B}_{\rm slab}^2 t^2}{2}.
  \label{eq:sigma_SLAB_SOQLT}
\end{equation}
By combining Eqs.~(\ref{eq:Dmumu_SLAB_SOQLT}--\ref{eq:sigma_SLAB_SOQLT}) and by adopting the spectrum of Eq.~(\ref{eq:S_slab}), it is possible to numerically compute $D_{\mu\mu}$ and, thus, $\lambda_\parallel$.

\subsection{Physics of perpendicular scattering}

The transport perpendicular to the background magnetic field is usually interpreted,
in the absence of Coulomb collisions, as being 
the result of the magnetic field line random walk (FLRW), under the quasilinear assumption that particle gyrocenters follow magnetic field lines \citep{jokipii1966cosmicray}. Although providing a simple physical picture, even qualitative predictions turn out to be not consistent with numerical experiments \citep{giacalone1999transport}. Indeed, the observed perpendicular diffusion is smaller than the FLRW expectation, especially at small rigidities (for example, see Fig.~7 of Ref.~\citep{giacalone1999transport}). The basic physics responsible 
for this is the profound effect of parallel scattering 
on the behavior particles as they move along field 
lines~\cite{Urch77,QinEA02-apj}.  

The failure of QLT for perpendicular transport has motivated the development of several nonlinear theories. Within the context of nonlinear closures approximations, we here describe the nonlinear guiding center theory~\citep{matthaeus2003nonlinear} that has showed both good agreement with numerical results in case of the slab/2D turbulence, and served as a basis for later improvements and variations~\citep{shalchi2009}.

The NLGC theory allows to compute the perpendicular diffusion coefficient $D_\perp$ for a given parallel diffusion coefficient $D_\parallel$. Its derivation starts from the perpendicular diffusion coefficient, defined in the Tylor-Green-Kubo (TGK) formulation~\citep{kubo1957statistical}:
 \begin{equation}
 D_{xx} = \int_0^\infty dt \langle v_x(t) v_x(0)\rangle \, .
 \end{equation}

By starting from the particle motion equations and by assuming slow perturbations, one easily finds the perpendicular speed of the particle gyrocenters:
\begin{eqnarray}
 && \tilde{v}_x (t) = v_z(t) \frac{\delta B_x}{B_0} - v_x(t) \frac{\delta B_z}{B_0}, \\
  && \tilde{v}_y (t) = v_z(t) \frac{\delta B_y}{B_0} - v_y(t) \frac{\delta B_z}{B_0}, 
 \label{eq:gyrospeed}
\end{eqnarray}
being $\tilde{v}_i (t)=1/T \int_t^{t+T}d\tau\, v_i(\tau)$ ($i=x,y,z$) and $T=2\pi/\Omega_g$.

The standard NLGC theory, initially derived for the composite (slab/2D) case, neglects the component of the fluctuations parallel to $z$, setting 
$\delta B_z=0$. In the following, we apply such theoretical development to both slab/2D and 3D isotropic cases, although in the latter $\delta B_z\ne0$. In fact we have generalised the NLGC theory to the case in which $\delta B_z\ne0$ (not presented here) but this development has no practical impact on the results, in that its effect is totally negligible. In any case, the focus of this article is on the comparison of numerical results with previous theoretical formulations rather presenting new theories, which will be discussed elsewhere.

By neglecting $\delta B_z$ terms, we get:
 \begin{align}
 D_{xx} = \int_0^\infty dt &\langle \tilde{v}_x(t) \tilde{v}_x(0)\rangle = \nonumber \\ 
 =\frac{a^2}{B_0^2} \int_0^\infty dt & \langle \tilde{v}_z(t) \tilde{v}_z(0) \delta B_x({\bf r},t) \delta B_x({\bf r},0) \rangle
 \end{align}
where two integrals containing mixed correlators have been neglected ($\sim\tilde{v}_z\tilde{v}_x\delta B_z \delta B_x$).
The numerical parameter \(a\) has been introduced to describe the departure of the guiding centers from following field lines, which is \(a=1\) in case where field lines are being followed, while \(a = 1/\sqrt{3}\) is set to better match results of simulations in the slab/2D composite model of turbulence \cite{matthaeus2003nonlinear,shalchi2004nonlinear}.

The derivation proceeds as follows:
\begin{enumerate}
    \item Each fourth order correlator is separated into two second order correlators.
    \item The velocity correlation function is modeled as an exponential in time, 
    \begin{equation}
        \langle \tilde{v}_{z}(t) \tilde{v}_{z}(0)\rangle = \frac{v^2}{3} \exp\left[- \frac{vt}{\lambda_{\parallel}}\right],
        \label{eq:exp}
    \end{equation}
    an ans\"atz that automatically satisfies the TGK formula for parallel diffusion. 
    When needed, an analogous assumption is made for the perpendicular 
    velocity autocorrelation  
    $\langle \tilde{v}_{x}(t) \tilde{v}_{x}(0)\rangle$.
    \item Corrsin's hypothesis \citep{corrsin1959progress} and turbulence spatial homogeneity are invoked to relate the Lagrangian 
    magnetic field autocorrelation function to the appropriate element of
    the energy spectrum and the characteristic function of the random displacements
    ${\bf \Delta r}(t)$,
    as:
      \begin{equation}\begin{split}
        R_{xx} &= \langle \delta B_{x}({\bf r}(t),t) \delta B_{x}({\bf r}(0),0) \rangle = \\
        &\int d^3k\, S_{xx}({\bf k}) \langle e^{\ii {\bf k}\cdot \Delta {\bf r}}  \rangle
    \end{split}\end{equation}  
    where $S_{xx}({\bf k})$ is the spectral tensor 
    for the magnetostatic case.  This essentially is a closure for
    the Langrangian correlation in terms of the Eulerian correlation.
    \item The ensemble average containing spatial displacements is evaluated by assuming a Gaussian distribution of particles and a diffusive closure for the mean displacements.
\end{enumerate}
Taking all of these into account, one arrives to the final expression:
\begin{align}
    D_{xx} =& \frac{a^2 v^2}{3 B_0^2}\left. \int \dd^3 k \frac{S_{xx}(\bvec{k})}{\frac{v}{\lambda_\parallel} + D_{xx} k_x^2 + D_{yy} k_y^2 + D_{zz} k_z^2} \right. .
    \label{eq:NLGC}
\end{align}

In the composite slab/2D case, one gets:
\begin{align}
\frac{\lambda_\perp}{\lambda_\parallel} =& \frac{a^2}{2} \left[\left(\frac{\delta B_{\rm slab}}{B_0}\right)^2 K(s_{\rm slab},q_{\rm slab},\alpha_\parallel) \right. \nonumber \\
&\left. + \left(\frac{\delta B_{\rm 2D}}{B_0}\right)^2 K(s_{\rm 2D},q_{\rm 2D},\alpha_\perp)\right],
     \label{eq:SLAB2D-NLGC}
\end{align}
where $q_{\rm slab}=0$, $\alpha_\parallel = \lambda_\parallel^2/3l_{\rm slab}^2$, $
\alpha_\perp = \lambda_\parallel\lambda_\perp/3l_{\rm 2D}^2$, and
\begin{equation}
    K(s,q,\alpha) = \frac{s-1}{s+q}\, {}_2F_1 \left(1, \frac{q + 1}{2}; \frac{s + q}{2} + 1, 1 - \alpha\right).
\end{equation}

Note that the solution of Eq.~(\ref{eq:SLAB2D-NLGC}) requires to numerically calculate the roots of a transcendental expression. 

By employing the same procedure as for the 3D isotropic turbulence model [Eq.~(\ref{eq:S_iso})] and by neglecting $\delta B_z$, we find:
\begin{align}
   \frac{\lambda_\perp}{\lambda_\parallel} =& \frac{a^2 C(q=4,s)}{2} \left(\frac{\delta B_{\rm iso}}{B_0}\right)^2 \left.I\left(s,\alpha'_\parallel \frac{\lambda_\perp}{\lambda_\parallel}\alpha'_\parallel\right) \right.,
\label{eq:3DISO-NLGC}
\end{align}
being $\alpha'_\parallel = \lambda_\parallel^2/3 l_{\rm iso}^2$ and
\begin{align}
  &I\left(s,A_1,A_2\right) = \int_0^\infty d\eta_\perp \eta_\perp \int_{-\infty}^\infty d\eta_{\parallel} \times \nonumber \\
  &\quad \frac{\eta_\perp^2+ 2\eta_\parallel^2} {\left(1+\eta_\perp^2+\eta_\parallel^2\right)^{s/2+2}\left(1+A_1\eta_\perp^2+A_2\eta_\parallel^2\right)} \, .
\end{align}

We conclude this section by briefly revisiting the unified nonlinear theory (UNLT). The derivation of UNLT differs from the standard NLGC in that NLGC adopts an exponential functional form for the velocity autocorrelation as in Eq. (\ref{eq:exp}), while UNLT solves an auxiliary Fokker-Planck equation to determine this correlation.

As proposed in Ref. \cite{shalchi2010unified}, we compute the perpendicular diffusion coefficient as:
\begin{equation}
  D_{\perp} = \frac{a^2 v^2}{3 B_0^2} \int \dd^3 k \frac{S_{xx}(\bvec{k})}{\frac{v}{\lambda_\parallel} + \frac{4}{3}D_{\perp} k_{\perp}^2 + A(\bvec{k}) } \label{eq:UNLT},
\end{equation}
where $A(\bvec{k})=(v^2/3D_\perp)\left(k_\parallel/k_\perp\right)^2$ substitutes the term $D_\parallel k_\parallel^2$ of standard NLGC and there is a factor $4/3$ multiplying the term proportional to $D_\perp$. 

For the composite model of turbulence, UNLT provides:
\begin{equation}
  D_{\perp} = \frac{a^2 v^2}{3 B_0^2} \int \dd^3 k \frac{S^{\rm 2D}_{xx}(\bvec{k})}{\frac{v}{\lambda_\parallel} + \frac{4}{3}D_{\perp} k_{\perp}^2 + A(\bvec{k}) } \label{eq:SLAB2D-UNLT}
\end{equation}
that, after few algebraic steps, simplifies to:
\begin{equation}
\frac{\lambda_\perp}{\lambda_\parallel} = \frac{a^2}{2} \left(\frac{\delta B_{\rm 2D}}{B_0}\right)^2 K(s_{\rm 2D},q_{\rm 2D},\alpha_\perp'') ,
 \label{eq:SLAB2D-UNLT-def}
 \end{equation}
being $\alpha_\perp''=4/3\alpha_\perp$. Note that, in this case, the slab component contributes only in characterizing the parallel mean free path while, at variance with respect to NLGC, does not explicitly appear at the rhs of Eq. (\ref{eq:SLAB2D-UNLT-def}).

On the other hand, for the isotropic turbulence model [Eq. (\ref{eq:S_iso})], UNLT gives the following implicit equation:
\begin{equation}
1 = \frac{a^2}{2} \left(\frac{\delta B_{\rm iso}}{B_0}\right)^2 C(q=4,s) J'\left(s,\alpha_\parallel'',\frac{\lambda_\perp}{\lambda_\parallel}\right)
\label{eq:3DISO-UNLT}
\end{equation}
being $\alpha_\parallel''=4/3\alpha_\parallel'$ and
\begin{align}
  &J'\left(s,\alpha,r\right) = \int_0^\infty d\eta_\perp \eta_\perp^3 \int_{-\infty}^\infty d\eta_{\parallel} \times \nonumber\\
  &\quad \frac{\eta_\perp^2+2\eta_\parallel^2}{\left(1+\eta_\perp^2+\eta_\parallel^2\right)^{s/2+2}\left(r\eta_\perp^2(1+r\alpha\eta_\perp^2)+\eta_\parallel^2\right)} \, .
\end{align}

\subsection{The case $\delta B/B_0\rightarrow \infty$\label{sec:subedi_theory}}

The particle transport in case without any (globally) ordered magnetic field has been recently considered \citep{subedi2017charged}. The high-energy limit, i.e., when the particle gyroradius exceeds the correlation length, is known from literature (for example, \cite{aloisio2004diffusive}) and reads:
\begin{equation}
    D_{xx} = D_{yy} = D_{zz} = \frac{r_g^2 v}{2 l_{c,\rm iso}} \ .
    \label{eq:lambdaiso_highenergy}
\end{equation}
This expression reflects the fact that the particle's successive deflections are uncorrelated, hence, the deflection angle performs a Brownian motion. In this case, 
the decorrelation required for convergence of the TGK formula is accomplished 
due to straight line particle trajectory through the magnetic
fluctuations, which themselves 
decorrelate over a coherence length $l_c$.  

In the low-energy limit, although there is no regular magnetic field present at the largest scale given by the simulation box, particles with gyroradii smaller than the correlation length approximately gyrate along the local magnetic field. The direction/structure of the local field is determined by fluctuations near the scale of the correlation length $l_c$, where most of the turbulence energy resides. Both FLRW and resonant scattering in the local field affect particle diffusion. Here we quote the arguments of Ref. \citep{subedi2017charged} and introduce the final analytical expressions useful for practical proposes since they were omitted in the original work.

The starting point is the QLT pitch angle diffusion coefficient, properly modified to take into account the scattering at $\mu\sim 0$, that is,
\begin{equation}
D_{\mu\mu} = \frac{\pi \alpha^2 (1-\mu^2)}{v} E_y \left(k_z=\frac{\Omega_g}{v}\right).
\end{equation}
Here $E_y (k_z)=\int dk_x dk_y S_{yy}({\bf k})$, $\alpha=q/m\gamma c$ and $\Omega_g=\alpha B$, where $B$ is the local field. 
Note that the reduced spectrum is the Fourier transform of 
the corresponding spatial correlation function Eq. (\ref{eq:corr_tensor})
with respect to a single coordinate, the other 
spatial lags set to zero, i.e., 
$E_y(k_z) = \frac{1}{2\pi} \int dz R_{yy}(0,0,z)\exp{(-ik_z z)}$.
We remark that the 
one-dimensional reduced spectrum $E_y (k_z)$ represents the energy associated with $y$ fluctuations reduced by integrating on $k_x$ and $k_y$; 
this differs from the omni-directional spectral energy $E(k)=2\pi k^2 S^{\rm iso}(k)$, though they have the same dimensions and are functionally related \citep{batchelor1982}.  Moreover, as discussed above, the resonance is provided by the local field $B$. 

By adopting the spectrum in Eq. (\ref{eq:S_iso}) and by averaging the pitch angle diffusion coefficient on a Maxwellian distribution of the local magnetic field strength, we get:
\begin{equation}
    \bar{D}_{\mu\mu} = \frac{\sqrt{2\pi} C(4,s)
    (1-\mu^2)}{s(s+2)} \frac{v}{l_{\rm iso}} \left(\frac{l_{\rm iso}}{r_g}\right)^2 I\left(s,\frac{\sqrt{3}r_g}{l_{\rm iso}} \right),
\end{equation}
where the gyroradius $r_g$ is computed using $\delta B_{\rm rms}$ and
\begin{equation}
I(s,R) = \int_0^\infty d\xi\, \xi^2 e^{-\xi^2/2} \frac{1 + (1+s)\left(\frac{\xi}{R}\right)^2}{\left(1 + \left(\frac{\xi}{R}\right)^2 \right)^{\frac{s}{2}+1}} \, .
\end{equation}

\begin{figure}[!t]
    \centering
    \includegraphics[width=\columnwidth]{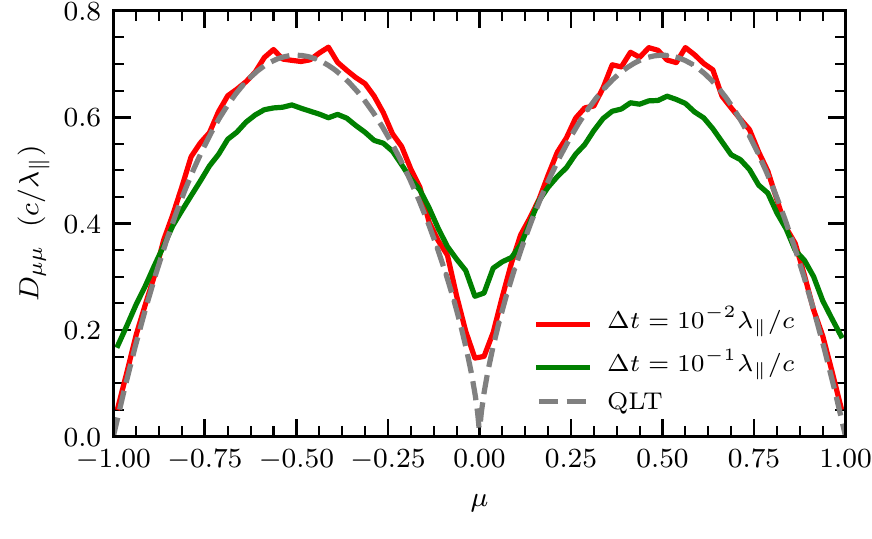}
    \caption{Pitch angle diffusion coefficient $D_{\mu\mu}$ for slab turbulence, as a function of ${\mu}$, evaluated at the time lag $\Delta t = 10^{-2} \lambda_\parallel/c$ (solid red) and $\Delta t = 10^{-1} \lambda_\parallel/c$ (solid green), for the case $r_L/l_{\rm slab}=0.02$ and ${\delta B}_{\rm slab}/B_0=10^{-2}$. The dashed gray line is the theoretical prediction of QLT (Eq.~\ref{eq:Dmumu_SLAB2}).}
    \label{fig:Dmumu-SLAB}
\end{figure}

Finally, by expanding $\bar{D}_{\mu\mu}$ in the limit $r_g/l_{\rm iso} \ll 1$ and inserting in Eq. (\ref{eq:lambda_par}), we obtain:
\begin{equation}
\lambda_{\rm iso} = \frac{3 l_{\rm iso}}{8} A(s_{\rm iso}) \left(\frac{r_g}{l_{\rm iso}}\right)^{2-s},
\label{eq:lambdaiso}
\end{equation}
where $\lambda_{\rm iso}=\lambda_{\parallel}=\lambda_{xx} = \lambda_{yy}=\lambda_{zz}$ and 
\begin{equation}
    A(s) = \left(\frac{2}{3}\right)^{s/2} \frac{s(s+2)}{s+1}\frac{\Gamma\left(\frac{s-1}{2}\right)}{\Gamma\left(\frac{s}{2} +2\right)\Gamma\left(\frac{3-s}{2}\right)} \, .
\end{equation}
In the case of Kolmogorov slope, we recover the scaling $\lambda_{\rm iso}\sim (r_g/l_{\rm iso})^{1/3}$.
For a Kraichnan inertial range with slope $-3/2$ this becomes
$\lambda_{\rm iso}\sim (r_g/l_{\rm iso})^{\frac{1}{2}}$.

In Ref. \citep{subedi2017charged}, authors also developed an extended low-energy theory, based on the idea that particles make an unperturbed orbit along the local mean magnetic field. Hence, the mean perpendicular displacement is about $r_g$ and $E_y (k_z)$ is evaluated as $E_y (k_z)=\int dk_x dk_y S_{yy}({\bf k})e^{-k_\perp^2 r_g^2/6}$. Again, after expanding the expression in the limit $r_g/l_{\rm iso} \ll 1$, we retrieve:
\begin{equation}
\lambda_{\rm iso} = \frac{3 l_{\rm iso}}{8}  \frac{A(s_{\rm iso})}{B(s_{\rm iso})} \left(\frac{r_g}{l_{\rm iso}}\right)^{2-s},
\label{eq:lambdaisoext}
\end{equation}
in which \(A(s)\) is the same as above, while \(B(s)\) reads:
\begin{equation}
\begin{split}
 B(s) = &\frac{3^2}{2^{10}}\frac{1}{(s-2)(s+1)}\left[2^5 s(4s-5) - 221 \right. + \\ 
& \left. 3^{1-s}(13+8s)\,{}_2F_1\left(-\frac{1}{2},\frac{3-s}{2},\frac{1}{2}, -8\right) \right].
\end{split}
\end{equation}
From \(B(s)\) one can easily check that Eqs.(\ref{eq:lambdaiso}) and  (\ref{eq:lambdaisoext}) differ by $15\%$ for \(s=5/3\) and \(20\%\) for \(s=3/2\).

The explicit expressions derived here for the low-energy theory Eq. (\ref{eq:lambdaiso}),
and for the extended low-energy theory, 
Eq. (\ref{eq:lambdaisoext}), 
may be used to develop transport models in astrophysical environments 
where $\delta B \gg B_0$. This would be appropriate 
close to CR sources such as supernovae shocks \citep{blasi2013origin,caprioli2013cosmicray}.

\section{Numerical results and discussion}
\label{sec:results}

In this Section we discuss the results of numerical simulations for several models of turbulence discussed above. 

\subsection{Slab model}

\begin{figure}
\centering
    \includegraphics[width=\columnwidth]{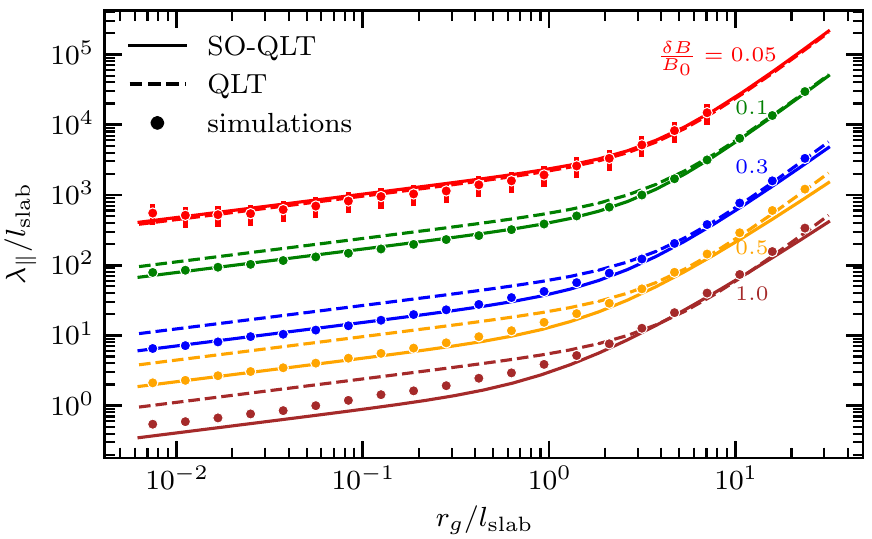}
    \caption{Parallel mean free path for slab turbulence as a function of rigidity. The points refer to the results of our test particle simulations for different values of \(\delta B/B_0\), as indicated [from \(0.05\) (top points) to \(1\) (bottom points)]. The dashed lines depict the corresponding predictions from QLT (Eq. \ref{eq:lambda_par_SLAB}). The solid lines come from the second order QLT (Eq. \ref{eq:Dmumu_SLAB_SOQLT}--\ref{eq:sigma_SLAB_SOQLT}). The spectral index of turbulence is \(s=5/3\).}
    \label{fig:diff_coef-SLAB}
\end{figure}

For the slab model of the turbulent magnetic field, the power spectrum tensor is given by Eq. (\ref{eq:S_slab}). In this case, our simulations are characterised by a grid size $N_z=2^{25}$, while $l_{\rm slab}=10^{-4}L_{z}$ (for 
box size $L_z = 400 {\rm kpc}$) and $B_0=1 \mu G$. The spectral slope in the inertial range is $s=5/3$. Note that we set $l_{\rm slab}\ll L_z$ in such a way that the parallel mean path $\lambda_\parallel = 3 D_\parallel/c$ is 
well contained inside the numerical box for most of the considered gyroradii and $\delta B_{\rm slab}/B_0$ values. 

Figure \ref{fig:Dmumu-SLAB} shows the pitch angle diffusion coefficient, $D_{\mu\mu}$, as a function of the pitch angle $\mu$, for the case $\delta B_{\rm slab}/B_0=10^{-2}$ and for the gyroradius $r_g=0.02 \, l_{\rm slab}$. For a given time-lag $\Delta t$, in order to evaluate $D_{\mu\mu}(\mu)$ we selected a temporal window of width $T=t_{\rm max}/5$, being $t_{\rm max}\simeq 10 \lambda_\parallel/c$ the maximum time of the simulation. Within the selected window, we increased the statistics by shifting the origin $t_0$ in $\Delta \mu = \mu (t_0 +\Delta t) -\mu (t_0)$. No differences in $D_{\mu\mu}$ are recovered by changing the window on which we apply this procedure. Finally, the histogram of $\Delta \mu$ with respect to $\mu (t)$ is calculated. 
\begin{figure}[!htb]
    \centering
    \includegraphics[width=\columnwidth]{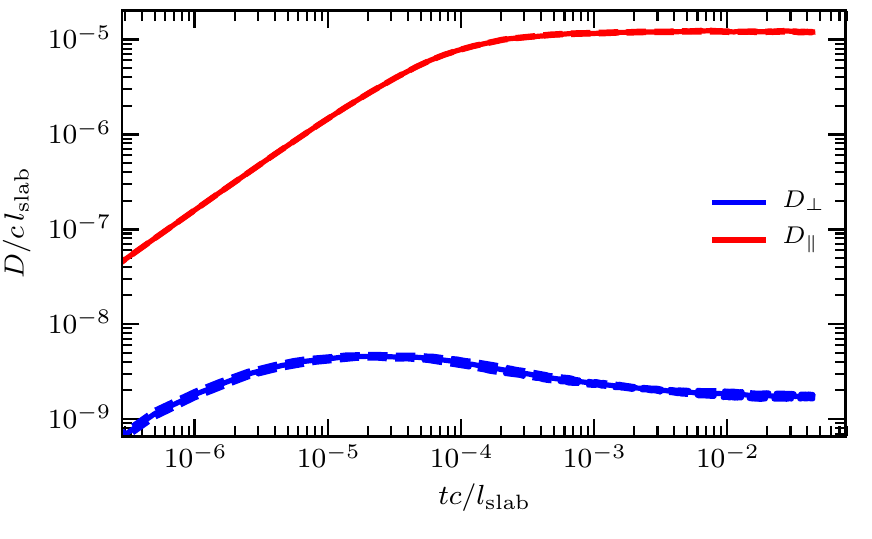}
    \caption{Time evolution of $D_\parallel$ (red) and $D_\perp$ (blue), for the case of slab/2D field turbulence where the 2D to slab power ratio is $80\%-20\%$, and $r_g/l_{\rm slab}=1.6 \times 10^{-2}$. The dashed lines refer to standard deviations.}
    \label{fig:timeev-SLAB2D}
\end{figure}

At a small time lag $\Delta t=10^{-2}\lambda_\parallel/c$, where  $\lambda_\parallel=3D_{zz}/c$ is evaluated from the running diffusion coefficient of the numerical simulation, simulations (red line) are in agreement with the QLT prediction, given by Eq.~(\ref{eq:Dmumu_SLAB2}) (dashed gray line). When increasing the time-lag $\Delta t=10^{-1}\lambda_\parallel/c$ (green line), a discrepancy at $\mu\sim 0$, typical of resonance broadening, is found. Such difference confirms that QLT validity is limited not only to the small amplitude case, but also to the small time-lag regime, in other words, the pitch angle has to accumulate small changes.

The next test of our numerical procedures is to compare numerical results of the spatial diffusion coefficient \(D_{zz}\) in the slab turbulence with the theoretical expectations of QLT, Eq.~(\ref{eq:lambda_par_SLAB}), and second order QLT (SO-QLT) which we obtained by inserting Eqs.~(\ref{eq:Dmumu_SLAB_SOQLT}--\ref{eq:sigma_SLAB_SOQLT}) in Eq.~(\ref{eq:lambda_par}).

\begin{figure*}[!htb]
    \centering
    \includegraphics[width=1\textwidth]{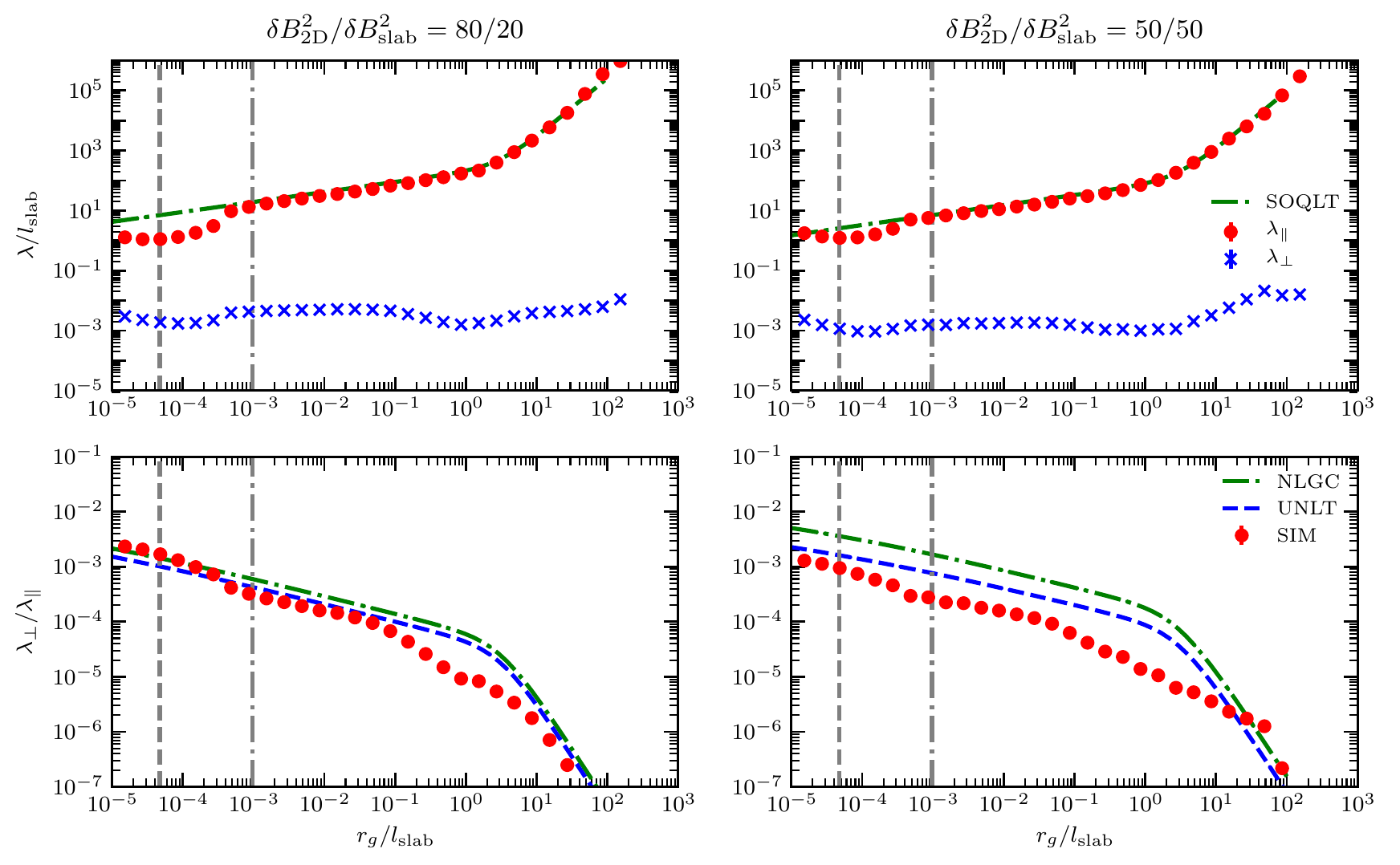}
    \caption{Slab/2D field model, assuming a slab power ratio of $80\%-20\%$ (left) and $50\%-50\%$ (right). The top rows show $\lambda_{\parallel}$ (red dots) and $\lambda_{\perp}$ (blue crosses) as a function of $r_g/l_{\rm slab}$. 
    Note that $l_{\rm 2D} = 0.1 l_{\rm slab}$.
    The green dashed lines correspond to the theoretical predictions for $\lambda_{\parallel}$ obtained with SO-QLT [Eqs.~(\ref{eq:Dmumu_SLAB_SOQLT}--\ref{eq:sigma_SLAB_SOQLT})]. The bottom rows display the ratio $\lambda_{\perp}/\lambda_{\parallel}$ as a function of $r_g/l_{\rm slab}$. The red symbols are values obtained from the simulations. The theoretical lines of the bottom panels are being calculated from the NLGC [Eq. (\ref{eq:SLAB2D-NLGC}), green] and UNL [Eq.~(\ref{eq:SLAB2D-UNLT-def}), blue] theories with \(a^2 = 1/3\) where, for the $D_{\parallel}$ input, SO-QLT is used.}
    \label{fig:NLGCsims-SLAB2D}
\end{figure*}

We adopted a grid size of \(N_z = 2^{30}\) grid points within \(L_z = 384 {\rm kpc}\), while \(l_{\rm slab} = 10^{-6} L_z\) to minimize the impact of periodicity of the box on the results. The mean field strength is set to \(1 {\rm \mu G}\) while the turbulent component ranges from $\delta B_{\rm slab} = 0.05$ to \(1 {\rm \mu G}\). Figure ~\ref{fig:diff_coef-SLAB} displays $\lambda_\parallel/l_{\rm slab}$ as a function of $r_g/l_{\rm slab}$, for the listed values of $\delta B_{\rm slab}/B_0$. The results from numerical simulations (points) are showed together with the QLT (dashed lines) and the SO-QLT (solid lines) predictions. 
The results are, as expected, in excellent agreement with both QLT and SO-QLT when the turbulent field fluctuations are small, while the SO-QLT performs better for the cases in which $\delta B_{\rm slab}/B_0$ is closer to unity. 
It is worth mentioning that, for larger values of $\delta B_{\rm slab}/B_0$, the simulation results tend to lie between the QLT and SO-QLT predictions. 

In the above results
we noticed a slight underestimate of the diffusion coefficient compared with the theory for the case $\delta B_{\rm slab}/B_0 = 0.05$, due to accumulation of numerical errors in the Boris method of the particle trajectory integration. When the Runge-Kutta method is used in the same setup, the diffusion coefficient is slightly overestimated. This difference in the integration method happens only in this stringent case in which \(\gtrsim 10^4 \Omega_g^{-1}\) are needed to reach the diffusion plateau. This uncertainty is represented through the error bars of the red points.

\begin{figure}[!htb]
    \includegraphics[width=\columnwidth]{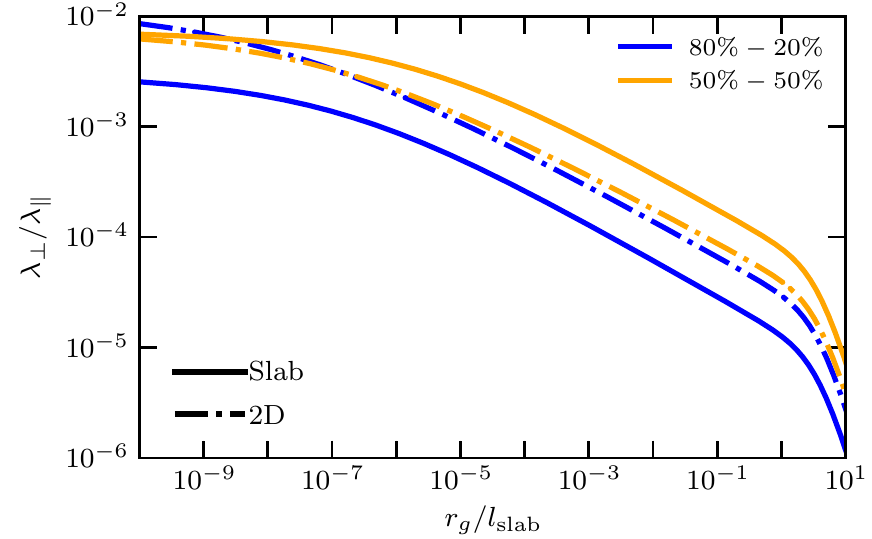}
    \caption{Energy dependence of the two terms in Eq. (\ref{eq:SLAB2D-NLGC}) for slab/2D turbulence with $\delta B/B_0=0.3$ and $l_{\rm 2D}=0.1 l_{\rm slab}$. The two terms are referred to as slab term (solid) and 2D term (dot dashed) respectively and are displayed for the $80\%-20\%$ (blue) and $50\%-50\%$ (orange) cases. The sum of the two terms coincides with the formal solution of the NLGC theory. The parallel diffusion coefficient has been computed by means of QLT.}
    \label{fig:NLGCtheory-SLAB2D_terms}
\end{figure}

\subsection{Slab/2D (composite) model}
Here we focus on simulations of particle transport in slab/2D models of turbulence [Eq. (\ref{eq:Plm_composite})]. The interest in this case is twofold. First, it represents an additional   test of our numerical approach to cases that have been studied in the literature both with simulations and with NLGC \citep{matthaeus2003nonlinear}. Second, and more importantly, by means of these simulations we address the issue of whether the ratio $\lambda_{\perp}/\lambda_{\parallel}$ becomes constant in the limit of small particle rigidity. Indeed, a constant ratio $\lambda_{\perp}/\lambda_{\parallel}$ is expected both in NLGC and {\it universal nonlinear} theories \citep{shalchi2010unified,shalchi2014universality,shalchi2015perpendicular}, but the limited dynamical range of existing simulations has limited the validation of this theoretical prediction \citep{de2007numerical, hussein2015influence}. 

The 3D numerical box size has been chosen to have $L_z = 400 {\rm kpc}$, $L_x=L_y=10^{-2}L_z$, and it has been discretized with $N_x=N_y=2^{14}$ and $N_z=2^{25}$ grid points. The magnetic perturbation amplitude is  ${\delta B}/B_0=0.3$, where 
${\delta B}^2 = \delta B_{\rm slab}^2 + \delta B_{\rm 2D}^2$ and $B_0=1 \mu G$. The slab bend-over length is $l_{\rm slab}/L_{z} = 10^{-3}$, while $l_{\rm 2D}/L_{x}=10^{-2}$, i.e. $l_{\rm 2D}=0.1 l_{\rm slab}$. The inertial range spectral slope is $s_{\rm slab}=s_{\rm 2D}=5/3$, while $q_{\rm 2D}=3$. 

Figure \ref{fig:timeev-SLAB2D} displays the time evolution of the running diffusion coefficients
$D_\perp$ and $D_\parallel$.
For the sake of simplicity, we only report the case $r_g/l_{\rm slab}=1.6 \times 10^{-2}$, that properly falls in the inertial range for both slab and 2D models. It is interesting to notice that the saturation in the perpendicular direction is slower with respect to the parallel direction. In particular, in the time range when parallel diffusion has already reached saturation, perpendicular diffusion undergoes a subdiffusive phase, followed by the proper diffusive behavior at later times \citep{QinEA02-apj,pucci2016energetic}. This is consistent with the naive expectation that perpendicular diffusion requires a time $(\delta B/B_{0})^{2}$ times longer to reach saturation. 

Figure~\ref{fig:NLGCsims-SLAB2D} shows the results of the two cases in which the 2D to slab power ratio is set either to $80\%/20\%$ (left column) or to $50\%/50\%$ (right column). The top-row panels show $\lambda_\parallel$ (red dots) and $\lambda_\perp$ (blue crosses) as a function of the normalized gyroradius $r_g/l_{\rm slab}$. The two vertical gray lines indicate the minimum rigidity for which the resonance can be considered as well resolved in our simulations. They are related to the minimum gyroradius in the parallel grid (dashed line) and the perpendicular grid (dot dashed line), evaluated in such a way that the minimum Larmor radius is described with five grid points in the parallel or perpendicular grid. 

In both $80\%/20\%$ and $50\%/50\%$ cases, the parallel mean free path is in quite good agreement with and SO-QLT [Eqs.~(\ref{eq:Dmumu_SLAB_SOQLT}--\ref{eq:sigma_SLAB_SOQLT}) inserted in Eq.~(\ref{eq:lambda_par}), green dashed line] predictions, showing a $r_g^2$ trend at high energies and the usual $r_g^{1/3}$ at smaller energies. Notice also that, since we do not see any transition at $l_{\rm 2D}=0.1 l_{\rm slab}$ in parallel diffusion within simulations, we conclude that, as expected from QLT considerations, parallel diffusion is not affected by the 2D component. Moreover, perpendicular diffusion is roughly constant at small gyroradius and shows a transition around $r_g\sim l_{\rm 2D}$ and then it increases above $r_g\sim l_{\rm slab}$.

It is interesting to check whether the ratio $\lambda_\perp/\lambda_\parallel$ is well described by NLGC and UNL theories. Bottom panels of Fig.~\ref{fig:NLGCsims-SLAB2D} show the ratio $\lambda_\perp/\lambda_\parallel$ as a function of $r_g/l_{\rm slab}$. Red points refer to the ratio directly computed from numerical simulations, while green dot dashed and blue dashed lines, respectively, display the theoretical calculations within NLGC [Eq.~(\ref{eq:SLAB2D-NLGC})] and UNL [Eq.~(\ref{eq:SLAB2D-UNLT-def})] theories, being the parallel path length evaluated through SO-QLT.
We set $a^2=1/3$ as in previous works \citep{matthaeus2003nonlinear,hussein2015influence}. A good agreement between numerical simulations and theories is recovered in the $80\%-20\%$ case. A larger discrepancy (about a factor of 5) is instead found in the $50\%-50\%$ case for both theories. In the two cases, UNLT provides slightly better results with respect to NLGC one. Moreover, NLGC and UNL theories depend on both $l_{\rm 2D}$ and $l_{\rm slab}$. Hence, one might expect 
breaks in the energy dependence at both of these scales in the simulations, since here $l_{\rm 2D}\neq l_{\rm slab}$.
However, as seen in the top panels of Fig.~\ref{fig:NLGCsims-SLAB2D}, there is no apparent transition at $l_{\rm 2D}$. On this basis, we suggest that the behaviour of $\lambda_\parallel$ influences the most the shape of the $\lambda_\perp/\lambda_\parallel$ ratio.

Within the current framework, in order to achieve a better agreement of theory and simulation when increasing the power in the slab component, one would 
require a smaller value of the parameter $a^2$. This different behaviour can be better understood with the help of Fig.~ \ref{fig:NLGCtheory-SLAB2D_terms}, where we show the relative importance of the two terms in Eq. (\ref{eq:SLAB2D-NLGC}), that here we refer to as the slab term and the 2D term. The lines show the two terms for the $50\%/50\%$ case (orange lines in Fig.~\ref{fig:NLGCtheory-SLAB2D_terms}) and the $80\%/20\%$ case (blue lines in Fig.~\ref{fig:NLGCtheory-SLAB2D_terms}). One can see that the 2D contribution to the ratio 
$\lambda_\perp/\lambda_\parallel$ is dominant for the $80\%/20\%$ case while the slab term slightly dominates in the $50\%/50\%$ case. In other words, it seems that the NLGC theory works best when the 2D component is dominant. This reason for this latter feature is at least twofold. The more intense slab component makes field line direction less parallel to the background magnetic field, hence the assumption of particle guiding centers following magnetic field lines is less satisfied with respect to the 2D dominant case. Moreover, for a more significant slab component, the parallel mean free path diminishes, thus reducing the distance over which particle guiding centers follow magnetic field lines \citep{minnie2009when}.

\begin{figure*}[!htb]
\centering
\begin{minipage}[t]{\columnwidth}
\includegraphics[width=\textwidth]{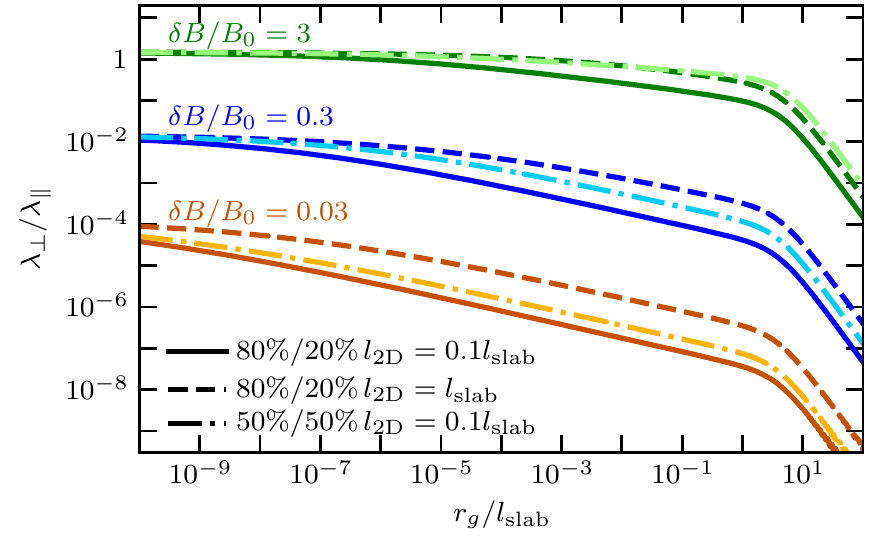}
\end{minipage}
\begin{minipage}[t]{\columnwidth}
\includegraphics[width=\textwidth]{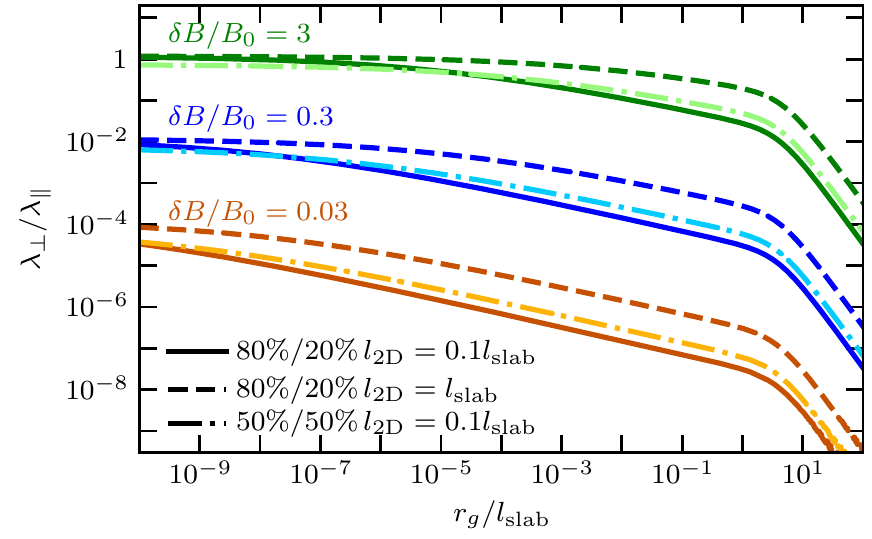}
\end{minipage}
    \caption{Solution of NLGC [Eq. (\ref{eq:SLAB2D-NLGC}), left] and UNL [Eq. (\ref{eq:SLAB2D-UNLT-def}), right] theories for different $\delta B/B_0=0.03,0.3,3$ (orange, blue and green respectively). Solid, dashed and dot dashed lines refer to $80\%/20\%$ and $l_{\rm 2D}=0.1 l_{\rm slab}$; $80\%/20\%$ and $l_{\rm 2D}=1 l_{\rm slab}$; and $50\%/50\%$ and $l_{\rm 2D}=0.1 l_{\rm slab}$, respectively.}
    \label{fig:theory-SLAB2D}
\end{figure*}

We remark here that, in the analyzed gyroradius range, the ratio $\lambda_\perp/\lambda_\parallel$ does not become constant when decreasing particle energy. In Fig.~\ref{fig:theory-SLAB2D} we show $\lambda_\perp/\lambda_\parallel$, evaluated from both NLGC [Eq.~(\ref{eq:SLAB2D-NLGC}), left] and UNL [Eq.~(\ref{eq:SLAB2D-UNLT-def}), right] theories, as a function of the gyroradius $r_g/l_{\rm slab}$, for different values of the parameters ($\delta B/B_0$, 2D to slab power ratio, and 2D to slab bend-over lengths ratio). $D_\parallel$ has been evaluated using SO-QLT. Within both theoretical formulations, the flatness of the ratio $\lambda_\perp/\lambda_\parallel$ only occurs at very small values of $r_g/l_{\rm slab}$, unless $\delta B\gtrsim B_0$. This behaviour is easily understood: from NLGC and UNL theories one concludes that the ratio becomes constant occurs when $\lambda_\parallel\sim l_{\rm slab}$. 
But in QLT one can roughly approximate the parallel mean free path as $\lambda_{\parallel}\approx r_{g}^{1/3} l_{\rm slab}^{2/3} (\delta B/B_{0})^{-2}$, so that the 
condition $\lambda_\parallel\sim l_{\rm slab}$ is obtained when 
\begin{equation}
    \frac{r_g}{l_{\rm slab}} \sim \left(\frac{\delta B_{\rm slab}}{B_0}\right)^6 \,.
\label{eq:condition}
\end{equation}
Therefore, the ratio $\lambda_\perp/\lambda_\parallel$ becomes constant at ever smaller energies 
as the slab turbulence amplitude 
$\delta B_{\rm slab}/B_0$ is reduced. 
Note also that, since $l_{\rm slab}\sim l_c$, the same line of reasoning 
applies for $r_{g}/l_c$. 

For typical values of parameters consistent with properties of the interstellar medium (ISM) ($\delta B/B_0\sim 0.1$), the condition 
in Eq. \ref{eq:condition}
is fulfilled only for very low energy particles.
This condition can even correspond to
nonrelativistic protons if the Larmor radius equals the coherence scale at PeV energies, as typically happens. In other words, for most energies the parallel and perpendicular diffusion coefficients are expected to have a different energy dependence. 
Notably, as visible in Fig.~\ref{fig:theory-SLAB2D}, the ratio $\lambda_\perp/\lambda_\parallel$ is a decreasing function of rigidity for most of the relevant values of rigidity, for the present case of slab/2D model of turbulence. 
As we discuss below, this conclusion appears to be reversed in the case of isotropic turbulence in the presence of an ordered field $B_{0}$.

\subsection{Isotropic model without \(B_0\)}

Here we focus on the three-dimensional isotropic model of turbulence. These simulations have been performed with $N_{x,y,z}=2048$ grid points. The bend-over scale is $l_{\rm iso}=L_{\rm box}/8$, 
where
$L_{\rm box}=512\,{\rm pc}$. The magnetic field fluctuation amplitude is 
set to $\delta B= \delta B_{\rm iso}=1 {\rm \mu G}$. In general, in this section, we show results for both the Kolmogorov and Kraichnan slopes.

\begin{figure}[t]
\begin{minipage}{\columnwidth}
\includegraphics[width=\columnwidth]{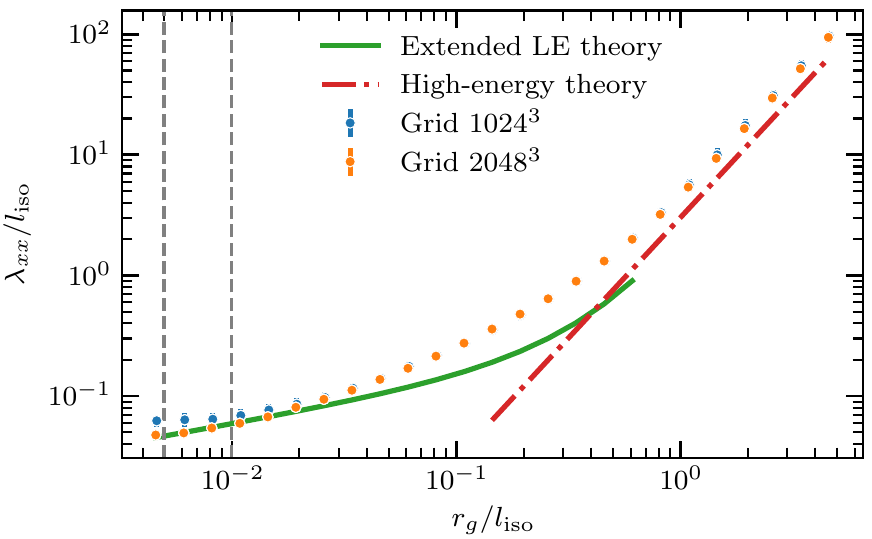}
\end{minipage} \hspace{0.02\textwidth}
\begin{minipage}[h]{\columnwidth}
\includegraphics[width=\columnwidth]{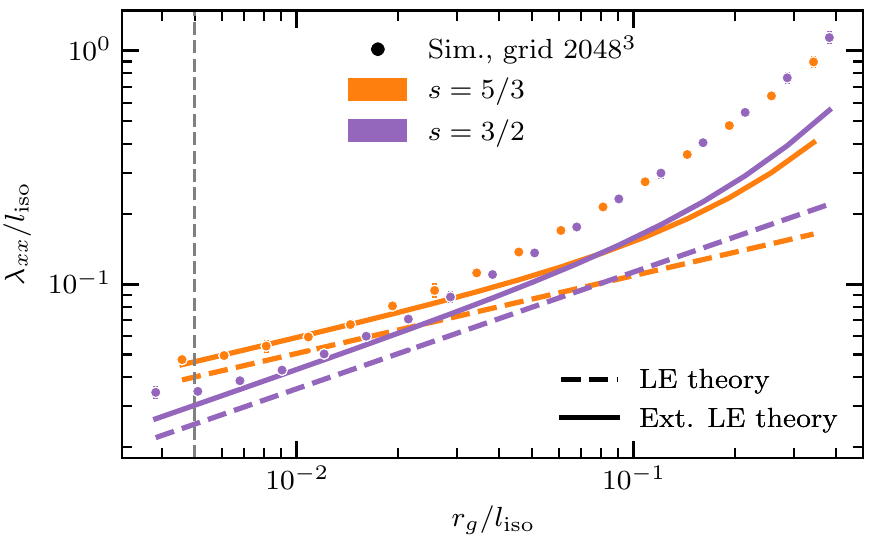}
\end{minipage}
\caption{Path length for CR transport in isotropic turbulence without mean field ($B_0=0$) as a function of rigidity. The simulation parameters are ${\delta B} = 1{\rm {\mu G}}$, $L_{\rm box}/l_{\rm iso} = 8$, $N_x=N_y=N_z=\{1024,2048\}$, $L_{\rm box} = 512 {\rm pc}$. The blue and orange points refer to simulation results for the Kolmogorov spectrum in the \(1024\) and \(2048\) grid size boxes, respectively. The gray vertical lines represents the so-called grid limit for  \(1024\) and \(2048\) (see text). The green solid line shows the extended low-energy (LE) theory given in Eq. (\ref{eq:lambdaisoext}), while the red dashed-dotted one is the high energy theory, Eq. (\ref{eq:lambdaiso_highenergy}). The bottom panel is the enlarged upper panel to emphasize the low-energy region in which the Kraichnan simulation points are added (purple points) together with its theoretical line for extended LE theory (solid lines). The dashed lines show the nonextended LE theory, Eq. (\ref{eq:lambdaiso}). The orange color refers to the Kolmogorov spectrum.}
\label{fig:3Diso_B0=0}
\end{figure}

We first revisit results of Ref. \citep{subedi2017charged} in which the diffusion coefficient is studied as a special case where $\delta B/B_0 \rightarrow \infty$ for high- and low-rigidity range. Here we investigate its agreement with our simulations when $B_0=0$ and $\delta B_{\rm iso}=1 {\rm \mu G}$. The top panel of Fig.~\ref{fig:3Diso_B0=0} displays the behaviour of the isotropic mfp $\lambda_{\rm iso}/l_{\rm iso}$ as a function of $r_g/l_{\rm iso}$, focusing on the region just below $r_g/l_{\rm iso} = 1$ where the transition between high rigidity and low rigidity asymptotic forms occurs. For high rigidities, i.e., $r_g/l_{\rm iso} > 1$, the expected behavior $\lambda_{\rm iso} \sim r_g^2$ is recovered. The small offset in the normalization of the high-energy line is due to the difference between the theoretical correlation length given in Eq. (\ref{eq:corr_length_isotropic}), and one actually present in the simulated turbulent field. With the ratio $l_{\rm iso}=L_{\rm box}/8$ a small fraction of turbulence energy (approximately \(\sim 10\%\)) is not contained in the box, and that leads to slightly different \(L_{\rm c, iso}\) (see Sec. \ref{sec:numerical_methods}) which reflects on the normalisation of the high-energy points. The correlation length can be calculated for the finite integral of \(k\) in Eq. (\ref{eq:corr_length_definition}), from \(k_{\rm min}\) to \(k_{\rm max}\) which would give a better agreement with the simulation results.

In the opposite regime, for small gyroradii compared to the bend-over scale of turbulence, $r_g/l_{\rm iso} < 1$, the simulation results also agree with the theoretical predictions, especially for the bigger box (\(2048^3\) grid points). We define a marker called the grid limit that represents the gyroradius that comprises at least five grid points. At scales above the grid limit, a particle can \textit{sense} the resonance in the inertial range of turbulence while for the smaller gyroradii, this cannot be guaranteed and leads to unphysical results. 

In the upper panel, only the Kolmogorov (\(s=5/3\)) points are shown for two grid sizes, \(1024\) and \(2048\) in blue and orange respectively. Here it can be seen that the blue points, as they 
approach the grid limit from above, 
begin to lose resonance at higher energies than the orange points. The orange points agree well with the extended low-energy theory (the solid green line) given in Eq.~(\ref{eq:lambdaisoext})
to lower rigidities, $r_g/l_{\rm iso}< 10^{-2}$. 
The lower panel enlarges in the lower-energy range of the upper panel, and adds the Kraichnan case for comparisons (dark gray points) and the accompanying theory line (the solid violet line). The dashed lines depict Eq.~(\ref{eq:lambdaiso}) for the Kolmogorov and Kraichnan cases and are colored the same way as their extended counterparts.

The detailed numerical simulations of this section confirm that the low-energy theory presented in \citep{subedi2017charged} indeed agrees well with the simulations. We also note that accurate numerical simulation in the low rigidity regime is considerably demanding in terms of computational resources.

\subsection{Isotropic model with \(B_0 \neq 0\)}\label{sec:isomol}

This section is concluded with results of the isotropic model with the background field present, i.e., $B_0 \neq 0$. We consider this as the most important result since it departs from expectations of known theoretical models.

Theoretical approaches in modelling the axisymmetric diffusion of particles normally consider \(B_0\) as the only component relevant for defining gyroradius \(r_g\), primarily because \(\delta B \ll B_0\) is assumed, such as in QLT. So far in the current work, we have adopted the same definition. However, with increasing \(\delta B\), and especially when \(\delta B \gtrsim B_0\), this definition of \(r_g\) cannot support any meaningful transition to the isotropic case without \(B_0\), shown in the previous subsection. That is why, hereafter, we use \(B_{\rm tot} \equiv \sqrt{B_0^2 + \delta B_{\rm iso}^2}\) in the definition of \(r_g\) when dealing with isotropic turbulence to seamlessly converge to the case of vanishing \(B_0\). If compared with results made with the previous definition, the difference is visually noticeable only for \(\delta B_{\rm iso}/B_0 > 0.5\).

\begin{figure}[t]
\centering
    \includegraphics[width=\columnwidth]{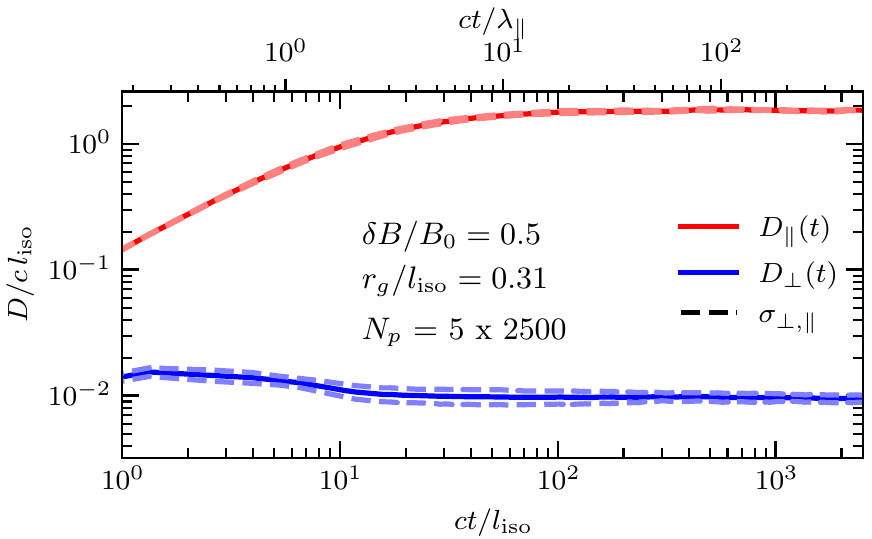}
    \caption{Sample running diffusion coefficients, \(D_\perp = (D_{xx} + D_{yy})/2\), \(D_\parallel = D_{zz}\), in the isotropic model with \(B_0 \neq 0\) represented as mean free paths over the bend-over scale for one simulation runs with parameters \(\delta B/B_0 = 0.5\) and \(r_g/l_\mathrm{iso} \sim 0.3\), while dashed lines refer to standard deviations. The upper scale is measured in the correlation length in the parallel direction, \(\lambda_{\parallel}\).}
    \label{fig:3Diso_running}
\end{figure}

The simulation parameters are the same as in the previous subsection except for the presence of the mean magnetic field. In Fig.~\ref{fig:3Diso_running} we show to convergence of the procedure for the calculation of the parallel and perpendicular diffusion coefficients in terms of a plateau in the running diffusion coefficients. In this case, the plot is obtained with 2500 particles in five different realisations of the magnetic field and, as one can see the result, is not affected by any systematics associated with any of the realizations. The same result is achieved with simulations that include only one realization.

\begin{figure}[t]
\begin{minipage}{\columnwidth}
    \includegraphics[width=\columnwidth]{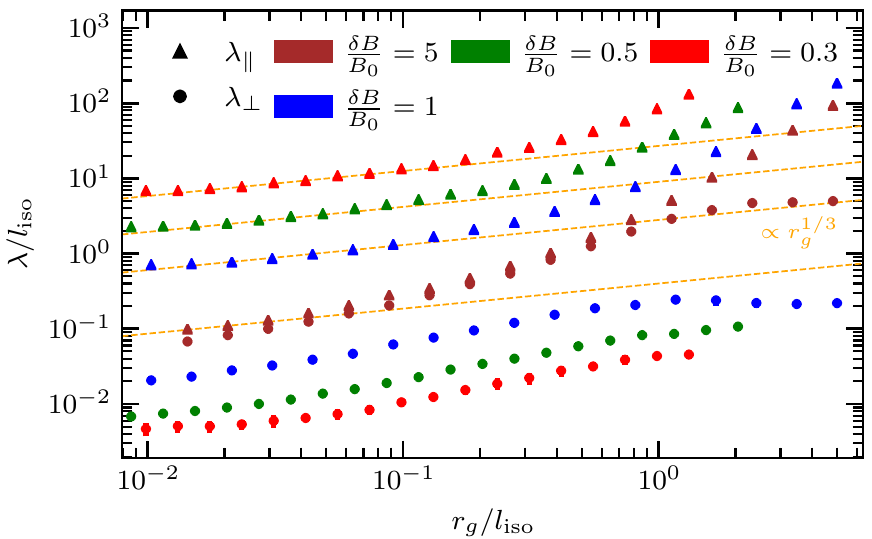}
\end{minipage}
\begin{minipage}{\columnwidth}
    \includegraphics[width=\columnwidth]{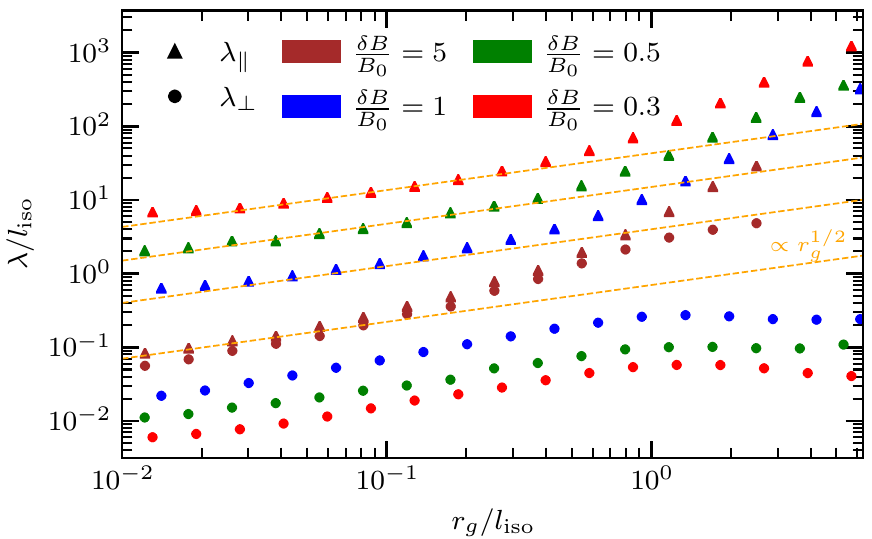}
\end{minipage}
\caption{Parallel (triangles) and perpendicular (circles) mean free paths as functions of \(r_g/l_{\rm iso}\) for several values of \(\delta B/B_0\) (different colours). The top panel refers to Kolmogorov turbulence and the bottom to Kraichnan turbulence. Parallel mean free path scales as \(r_g^{1/3}\) and \(r_g^{1/2}\) for Kolmogorov and Kraichnan cases, respectively (yellow dashed lines). The simulations were performed on 2048 and 1024 grids.}
    \label{fig:3Diso_Kolm_Krai}
\end{figure}

The energy dependence of the mean free paths, $\lambda_\parallel$ and $\lambda_\perp$ is displayed in Fig.~ \ref{fig:3Diso_Kolm_Krai}, for the Kolmogorov spectrum ($s_{\rm iso}=5/3$) in the upper panel and for the Kraichnan spectrum ($s_{\rm iso}=3/2$) in the lower panel. Each panel shows results for $\delta B_{\rm iso}/B_0=0.3$ (red), $\delta B_{\rm iso}/B_0=0.5$ (green), $\delta B_{\rm iso}/B_0=1$ (blue) and $\delta B_{\rm iso}/B_0=5$ (brown). The triangles and dots, respectively, refer to $\lambda_{\parallel}$ and $\lambda_\perp$. The parallel mean free paths clearly follow the slope expectation based upon QLT. 
The general trend is revealing.
For rigidity values 
corresponding to the inertial range of turbulence, 
and therefore to a regime in which there is resonant scattering, 
one finds that 
the parallel and perpendicular diffusion coefficients 
have different energy dependence. On the other hand, as expected, increasing the turbulence amplitude $\delta B_{\rm iso}/B_0$, 
the difference between $\lambda_{\parallel}$ and $\lambda_\perp$ reduces towards the limiting 
case of no background magnetic field.
In that case, there is, as expected,
no distinction between parallel and perpendicular directions. 

\begin{figure}[!thb]
    \centering
    \includegraphics[width=\columnwidth]{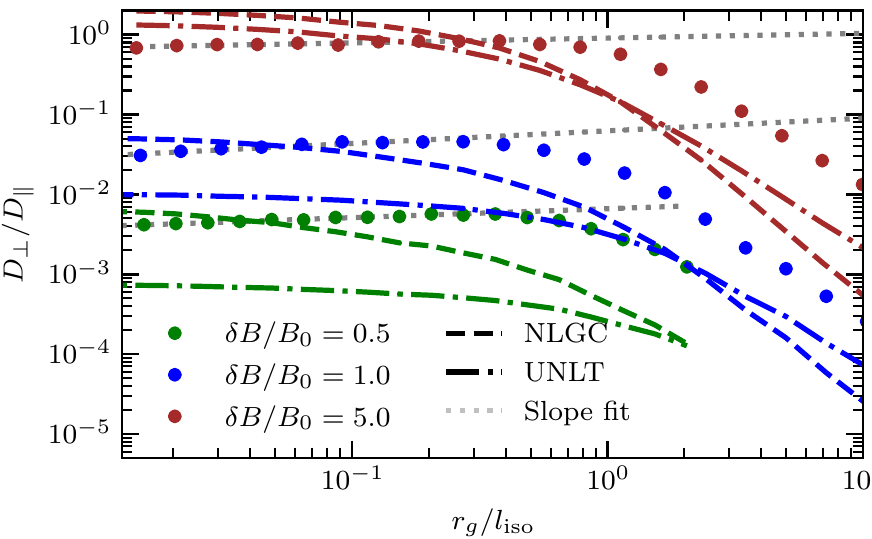}
    \caption{\(D_\perp / D_\parallel\) ratio as a function of gyroradius for different values of \({\delta B}/{B_0}\) with the Kolmogorov spectrum. {Dashed and dot dashed lines respectively represent predictions of  NLGC [Eq. (\ref{eq:3DISO-NLGC})] and UNL theories [Eq. \ref{eq:3DISO-UNLT}] for isotropic turbulence, both evaluated with $a^2=1/3$  {and where, for the $D_{\parallel}$ input, simulation points are used.}} Gray dotted lines are fits to the inertial range behavior as used in Fig.~\ref{fig:3Diso_slopes}.}
    \label{fig:3Diso_Kolm_ratio}
\end{figure}

Our simulations, having a large dynamical range compared with previous investigations of this problem, confirm previous hints \cite{de2007numerical} of a different energy dependence of the parallel and perpendicular diffusion coefficients for \(\delta B_{\rm iso}/B_0\lesssim 1\). Since there are no theoretical predictions for the case of particle transport in isotropic turbulence in the presence of an ordered magnetic field to compare our simulations against, we have derived these predictions by the straightforward application of NLGC and UNL theories, as shown above, although, with with the explicit assumption $\delta B_z = 0$ to simplify the presentation. Unfortunately, both theories return predictions that are at odds with the results of the simulations, suggesting that they miss some important pieces of physics of transport in this situation. This can be clearly observed from Figure ~\ref{fig:3Diso_Kolm_ratio}, where we show the energy dependence of the ratio of the perpendicular and parallel diffusion coefficients, especially in the region $r_g<l_c$. The dashed lines in the same figure represent the results of NLGC calculation for the 3D isotropic model of turbulence, Eq.~(\ref{eq:3DISO-NLGC}), while dot dashed lines refer to UNLT, Eq.~(\ref{eq:3DISO-UNLT}). Both theories have been evaluated with \(a^2=1/3\) and simulations points have been used for the \(D_\parallel\) input.
Fig.~\ref{fig:3Diso_Kolm_ratio} displays results only for the Kolmogorov case, for the sake of clarity, but the same trend is also found in the Kraichnan case. We stress once more that the problem is not related to the fact that we neglected the contribution due to non vanishing $\delta B_{z}$, which can be shown to be negligible in terms of perpendicular diffusion. 

There is no doubt that further theoretical investigation is needed to understand the physical reason for the trend in the ratio. We stress once more that the increasing trend as a function of energy is opposite to that found in the case of slab/2D turbulence model discussed above, where the NLGC theory provides at least a qualitatively correct description of the results of simulations. 

\begin{figure}[t]
    \centering
    \includegraphics[width=\columnwidth]{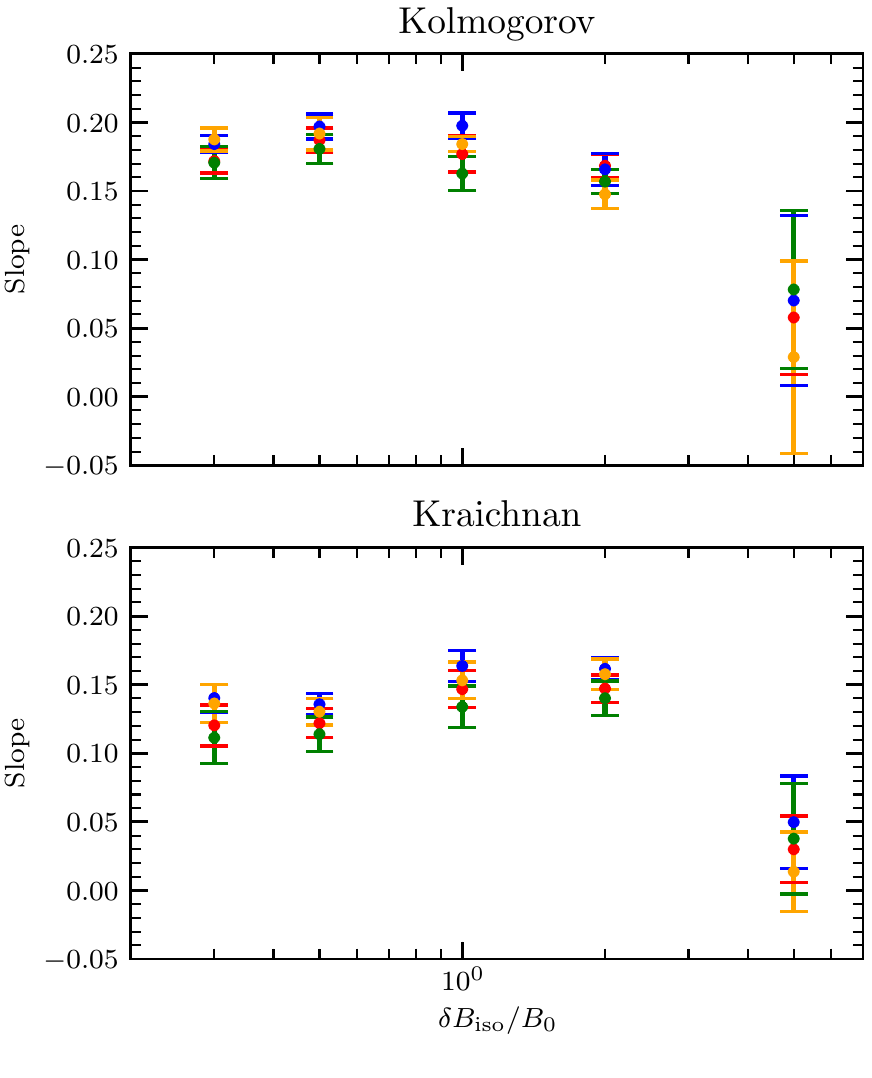}
    \caption{Slope of \(D_\perp/D_\parallel\) as a function of \(\delta B_{\rm iso}/B_0\) for the Kolmogorov (top) and Kraichnan (bottom) spectra. Different colours correspond to different ranges of Larmor radii over which the fitting procedure has been implemented as described at the end of Sec.~\ref{sec:isomol}.}
    \label{fig:3Diso_slopes}
\end{figure}

The change of slope of the \(D_\perp/D_\parallel\) ratios in the inertial range is also demonstrated in Fig.~ \ref{fig:3Diso_slopes}. It displays the slope for both Kolmogorov (top) and Kraichnan (bottom) models of turbulence as a function of \(\delta B_{\rm iso}/B_0\). 
The different colours of the points correspond to the different fitting procedures for obtaining the slope, performed as follows. For each value of \(\delta B_{\rm iso}/B_0\), we have first selected the optimal range of Larmor radii, within the inertial range defined as where \(D_\parallel \propto r_g^{2-s}\) where \(s\) refers to the spectral index. The results of fits performed on this ``best'' range are displayed as the red points. Then, to verify if the selection procedure of the fit range is not determining the results, we have repeated the procedures by reducing the fit range, i.e., by removing edge points. In particular, we have excluded one point at high-energy (blue); one point at low-energy (green); or one point at high-energy and another point at low-energy (orange). The results of fitting procedures in different ranges are consistent among themselves, demonstrating the robustness of the slope results.

\section{Discussion and conclusions}
\label{sec:discuss}

We performed numerical simulations of CR transport in synthetic turbulence in the case of slab, slab/2D and isotropic turbulence. The dynamical range achieved in these simulations allows us to reach solid conclusions concerning the energy dependence of the diffusion coefficient in the directions parallel and perpendicular to the large scale ordered magnetic field \(\bvec{B_{0}}\). For the case of isotropic turbulence ($\delta B_{z} \neq 0$), we also extend previous simulations \cite{de2007numerical,subedi2017charged} to cover a larger dynamical range. 

The numerical approach has been tested versus the case of slab turbulence with different levels of turbulence $\delta B/B_{0}$: the results of numerical simulations for the pitch angle diffusion coefficient and the spatial diffusion coefficient are in excellent agreement with QLT as long as $\delta B/B_{0}\lesssim 0.05$. For larger levels of turbulence, $0.05\lesssim \delta B/B_{0}\lesssim 0.5$, the second order QLT provides a better description of numerical results. For even larger levels of turbulence, the parallel mean free path derived numerically seems to be in between the predictions of QLT and SO-QLT. 
The behaviour of the pitch angle diffusion coefficient derived from simulations clearly shows the resonance broadening around $\mu\sim 0$ where QLT becomes problematic (the $90^{o}$ problem).

The slope of $D_{\parallel}$ as a function of energy, at low energies is always consistent with what would be naively expected from QLT for the given spectrum of turbulence (for instance the slope is 1/3 for Kolmogorov turbulence), provided the resonant scale $k$ is resolved by the numerical simulation. Typically 
in the simulations
the latter scale is between two and three orders of magnitude below the correlation length, $l_{\rm slab}$, for slab turbulence. At high energies, for which $r_{g}\gg l_{\rm slab}$, we find $D_{\parallel}\propto r_{g}^{2}$, as expected.

For slab/2D turbulence, the situation is more complex. We investigated two configurations in which energy is shared between 2D and slab turbulence as 80\%/20\% and 50\%/50\% respectively. The case we investigated has $\delta B/B_{0}=0.3$, with $\delta B^{2}=\delta B_{\rm slab}^{2}+\delta B_{\rm 2D}^{2}$. In both cases and for the whole range of Larmor radii here considered, the parallel path length is in good agreement with SO-QLT, thus confirming the powerful nature of this theoretical approach, despite its perturbative origin. In the same conditions the perpendicular path length is basically energy independent in the low energy regime. 

Simulation results for the composite model of turbulence have been also compared with the predictions of the NLGC and UNL theories for the $D_{\perp}/D_{\parallel}$ ratio. Since both theories require as an input the parallel path length, we adopted $\lambda_{\parallel}$ resulting from SO-QLT, while we considered $a^2=1/3$ \citep{matthaeus2003nonlinear}. A good agreement between numerical simulations and theories is recovered in the $80\%-20\%$ case, while a more significant disagreement is found in the $50\%-50\%$ case. Moreover UNLT yields slightly more accurate results with respect to the NLGC theory. We also remark that, the single transition recovered, for both NLGC and UNL theories, in the ration $\lambda_\perp/\lambda_\parallel$ occurs at $r_g\sim l_{\rm slab}$, while no transitions are found at $r_g\sim l_{\rm 2D}=0.1l_{\rm slab}$. This may suggest that, within such theories, the qualitative behavior of the ratio $\lambda_\perp/\lambda_\parallel$ is mainly governed by $\lambda_\parallel$.

One should also keep in mind that UNL and NLGC theories make use of a free parameter $a^{2}$ which is somehow tuned to fit data or simulations, which is a weak point of this approach. In our calculations for the slab/2D case, as in much of the literature, we assume $a^{2}=1/3$, but it is possible that a different choice of $a^2$, perhaps depending on the turbulence intensity, might provide a better agreement with the simulated $D_{\perp}/D_{\parallel}$. A rule of thumb seems that the agreement of the NLGC results with simulations is better when most of the power is concentrated in the 2D component of the turbulence. In this perspective note that \citet{shalchi2019heuristic} recently claimed to prove the physical meaning of setting $a^2=1/3$. However simulations results (e.g., results of \citet{arendt2020detailed} suggest that this parameter depends on both particle energy and fluctuations level) indicate that this feature is still poorly understood and an intense effort is required to deeply comprehend it.

Both simulations and NLGC results agree on predicting that the $D_{\perp}/D_{\parallel}$ is a decreasing function of energy for slab/2D turbulence. Due to limited computational resources, it is not possible to extend the simulations to energies much below the ones considered here. However we can tentatively trust the NLGC theory in predicting the trend of the $D_{\perp}/D_{\parallel}$ ratio at low energies, so as to check claims of universality \cite{hussein2015influence}, namely constancy of the ratio at low energy. We proved that such universality is achieved when $r_{g}/l_{\rm slab}\sim (\delta B_{\rm slab}/B_{0})^{6}$ for a Kolmogorov shape of the spectrum. For conditions that are typically adopted for the ISM in our Galaxy ($l_{\rm slab}\sim 10$ pc and $\delta B_{\rm slab}/B_{0}\sim 0.1$), one can clearly see that the $D_{\perp}/D_{\parallel}$ ratio should become constant only at energies below $\sim 10$ GeV, while at higher energies the ratio should be a decreasing function of energy. 

Finally, we discuss the case of isotropic turbulence ($\delta B_{z} \neq 0$). In the presence of an ordered field \(\bvec{B}_0\), we show that the diffusion coefficients parallel and perpendicular to \(\bvec{B}_0\) have slopes that differ from one another in the low energy regime. This behavior was found for both Kolmogorov and Kraichnan spectra of the turbulence.
and confirms previous hints \cite{de2007numerical}, obtained with smaller boxes of the simulated field. At low energies, the parallel mean free path has the slope that would be expected based on QLT, for all levels of turbulence, and for the turbulent spectra considered here, contrary to what has recently been claimed in \cite{2020MNRAS.498.5051R}. Their result is probably simply due to the fact that in the lowest energy bins their simulations are unable to capture the resonances responsible for particle scattering. This physical ingredient is instead correctly described in our simulations, due to the very large box adopted here. By qualitatively comparing results described here and the ones obtained with the plane-waves method in \citet{casse2001transport}, we conclude that the dynamical range in Ref. \citep{casse2001transport} was not extended enough in the regime of small gyroradii (see, e.g., Fig. 4 of \citep{casse2001transport}) to observe the energy dependence of the $D_\perp/D_\parallel$ ratio. More powerful computational resources available at the present time, have allowed us to clarify this issue, because of the access to a wider range of energies and because of the capability to have numerical accuracy under control. We aim to directly compare our method with plane-wave methods in a future work.

The ratio $D_{\perp}/D_{\parallel}$ changes with energy in a different way depending on the level of turbulence, $\delta B/B_{0}$ and, as expected, becomes constant for $\delta B/B_{0}\to \infty$ (or $B_{0}\to 0$). For $\delta B/B_{0}<1$ the ratio grows with energy for $r_{g}/l_{c}<1$, at odds with the case of slab/2D turbulence, while it decreases with energy for $r_{g}/l_{c}<1$, where the parallel scattering is no longer dominated by resonances. Interestingly, no theory exists for the case of isotropic turbulence in the presence of an ordered field \(\bvec{B}_0\), while a generalisation of the NLGC approach for the strong turbulence case ($B_{0}\to 0$) was previously presented in~\cite{subedi2017charged}. 
In the present article we provided a simple application of the NLGC and the UNL theories to the isotropic case, in order to estimate the perpendicular diffusion coefficient, but the results of such calculation do not match the $D_{\perp}/D_{\parallel}$ ratio obtained from our numerical simulations. The results collected here suggest that the breaking of universality emerges due to the energy dependence of the perpendicular mean free path, since the parallel mean free path is consistent with standard theoretical approaches (QLT/SO-QLT). A theoretical approach to the description of particle transport in isotropic turbulence with an ordered magnetic field is definitely still missing and will be the subject of future efforts. 

\begin{acknowledgments}
The authors are grateful to  A. P. Snodin and P. Reichherzer for several discussions on the topics illustrated here. Numerical simulations discussed here have been performed on the CNAF cluster of the National Institute of Nuclear Physics (INFN) and on the Newton cluster at the University of Calabria (Italy). O.P. thanks Dr J. Settino for his friendly support in the transition to Python. W.H.M. was partially supported by the Parker Solar Probe ISOIS project under subcontract SUB0000165 from Princeton University, and by NASA Heliospheric Supporting Research Grant No. NNX17AB79G.
\end{acknowledgments}

\bibliographystyle{apsrev4-2} 
\bibliography{bibliography}

\begin{thebibliography}{79}%
\makeatletter
\providecommand \@ifxundefined [1]{%
 \@ifx{#1\undefined}
}%
\providecommand \@ifnum [1]{%
 \ifnum #1\expandafter \@firstoftwo
 \else \expandafter \@secondoftwo
 \fi
}%
\providecommand \@ifx [1]{%
 \ifx #1\expandafter \@firstoftwo
 \else \expandafter \@secondoftwo
 \fi
}%
\providecommand \natexlab [1]{#1}%
\providecommand \enquote  [1]{``#1''}%
\providecommand \bibnamefont  [1]{#1}%
\providecommand \bibfnamefont [1]{#1}%
\providecommand \citenamefont [1]{#1}%
\providecommand \href@noop [0]{\@secondoftwo}%
\providecommand \href [0]{\begingroup \@sanitize@url \@href}%
\providecommand \@href[1]{\@@startlink{#1}\@@href}%
\providecommand \@@href[1]{\endgroup#1\@@endlink}%
\providecommand \@sanitize@url [0]{\catcode `\\12\catcode `\$12\catcode
  `\&12\catcode `\#12\catcode `\^12\catcode `\_12\catcode `\%12\relax}%
\providecommand \@@startlink[1]{}%
\providecommand \@@endlink[0]{}%
\providecommand \url  [0]{\begingroup\@sanitize@url \@url }%
\providecommand \@url [1]{\endgroup\@href {#1}{\urlprefix }}%
\providecommand \urlprefix  [0]{URL }%
\providecommand \Eprint [0]{\href }%
\providecommand \doibase [0]{https://doi.org/}%
\providecommand \selectlanguage [0]{\@gobble}%
\providecommand \bibinfo  [0]{\@secondoftwo}%
\providecommand \bibfield  [0]{\@secondoftwo}%
\providecommand \translation [1]{[#1]}%
\providecommand \BibitemOpen [0]{}%
\providecommand \bibitemStop [0]{}%
\providecommand \bibitemNoStop [0]{.\EOS\space}%
\providecommand \EOS [0]{\spacefactor3000\relax}%
\providecommand \BibitemShut  [1]{\csname bibitem#1\endcsname}%
\let\auto@bib@innerbib\@empty
\bibitem [{\citenamefont {{Jokipii}}(1966)}]{jokipii1966cosmicray}%
  \BibitemOpen
  \bibfield  {author} {\bibinfo {author} {\bibfnamefont {J.~R.}\ \bibnamefont
  {{Jokipii}}},\ }\href {https://doi.org/10.1086/148912} {\bibfield  {journal}
  {\bibinfo  {journal} {\apj}\ }\textbf {\bibinfo {volume} {146}},\ \bibinfo
  {pages} {480} (\bibinfo {year} {1966})}\BibitemShut {NoStop}%
\bibitem [{\citenamefont {{Shalchi}}(2005)}]{shalchi2005second}%
  \BibitemOpen
  \bibfield  {author} {\bibinfo {author} {\bibfnamefont {A.}~\bibnamefont
  {{Shalchi}}},\ }\href {https://doi.org/10.1063/1.1895805} {\bibfield
  {journal} {\bibinfo  {journal} {Physics of Plasmas}\ }\textbf {\bibinfo
  {volume} {12}},\ \bibinfo {pages} {052905} (\bibinfo {year}
  {2005})}\BibitemShut {NoStop}%
\bibitem [{\citenamefont {{Blasi}}(2013)}]{blasi2013origin}%
  \BibitemOpen
  \bibfield  {author} {\bibinfo {author} {\bibfnamefont {P.}~\bibnamefont
  {{Blasi}}},\ }\href {https://doi.org/10.1007/s00159-013-0070-7} {\bibfield
  {journal} {\bibinfo  {journal} {\aapr}\ }\textbf {\bibinfo {volume} {21}},\
  \bibinfo {eid} {70} (\bibinfo {year} {2013})},\ \Eprint
  {https://arxiv.org/abs/1311.7346} {arXiv:1311.7346 [astro-ph.HE]}
  \BibitemShut {NoStop}%
\bibitem [{\citenamefont {{Blasi}}(2019{\natexlab{a}})}]{2019NCimR..42..549B}%
  \BibitemOpen
  \bibfield  {author} {\bibinfo {author} {\bibfnamefont {P.}~\bibnamefont
  {{Blasi}}},\ }\href {https://doi.org/10.1393/ncr/i2019-10166-0} {\bibfield
  {journal} {\bibinfo  {journal} {Nuovo Cimento Rivista Serie}\ }\textbf
  {\bibinfo {volume} {42}},\ \bibinfo {pages} {549} (\bibinfo {year}
  {2019}{\natexlab{a}})}\BibitemShut {NoStop}%
\bibitem [{\citenamefont {{Evoli}}\ \emph {et~al.}(2019)\citenamefont
  {{Evoli}}, \citenamefont {{Aloisio}},\ and\ \citenamefont
  {{Blasi}}}]{2019PhRvD..99j3023E}%
  \BibitemOpen
  \bibfield  {author} {\bibinfo {author} {\bibfnamefont {C.}~\bibnamefont
  {{Evoli}}}, \bibinfo {author} {\bibfnamefont {R.}~\bibnamefont {{Aloisio}}},\
  and\ \bibinfo {author} {\bibfnamefont {P.}~\bibnamefont {{Blasi}}},\ }\href
  {https://doi.org/10.1103/PhysRevD.99.103023} {\bibfield  {journal} {\bibinfo
  {journal} {Phys. Rev. D}\ }\textbf {\bibinfo {volume} {99}},\ \bibinfo {eid}
  {103023} (\bibinfo {year} {2019})},\ \Eprint
  {https://arxiv.org/abs/1904.10220} {arXiv:1904.10220 [astro-ph.HE]}
  \BibitemShut {NoStop}%
\bibitem [{\citenamefont {{Evoli}}\ \emph {et~al.}(2020)\citenamefont
  {{Evoli}}, \citenamefont {{Morlino}}, \citenamefont {{Blasi}},\ and\
  \citenamefont {{Aloisio}}}]{2020PhRvD.101b3013E}%
  \BibitemOpen
  \bibfield  {author} {\bibinfo {author} {\bibfnamefont {C.}~\bibnamefont
  {{Evoli}}}, \bibinfo {author} {\bibfnamefont {G.}~\bibnamefont {{Morlino}}},
  \bibinfo {author} {\bibfnamefont {P.}~\bibnamefont {{Blasi}}},\ and\ \bibinfo
  {author} {\bibfnamefont {R.}~\bibnamefont {{Aloisio}}},\ }\href
  {https://doi.org/10.1103/PhysRevD.101.023013} {\bibfield  {journal} {\bibinfo
   {journal} {Phys. Rev. D}\ }\textbf {\bibinfo {volume} {101}},\ \bibinfo
  {eid} {023013} (\bibinfo {year} {2020})},\ \Eprint
  {https://arxiv.org/abs/1910.04113} {arXiv:1910.04113 [astro-ph.HE]}
  \BibitemShut {NoStop}%
\bibitem [{\citenamefont {{DeMarco}}\ \emph {et~al.}(2007)\citenamefont
  {{DeMarco}}, \citenamefont {{Blasi}},\ and\ \citenamefont
  {{Stanev}}}]{de2007numerical}%
  \BibitemOpen
  \bibfield  {author} {\bibinfo {author} {\bibfnamefont {D.}~\bibnamefont
  {{DeMarco}}}, \bibinfo {author} {\bibfnamefont {P.}~\bibnamefont {{Blasi}}},\
  and\ \bibinfo {author} {\bibfnamefont {T.}~\bibnamefont {{Stanev}}},\ }\href
  {https://doi.org/10.1088/1475-7516/2007/06/027} {\bibfield  {journal}
  {\bibinfo  {journal} {\jcap}\ }\textbf {\bibinfo {volume} {2007}},\ \bibinfo
  {eid} {027} (\bibinfo {year} {2007})},\ \Eprint
  {https://arxiv.org/abs/0705.1972} {arXiv:0705.1972 [astro-ph]} \BibitemShut
  {NoStop}%
\bibitem [{\citenamefont {{Casse}}\ \emph {et~al.}(2001)\citenamefont
  {{Casse}}, \citenamefont {{Lemoine}},\ and\ \citenamefont
  {{Pelletier}}}]{casse2001transport}%
  \BibitemOpen
  \bibfield  {author} {\bibinfo {author} {\bibfnamefont {F.}~\bibnamefont
  {{Casse}}}, \bibinfo {author} {\bibfnamefont {M.}~\bibnamefont {{Lemoine}}},\
  and\ \bibinfo {author} {\bibfnamefont {G.}~\bibnamefont {{Pelletier}}},\
  }\href {https://doi.org/10.1103/PhysRevD.65.023002} {\bibfield  {journal}
  {\bibinfo  {journal} {\prd}\ }\textbf {\bibinfo {volume} {65}},\ \bibinfo
  {eid} {023002} (\bibinfo {year} {2001})},\ \Eprint
  {https://arxiv.org/abs/astro-ph/0109223} {arXiv:astro-ph/0109223 [astro-ph]}
  \BibitemShut {NoStop}%
\bibitem [{\citenamefont {{Evoli}}\ \emph {et~al.}(2012)\citenamefont
  {{Evoli}}, \citenamefont {{Gaggero}}, \citenamefont {{Grasso}},\ and\
  \citenamefont {{Maccione}}}]{2012PhRvL.108u1102E}%
  \BibitemOpen
  \bibfield  {author} {\bibinfo {author} {\bibfnamefont {C.}~\bibnamefont
  {{Evoli}}}, \bibinfo {author} {\bibfnamefont {D.}~\bibnamefont {{Gaggero}}},
  \bibinfo {author} {\bibfnamefont {D.}~\bibnamefont {{Grasso}}},\ and\
  \bibinfo {author} {\bibfnamefont {L.}~\bibnamefont {{Maccione}}},\ }\href
  {https://doi.org/10.1103/PhysRevLett.108.211102} {\bibfield  {journal}
  {\bibinfo  {journal} {Phys. Rev. Lett.}\ }\textbf {\bibinfo {volume} {108}},\
  \bibinfo {eid} {211102} (\bibinfo {year} {2012})},\ \Eprint
  {https://arxiv.org/abs/1203.0570} {arXiv:1203.0570 [astro-ph.HE]}
  \BibitemShut {NoStop}%
\bibitem [{\citenamefont {{Cerri}}\ \emph {et~al.}(2017)\citenamefont
  {{Cerri}}, \citenamefont {{Gaggero}}, \citenamefont {{Vittino}},
  \citenamefont {{Evoli}},\ and\ \citenamefont
  {{Grasso}}}]{2017JCAP...10..019C}%
  \BibitemOpen
  \bibfield  {author} {\bibinfo {author} {\bibfnamefont {S.~S.}\ \bibnamefont
  {{Cerri}}}, \bibinfo {author} {\bibfnamefont {D.}~\bibnamefont {{Gaggero}}},
  \bibinfo {author} {\bibfnamefont {A.}~\bibnamefont {{Vittino}}}, \bibinfo
  {author} {\bibfnamefont {C.}~\bibnamefont {{Evoli}}},\ and\ \bibinfo {author}
  {\bibfnamefont {D.}~\bibnamefont {{Grasso}}},\ }\href
  {https://doi.org/10.1088/1475-7516/2017/10/019} {\bibfield  {journal}
  {\bibinfo  {journal} {\jcap}\ }\textbf {\bibinfo {volume} {2017}},\ \bibinfo
  {eid} {019} (\bibinfo {year} {2017})},\ \Eprint
  {https://arxiv.org/abs/1707.07694} {arXiv:1707.07694 [astro-ph.HE]}
  \BibitemShut {NoStop}%
\bibitem [{\citenamefont {{Giacinti}}\ \emph {et~al.}(2018)\citenamefont
  {{Giacinti}}, \citenamefont {{Kachelrie\ss}},\ and\ \citenamefont
  {{Semikoz}}}]{2018JCAP...07..051G}%
  \BibitemOpen
  \bibfield  {author} {\bibinfo {author} {\bibfnamefont {G.}~\bibnamefont
  {{Giacinti}}}, \bibinfo {author} {\bibfnamefont {M.}~\bibnamefont
  {{Kachelrie\ss}}},\ and\ \bibinfo {author} {\bibfnamefont {D.~V.}\
  \bibnamefont {{Semikoz}}},\ }\href
  {https://doi.org/10.1088/1475-7516/2018/07/051} {\bibfield  {journal}
  {\bibinfo  {journal} {\jcap}\ }\textbf {\bibinfo {volume} {2018}},\ \bibinfo
  {eid} {051} (\bibinfo {year} {2018})},\ \Eprint
  {https://arxiv.org/abs/1710.08205} {arXiv:1710.08205 [astro-ph.HE]}
  \BibitemShut {NoStop}%
\bibitem [{\citenamefont {{Shebalin}}\ \emph
  {et~al.}(1983{\natexlab{a}})\citenamefont {{Shebalin}}, \citenamefont
  {{Matthaeus}},\ and\ \citenamefont {{Montgomery}}}]{ShebalinEA83}%
  \BibitemOpen
  \bibfield  {author} {\bibinfo {author} {\bibfnamefont {J.~V.}\ \bibnamefont
  {{Shebalin}}}, \bibinfo {author} {\bibfnamefont {W.~H.}\ \bibnamefont
  {{Matthaeus}}},\ and\ \bibinfo {author} {\bibfnamefont {D.}~\bibnamefont
  {{Montgomery}}},\ }\href {https://doi.org/10.1017/S0022377800000933}
  {\bibfield  {journal} {\bibinfo  {journal} {Journal of Plasma Physics}\
  }\textbf {\bibinfo {volume} {29}},\ \bibinfo {pages} {525} (\bibinfo {year}
  {1983}{\natexlab{a}})}\BibitemShut {NoStop}%
\bibitem [{\citenamefont {{Goldreich}}\ and\ \citenamefont
  {{Sridhar}}(1995)}]{goldreich1995toward}%
  \BibitemOpen
  \bibfield  {author} {\bibinfo {author} {\bibfnamefont {P.}~\bibnamefont
  {{Goldreich}}}\ and\ \bibinfo {author} {\bibfnamefont {S.}~\bibnamefont
  {{Sridhar}}},\ }\href {https://doi.org/10.1086/175121} {\bibfield  {journal}
  {\bibinfo  {journal} {\apj}\ }\textbf {\bibinfo {volume} {438}},\ \bibinfo
  {pages} {763} (\bibinfo {year} {1995})}\BibitemShut {NoStop}%
\bibitem [{\citenamefont {{Oughton}}\ \emph {et~al.}(2015)\citenamefont
  {{Oughton}}, \citenamefont {{Matthaeus}}, \citenamefont {{Wan}},\ and\
  \citenamefont {{Osman}}}]{OughtonEA15}%
  \BibitemOpen
  \bibfield  {author} {\bibinfo {author} {\bibfnamefont {S.}~\bibnamefont
  {{Oughton}}}, \bibinfo {author} {\bibfnamefont {W.~H.}\ \bibnamefont
  {{Matthaeus}}}, \bibinfo {author} {\bibfnamefont {M.}~\bibnamefont {{Wan}}},\
  and\ \bibinfo {author} {\bibfnamefont {K.~T.}\ \bibnamefont {{Osman}}},\
  }\href {https://doi.org/10.1098/rsta.2014.0152} {\bibfield  {journal}
  {\bibinfo  {journal} {Philosophical Transactions of the Royal Society of
  London Series A}\ }\textbf {\bibinfo {volume} {373}},\ \bibinfo {pages}
  {20140152} (\bibinfo {year} {2015})}\BibitemShut {NoStop}%
\bibitem [{\citenamefont {{Bieber}}\ \emph
  {et~al.}(1994{\natexlab{a}})\citenamefont {{Bieber}}, \citenamefont
  {{Matthaeus}}, \citenamefont {{Smith}}, \citenamefont {{Wanner}},
  \citenamefont {{Kallenrode}},\ and\ \citenamefont
  {{Wibberenz}}}]{BieberEA94}%
  \BibitemOpen
  \bibfield  {author} {\bibinfo {author} {\bibfnamefont {J.~W.}\ \bibnamefont
  {{Bieber}}}, \bibinfo {author} {\bibfnamefont {W.~H.}\ \bibnamefont
  {{Matthaeus}}}, \bibinfo {author} {\bibfnamefont {C.~W.}\ \bibnamefont
  {{Smith}}}, \bibinfo {author} {\bibfnamefont {W.}~\bibnamefont {{Wanner}}},
  \bibinfo {author} {\bibfnamefont {M.-B.}\ \bibnamefont {{Kallenrode}}},\ and\
  \bibinfo {author} {\bibfnamefont {G.}~\bibnamefont {{Wibberenz}}},\ }\href
  {https://doi.org/10.1086/173559} {\bibfield  {journal} {\bibinfo  {journal}
  {Astrophys. J.}\ }\textbf {\bibinfo {volume} {420}},\ \bibinfo {pages} {294}
  (\bibinfo {year} {1994}{\natexlab{a}})}\BibitemShut {NoStop}%
\bibitem [{\citenamefont {{Yan}}\ and\ \citenamefont
  {{Lazarian}}(2004)}]{2004ApJ...614..757Y}%
  \BibitemOpen
  \bibfield  {author} {\bibinfo {author} {\bibfnamefont {H.}~\bibnamefont
  {{Yan}}}\ and\ \bibinfo {author} {\bibfnamefont {A.}~\bibnamefont
  {{Lazarian}}},\ }\href {https://doi.org/10.1086/423733} {\bibfield  {journal}
  {\bibinfo  {journal} {\apj}\ }\textbf {\bibinfo {volume} {614}},\ \bibinfo
  {pages} {757} (\bibinfo {year} {2004})},\ \Eprint
  {https://arxiv.org/abs/astro-ph/0408172} {arXiv:astro-ph/0408172 [astro-ph]}
  \BibitemShut {NoStop}%
\bibitem [{\citenamefont {{Evoli}}\ and\ \citenamefont
  {{Yan}}(2014)}]{2014ApJ...782...36E}%
  \BibitemOpen
  \bibfield  {author} {\bibinfo {author} {\bibfnamefont {C.}~\bibnamefont
  {{Evoli}}}\ and\ \bibinfo {author} {\bibfnamefont {H.}~\bibnamefont
  {{Yan}}},\ }\href {https://doi.org/10.1088/0004-637X/782/1/36} {\bibfield
  {journal} {\bibinfo  {journal} {\apj}\ }\textbf {\bibinfo {volume} {782}},\
  \bibinfo {eid} {36} (\bibinfo {year} {2014})},\ \Eprint
  {https://arxiv.org/abs/1310.5732} {arXiv:1310.5732 [astro-ph.HE]}
  \BibitemShut {NoStop}%
\bibitem [{\citenamefont {{Xu}}\ \emph {et~al.}(2016)\citenamefont {{Xu}},
  \citenamefont {{Yan}},\ and\ \citenamefont
  {{Lazarian}}}]{2016ApJ...826..166X}%
  \BibitemOpen
  \bibfield  {author} {\bibinfo {author} {\bibfnamefont {S.}~\bibnamefont
  {{Xu}}}, \bibinfo {author} {\bibfnamefont {H.}~\bibnamefont {{Yan}}},\ and\
  \bibinfo {author} {\bibfnamefont {A.}~\bibnamefont {{Lazarian}}},\ }\href
  {https://doi.org/10.3847/0004-637X/826/2/166} {\bibfield  {journal} {\bibinfo
   {journal} {\apj}\ }\textbf {\bibinfo {volume} {826}},\ \bibinfo {eid} {166}
  (\bibinfo {year} {2016})},\ \Eprint {https://arxiv.org/abs/1506.05585}
  {arXiv:1506.05585 [astro-ph.HE]} \BibitemShut {NoStop}%
\bibitem [{\citenamefont {{Beresnyak}}\ \emph {et~al.}(2011)\citenamefont
  {{Beresnyak}}, \citenamefont {{Yan}},\ and\ \citenamefont
  {{Lazarian}}}]{2011ApJ...728...60B}%
  \BibitemOpen
  \bibfield  {author} {\bibinfo {author} {\bibfnamefont {A.}~\bibnamefont
  {{Beresnyak}}}, \bibinfo {author} {\bibfnamefont {H.}~\bibnamefont {{Yan}}},\
  and\ \bibinfo {author} {\bibfnamefont {A.}~\bibnamefont {{Lazarian}}},\
  }\href {https://doi.org/10.1088/0004-637X/728/1/60} {\bibfield  {journal}
  {\bibinfo  {journal} {\apj}\ }\textbf {\bibinfo {volume} {728}},\ \bibinfo
  {eid} {60} (\bibinfo {year} {2011})},\ \Eprint
  {https://arxiv.org/abs/1002.2646} {arXiv:1002.2646 [astro-ph.GA]}
  \BibitemShut {NoStop}%
\bibitem [{\citenamefont {{Cohet}}\ and\ \citenamefont
  {{Marcowith}}(2016{\natexlab{a}})}]{2016A&A...588A..73C}%
  \BibitemOpen
  \bibfield  {author} {\bibinfo {author} {\bibfnamefont {R.}~\bibnamefont
  {{Cohet}}}\ and\ \bibinfo {author} {\bibfnamefont {A.}~\bibnamefont
  {{Marcowith}}},\ }\href {https://doi.org/10.1051/0004-6361/201527376}
  {\bibfield  {journal} {\bibinfo  {journal} {\aap}\ }\textbf {\bibinfo
  {volume} {588}},\ \bibinfo {eid} {A73} (\bibinfo {year}
  {2016}{\natexlab{a}})},\ \Eprint {https://arxiv.org/abs/1601.04971}
  {arXiv:1601.04971 [astro-ph.HE]} \BibitemShut {NoStop}%
\bibitem [{\citenamefont {{Bieber}}\ \emph
  {et~al.}(1994{\natexlab{b}})\citenamefont {{Bieber}}, \citenamefont
  {{Matthaeus}}, \citenamefont {{Smith}}, \citenamefont {{Wanner}},
  \citenamefont {{Kallenrode}},\ and\ \citenamefont
  {{Wibberenz}}}]{bieber1994proton}%
  \BibitemOpen
  \bibfield  {author} {\bibinfo {author} {\bibfnamefont {J.~W.}\ \bibnamefont
  {{Bieber}}}, \bibinfo {author} {\bibfnamefont {W.~H.}\ \bibnamefont
  {{Matthaeus}}}, \bibinfo {author} {\bibfnamefont {C.~W.}\ \bibnamefont
  {{Smith}}}, \bibinfo {author} {\bibfnamefont {W.}~\bibnamefont {{Wanner}}},
  \bibinfo {author} {\bibfnamefont {M.-B.}\ \bibnamefont {{Kallenrode}}},\ and\
  \bibinfo {author} {\bibfnamefont {G.}~\bibnamefont {{Wibberenz}}},\ }\href
  {https://doi.org/10.1086/173559} {\bibfield  {journal} {\bibinfo  {journal}
  {\apj}\ }\textbf {\bibinfo {volume} {420}},\ \bibinfo {pages} {294} (\bibinfo
  {year} {1994}{\natexlab{b}})}\BibitemShut {NoStop}%
\bibitem [{\citenamefont {{Subedi}}\ \emph {et~al.}(2017)\citenamefont
  {{Subedi}}, \citenamefont {{Sonsrettee}}, \citenamefont {{Blasi}},
  \citenamefont {{Ruffolo}}, \citenamefont {{Matthaeus}}, \citenamefont
  {{Montgomery}}, \citenamefont {{Chuychai}}, \citenamefont {{Dmitruk}},
  \citenamefont {{Wan}}, \citenamefont {{Parashar}},\ and\ \citenamefont
  {{Chhiber}}}]{subedi2017charged}%
  \BibitemOpen
  \bibfield  {author} {\bibinfo {author} {\bibfnamefont {P.}~\bibnamefont
  {{Subedi}}}, \bibinfo {author} {\bibfnamefont {W.}~\bibnamefont
  {{Sonsrettee}}}, \bibinfo {author} {\bibfnamefont {P.}~\bibnamefont
  {{Blasi}}}, \bibinfo {author} {\bibfnamefont {D.}~\bibnamefont {{Ruffolo}}},
  \bibinfo {author} {\bibfnamefont {W.~H.}\ \bibnamefont {{Matthaeus}}},
  \bibinfo {author} {\bibfnamefont {D.}~\bibnamefont {{Montgomery}}}, \bibinfo
  {author} {\bibfnamefont {P.}~\bibnamefont {{Chuychai}}}, \bibinfo {author}
  {\bibfnamefont {P.}~\bibnamefont {{Dmitruk}}}, \bibinfo {author}
  {\bibfnamefont {M.}~\bibnamefont {{Wan}}}, \bibinfo {author} {\bibfnamefont
  {T.~N.}\ \bibnamefont {{Parashar}}},\ and\ \bibinfo {author} {\bibfnamefont
  {R.}~\bibnamefont {{Chhiber}}},\ }\href
  {https://doi.org/10.3847/1538-4357/aa603a} {\bibfield  {journal} {\bibinfo
  {journal} {\apj}\ }\textbf {\bibinfo {volume} {837}},\ \bibinfo {eid} {140}
  (\bibinfo {year} {2017})},\ \Eprint {https://arxiv.org/abs/1612.09507}
  {arXiv:1612.09507 [physics.space-ph]} \BibitemShut {NoStop}%
\bibitem [{\citenamefont {{Robinson}}\ and\ \citenamefont
  {{Rusbridge}}(1971)}]{RobinsonRusbridge71}%
  \BibitemOpen
  \bibfield  {author} {\bibinfo {author} {\bibfnamefont {D.~C.}\ \bibnamefont
  {{Robinson}}}\ and\ \bibinfo {author} {\bibfnamefont {M.~G.}\ \bibnamefont
  {{Rusbridge}}},\ }\href {https://doi.org/10.1063/1.1693359} {\bibfield
  {journal} {\bibinfo  {journal} {Physics of Fluids}\ }\textbf {\bibinfo
  {volume} {14}},\ \bibinfo {pages} {2499} (\bibinfo {year}
  {1971})}\BibitemShut {NoStop}%
\bibitem [{\citenamefont {{Shebalin}}\ \emph
  {et~al.}(1983{\natexlab{b}})\citenamefont {{Shebalin}}, \citenamefont
  {{Matthaeus}},\ and\ \citenamefont {{Montgomery}}}]{shebalin1983anisotropy}%
  \BibitemOpen
  \bibfield  {author} {\bibinfo {author} {\bibfnamefont {J.~V.}\ \bibnamefont
  {{Shebalin}}}, \bibinfo {author} {\bibfnamefont {W.~H.}\ \bibnamefont
  {{Matthaeus}}},\ and\ \bibinfo {author} {\bibfnamefont {D.}~\bibnamefont
  {{Montgomery}}},\ }\href {https://doi.org/10.1017/S0022377800000933}
  {\bibfield  {journal} {\bibinfo  {journal} {Journal of Plasma Physics}\
  }\textbf {\bibinfo {volume} {29}},\ \bibinfo {pages} {525} (\bibinfo {year}
  {1983}{\natexlab{b}})}\BibitemShut {NoStop}%
\bibitem [{\citenamefont {{Oughton}}\ \emph {et~al.}(2017)\citenamefont
  {{Oughton}}, \citenamefont {{Matthaeus}},\ and\ \citenamefont
  {{Dmitruk}}}]{OughtonEA17-rmhd}%
  \BibitemOpen
  \bibfield  {author} {\bibinfo {author} {\bibfnamefont {S.}~\bibnamefont
  {{Oughton}}}, \bibinfo {author} {\bibfnamefont {W.~H.}\ \bibnamefont
  {{Matthaeus}}},\ and\ \bibinfo {author} {\bibfnamefont {P.}~\bibnamefont
  {{Dmitruk}}},\ }\href {https://doi.org/10.3847/1538-4357/aa67e2} {\bibfield
  {journal} {\bibinfo  {journal} {\apj}\ }\textbf {\bibinfo {volume} {839}},\
  \bibinfo {eid} {2} (\bibinfo {year} {2017})}\BibitemShut {NoStop}%
\bibitem [{\citenamefont {{Oughton}}\ and\ \citenamefont
  {{Matthaeus}}(2020)}]{OughtonMatthaeus20}%
  \BibitemOpen
  \bibfield  {author} {\bibinfo {author} {\bibfnamefont {S.}~\bibnamefont
  {{Oughton}}}\ and\ \bibinfo {author} {\bibfnamefont {W.~H.}\ \bibnamefont
  {{Matthaeus}}},\ }\href {https://doi.org/10.3847/1538-4357/ab8f2a} {\bibfield
   {journal} {\bibinfo  {journal} {\apj}\ }\textbf {\bibinfo {volume} {897}},\
  \bibinfo {eid} {37} (\bibinfo {year} {2020})},\ \Eprint
  {https://arxiv.org/abs/2006.04677} {arXiv:2006.04677 [physics.plasm-ph]}
  \BibitemShut {NoStop}%
\bibitem [{\citenamefont {{Kulsrud}}\ and\ \citenamefont
  {{Pearce}}(1969)}]{1969ApJ...156..445K}%
  \BibitemOpen
  \bibfield  {author} {\bibinfo {author} {\bibfnamefont {R.}~\bibnamefont
  {{Kulsrud}}}\ and\ \bibinfo {author} {\bibfnamefont {W.~P.}\ \bibnamefont
  {{Pearce}}},\ }\href {https://doi.org/10.1086/149981} {\bibfield  {journal}
  {\bibinfo  {journal} {\apj}\ }\textbf {\bibinfo {volume} {156}},\ \bibinfo
  {pages} {445} (\bibinfo {year} {1969})}\BibitemShut {NoStop}%
\bibitem [{\citenamefont {{Holmes}}(1975)}]{1975MNRAS.170..251H}%
  \BibitemOpen
  \bibfield  {author} {\bibinfo {author} {\bibfnamefont {J.~A.}\ \bibnamefont
  {{Holmes}}},\ }\href {https://doi.org/10.1093/mnras/170.2.251} {\bibfield
  {journal} {\bibinfo  {journal} {MNRAS}\ }\textbf {\bibinfo {volume} {170}},\
  \bibinfo {pages} {251} (\bibinfo {year} {1975})}\BibitemShut {NoStop}%
\bibitem [{\citenamefont {{Bell}}(2004)}]{2004MNRAS.353..550B}%
  \BibitemOpen
  \bibfield  {author} {\bibinfo {author} {\bibfnamefont {A.~R.}\ \bibnamefont
  {{Bell}}},\ }\href {https://doi.org/10.1111/j.1365-2966.2004.08097.x}
  {\bibfield  {journal} {\bibinfo  {journal} {MNRAS}\ }\textbf {\bibinfo
  {volume} {353}},\ \bibinfo {pages} {550} (\bibinfo {year}
  {2004})}\BibitemShut {NoStop}%
\bibitem [{\citenamefont {{Amato}}\ and\ \citenamefont
  {{Blasi}}(2018)}]{2018AdSpR..62.2731A}%
  \BibitemOpen
  \bibfield  {author} {\bibinfo {author} {\bibfnamefont {E.}~\bibnamefont
  {{Amato}}}\ and\ \bibinfo {author} {\bibfnamefont {P.}~\bibnamefont
  {{Blasi}}},\ }\href {https://doi.org/10.1016/j.asr.2017.04.019} {\bibfield
  {journal} {\bibinfo  {journal} {Advances in Space Research}\ }\textbf
  {\bibinfo {volume} {62}},\ \bibinfo {pages} {2731} (\bibinfo {year}
  {2018})},\ \Eprint {https://arxiv.org/abs/1704.05696} {arXiv:1704.05696
  [astro-ph.HE]} \BibitemShut {NoStop}%
\bibitem [{\citenamefont {{Blasi}}(2019{\natexlab{b}})}]{2019Galax...7...64B}%
  \BibitemOpen
  \bibfield  {author} {\bibinfo {author} {\bibfnamefont {P.}~\bibnamefont
  {{Blasi}}},\ }\href {https://doi.org/10.3390/galaxies7020064} {\bibfield
  {journal} {\bibinfo  {journal} {Galaxies}\ }\textbf {\bibinfo {volume} {7}},\
  \bibinfo {pages} {64} (\bibinfo {year} {2019}{\natexlab{b}})},\ \Eprint
  {https://arxiv.org/abs/1905.11149} {arXiv:1905.11149 [astro-ph.HE]}
  \BibitemShut {NoStop}%
\bibitem [{\citenamefont {{Blasi}}\ and\ \citenamefont
  {{Amato}}(2019)}]{2019PhRvL.122e1101B}%
  \BibitemOpen
  \bibfield  {author} {\bibinfo {author} {\bibfnamefont {P.}~\bibnamefont
  {{Blasi}}}\ and\ \bibinfo {author} {\bibfnamefont {E.}~\bibnamefont
  {{Amato}}},\ }\href {https://doi.org/10.1103/PhysRevLett.122.051101}
  {\bibfield  {journal} {\bibinfo  {journal} {Phys. Rev. Lett.}\ }\textbf
  {\bibinfo {volume} {122}},\ \bibinfo {eid} {051101} (\bibinfo {year}
  {2019})},\ \Eprint {https://arxiv.org/abs/1901.03609} {arXiv:1901.03609
  [astro-ph.HE]} \BibitemShut {NoStop}%
\bibitem [{\citenamefont {{Hussein}}\ \emph {et~al.}(2015)\citenamefont
  {{Hussein}}, \citenamefont {{Tautz}},\ and\ \citenamefont
  {{Shalchi}}}]{hussein2015influence}%
  \BibitemOpen
  \bibfield  {author} {\bibinfo {author} {\bibfnamefont {M.}~\bibnamefont
  {{Hussein}}}, \bibinfo {author} {\bibfnamefont {R.~C.}\ \bibnamefont
  {{Tautz}}},\ and\ \bibinfo {author} {\bibfnamefont {A.}~\bibnamefont
  {{Shalchi}}},\ }\href {https://doi.org/10.1002/2015JA021060} {\bibfield
  {journal} {\bibinfo  {journal} {Journal of Geophysical Research (Space
  Physics)}\ }\textbf {\bibinfo {volume} {120}},\ \bibinfo {pages} {4095}
  (\bibinfo {year} {2015})},\ \Eprint {https://arxiv.org/abs/1505.05099}
  {arXiv:1505.05099 [astro-ph.SR]} \BibitemShut {NoStop}%
\bibitem [{\citenamefont {{Snodin}}\ \emph {et~al.}(2016)\citenamefont
  {{Snodin}}, \citenamefont {{Shukurov}}, \citenamefont {{Sarson}},
  \citenamefont {{Bushby}},\ and\ \citenamefont
  {{Rodrigues}}}]{2016MNRAS.457.3975S}%
  \BibitemOpen
  \bibfield  {author} {\bibinfo {author} {\bibfnamefont {A.~P.}\ \bibnamefont
  {{Snodin}}}, \bibinfo {author} {\bibfnamefont {A.}~\bibnamefont
  {{Shukurov}}}, \bibinfo {author} {\bibfnamefont {G.~R.}\ \bibnamefont
  {{Sarson}}}, \bibinfo {author} {\bibfnamefont {P.~J.}\ \bibnamefont
  {{Bushby}}},\ and\ \bibinfo {author} {\bibfnamefont {L.~F.~S.}\ \bibnamefont
  {{Rodrigues}}},\ }\href {https://doi.org/10.1093/mnras/stw217} {\bibfield
  {journal} {\bibinfo  {journal} {\mnras}\ }\textbf {\bibinfo {volume} {457}},\
  \bibinfo {pages} {3975} (\bibinfo {year} {2016})},\ \Eprint
  {https://arxiv.org/abs/1509.03766} {arXiv:1509.03766 [astro-ph.HE]}
  \BibitemShut {NoStop}%
\bibitem [{\citenamefont {{Cohet}}\ and\ \citenamefont
  {{Marcowith}}(2016{\natexlab{b}})}]{cohet2016cosmic}%
  \BibitemOpen
  \bibfield  {author} {\bibinfo {author} {\bibfnamefont {R.}~\bibnamefont
  {{Cohet}}}\ and\ \bibinfo {author} {\bibfnamefont {A.}~\bibnamefont
  {{Marcowith}}},\ }\href {https://doi.org/10.1051/0004-6361/201527376}
  {\bibfield  {journal} {\bibinfo  {journal} {\aap}\ }\textbf {\bibinfo
  {volume} {588}},\ \bibinfo {eid} {A73} (\bibinfo {year}
  {2016}{\natexlab{b}})},\ \Eprint {https://arxiv.org/abs/1601.04971}
  {arXiv:1601.04971 [astro-ph.HE]} \BibitemShut {NoStop}%
\bibitem [{\citenamefont {{Mertsch}}(2019)}]{mertsch2019test}%
  \BibitemOpen
  \bibfield  {author} {\bibinfo {author} {\bibfnamefont {P.}~\bibnamefont
  {{Mertsch}}},\ }\href@noop {} {\bibfield  {journal} {\bibinfo  {journal}
  {arXiv e-prints}\ ,\ \bibinfo {eid} {arXiv:1910.01172}} (\bibinfo {year}
  {2019})},\ \Eprint {https://arxiv.org/abs/1910.01172} {arXiv:1910.01172
  [astro-ph.HE]} \BibitemShut {NoStop}%
\bibitem [{\citenamefont {{Batchelor}}(1982)}]{batchelor1982}%
  \BibitemOpen
  \bibfield  {author} {\bibinfo {author} {\bibfnamefont {G.~K.}\ \bibnamefont
  {{Batchelor}}},\ }\href@noop {} {\emph {\bibinfo {title} {{The Theory of
  Homogeneous Turbulence}}}}\ (\bibinfo {year} {1982})\BibitemShut {NoStop}%
\bibitem [{\citenamefont {{Montgomery}}\ and\ \citenamefont
  {{Turner}}(1981)}]{montgomery1981anisotropic}%
  \BibitemOpen
  \bibfield  {author} {\bibinfo {author} {\bibfnamefont {D.}~\bibnamefont
  {{Montgomery}}}\ and\ \bibinfo {author} {\bibfnamefont {L.}~\bibnamefont
  {{Turner}}},\ }\href {https://doi.org/10.1063/1.863455} {\bibfield  {journal}
  {\bibinfo  {journal} {Physics of Fluids}\ }\textbf {\bibinfo {volume} {24}},\
  \bibinfo {pages} {825} (\bibinfo {year} {1981})}\BibitemShut {NoStop}%
\bibitem [{\citenamefont {{Matthaeus}}\ and\ \citenamefont
  {{Smith}}(1981)}]{matthaeus1981structure}%
  \BibitemOpen
  \bibfield  {author} {\bibinfo {author} {\bibfnamefont {W.~H.}\ \bibnamefont
  {{Matthaeus}}}\ and\ \bibinfo {author} {\bibfnamefont {C.}~\bibnamefont
  {{Smith}}},\ }\href {https://doi.org/10.1103/PhysRevA.24.2135} {\bibfield
  {journal} {\bibinfo  {journal} {\pra}\ }\textbf {\bibinfo {volume} {24}},\
  \bibinfo {pages} {2135} (\bibinfo {year} {1981})}\BibitemShut {NoStop}%
\bibitem [{\citenamefont {Oughton}\ \emph {et~al.}(1997)\citenamefont
  {Oughton}, \citenamefont {R{\"a}dler},\ and\ \citenamefont
  {Matthaeus}}]{oughton1997general}%
  \BibitemOpen
  \bibfield  {author} {\bibinfo {author} {\bibfnamefont {S.}~\bibnamefont
  {Oughton}}, \bibinfo {author} {\bibfnamefont {K.-H.}\ \bibnamefont
  {R{\"a}dler}},\ and\ \bibinfo {author} {\bibfnamefont {W.~H.}\ \bibnamefont
  {Matthaeus}},\ }\href@noop {} {\bibfield  {journal} {\bibinfo  {journal}
  {Physical Review E}\ }\textbf {\bibinfo {volume} {56}},\ \bibinfo {pages}
  {2875} (\bibinfo {year} {1997})}\BibitemShut {NoStop}%
\bibitem [{\citenamefont {Monin}\ and\ \citenamefont
  {Yaglom}(2013)}]{monin2013statistical}%
  \BibitemOpen
  \bibfield  {author} {\bibinfo {author} {\bibfnamefont {A.~S.}\ \bibnamefont
  {Monin}}\ and\ \bibinfo {author} {\bibfnamefont {A.~M.}\ \bibnamefont
  {Yaglom}},\ }\href@noop {} {\emph {\bibinfo {title} {Statistical fluid
  mechanics, volume II: Mechanics of Turbulence}}},\ Vol.~\bibinfo {volume}
  {2}\ (\bibinfo  {publisher} {Courier Corporation},\ \bibinfo {year}
  {2013})\BibitemShut {NoStop}%
\bibitem [{\citenamefont {{Davidson}}(2015)}]{davidson2015turbulence}%
  \BibitemOpen
  \bibfield  {author} {\bibinfo {author} {\bibfnamefont {P.~A.}\ \bibnamefont
  {{Davidson}}},\ }\href@noop {} {\emph {\bibinfo {title} {Turbulence: An
  Introduction for Scientists and Engineers, 2nd ed.}}}\ (\bibinfo  {publisher}
  {Oxford University Press},\ \bibinfo {year} {2015})\BibitemShut {NoStop}%
\bibitem [{\citenamefont {{Matthaeus}}\ \emph {et~al.}(2007)\citenamefont
  {{Matthaeus}}, \citenamefont {{Bieber}}, \citenamefont {{Ruffolo}},
  \citenamefont {{Chuychai}},\ and\ \citenamefont
  {{Minnie}}}]{matthaeus2007spectral}%
  \BibitemOpen
  \bibfield  {author} {\bibinfo {author} {\bibfnamefont {W.~H.}\ \bibnamefont
  {{Matthaeus}}}, \bibinfo {author} {\bibfnamefont {J.~W.}\ \bibnamefont
  {{Bieber}}}, \bibinfo {author} {\bibfnamefont {D.}~\bibnamefont {{Ruffolo}}},
  \bibinfo {author} {\bibfnamefont {P.}~\bibnamefont {{Chuychai}}},\ and\
  \bibinfo {author} {\bibfnamefont {J.}~\bibnamefont {{Minnie}}},\ }\href
  {https://doi.org/10.1086/520924} {\bibfield  {journal} {\bibinfo  {journal}
  {\apj}\ }\textbf {\bibinfo {volume} {667}},\ \bibinfo {pages} {956} (\bibinfo
  {year} {2007})}\BibitemShut {NoStop}%
\bibitem [{\citenamefont {{Sonsrettee}}\ \emph {et~al.}(2015)\citenamefont
  {{Sonsrettee}}, \citenamefont {{Subedi}}, \citenamefont {{Ruffolo}},
  \citenamefont {{Matthaeus}}, \citenamefont {{Snodin}}, \citenamefont
  {{Wongpan}},\ and\ \citenamefont {{Chuychai}}}]{sonsrettee2015magnetic}%
  \BibitemOpen
  \bibfield  {author} {\bibinfo {author} {\bibfnamefont {W.}~\bibnamefont
  {{Sonsrettee}}}, \bibinfo {author} {\bibfnamefont {P.}~\bibnamefont
  {{Subedi}}}, \bibinfo {author} {\bibfnamefont {D.}~\bibnamefont {{Ruffolo}}},
  \bibinfo {author} {\bibfnamefont {W.~H.}\ \bibnamefont {{Matthaeus}}},
  \bibinfo {author} {\bibfnamefont {A.~P.}\ \bibnamefont {{Snodin}}}, \bibinfo
  {author} {\bibfnamefont {P.}~\bibnamefont {{Wongpan}}},\ and\ \bibinfo
  {author} {\bibfnamefont {P.}~\bibnamefont {{Chuychai}}},\ }\href
  {https://doi.org/10.1088/0004-637X/798/1/59} {\bibfield  {journal} {\bibinfo
  {journal} {\apj}\ }\textbf {\bibinfo {volume} {798}},\ \bibinfo {eid} {59}
  (\bibinfo {year} {2015})}\BibitemShut {NoStop}%
\bibitem [{\citenamefont {{Shalchi}}\ and\ \citenamefont
  {{Weinhorst}}(2009)}]{shalchi2009random}%
  \BibitemOpen
  \bibfield  {author} {\bibinfo {author} {\bibfnamefont {A.}~\bibnamefont
  {{Shalchi}}}\ and\ \bibinfo {author} {\bibfnamefont {B.}~\bibnamefont
  {{Weinhorst}}},\ }\href {https://doi.org/10.1016/j.asr.2008.12.022}
  {\bibfield  {journal} {\bibinfo  {journal} {Advances in Space Research}\
  }\textbf {\bibinfo {volume} {43}},\ \bibinfo {pages} {1429} (\bibinfo {year}
  {2009})}\BibitemShut {NoStop}%
\bibitem [{\citenamefont {{Bieber}}\ \emph {et~al.}(1993)\citenamefont
  {{Bieber}}, \citenamefont {{Chen}}, \citenamefont {{Matthaeus}},
  \citenamefont {{Smith}},\ and\ \citenamefont
  {{Pomerantz}}}]{bieber1993longterm}%
  \BibitemOpen
  \bibfield  {author} {\bibinfo {author} {\bibfnamefont {J.~W.}\ \bibnamefont
  {{Bieber}}}, \bibinfo {author} {\bibfnamefont {J.}~\bibnamefont {{Chen}}},
  \bibinfo {author} {\bibfnamefont {W.~H.}\ \bibnamefont {{Matthaeus}}},
  \bibinfo {author} {\bibfnamefont {C.~W.}\ \bibnamefont {{Smith}}},\ and\
  \bibinfo {author} {\bibfnamefont {M.~A.}\ \bibnamefont {{Pomerantz}}},\
  }\href {https://doi.org/10.1029/92JA02566} {\bibfield  {journal} {\bibinfo
  {journal} {\jgr}\ }\textbf {\bibinfo {volume} {98}},\ \bibinfo {pages} {3585}
  (\bibinfo {year} {1993})}\BibitemShut {NoStop}%
\bibitem [{\citenamefont {{Matthaeus}}\ \emph {et~al.}(1990)\citenamefont
  {{Matthaeus}}, \citenamefont {{Goldstein}},\ and\ \citenamefont
  {{Roberts}}}]{matthaeus1990evidence}%
  \BibitemOpen
  \bibfield  {author} {\bibinfo {author} {\bibfnamefont {W.~H.}\ \bibnamefont
  {{Matthaeus}}}, \bibinfo {author} {\bibfnamefont {M.~L.}\ \bibnamefont
  {{Goldstein}}},\ and\ \bibinfo {author} {\bibfnamefont {D.~A.}\ \bibnamefont
  {{Roberts}}},\ }\href {https://doi.org/10.1029/JA095iA12p20673} {\bibfield
  {journal} {\bibinfo  {journal} {\jgr}\ }\textbf {\bibinfo {volume} {95}},\
  \bibinfo {pages} {20673} (\bibinfo {year} {1990})}\BibitemShut {NoStop}%
\bibitem [{\citenamefont {{Shalchi}}(2009)}]{shalchi2009}%
  \BibitemOpen
  \bibfield  {author} {\bibinfo {author} {\bibfnamefont {A.}~\bibnamefont
  {{Shalchi}}},\ }\href {https://doi.org/10.1007/978-3-642-00309-7} {\emph
  {\bibinfo {title} {{Nonlinear Cosmic Ray Diffusion Theories}}}},\ Vol.\
  \bibinfo {volume} {362}\ (\bibinfo {year} {2009})\BibitemShut {NoStop}%
\bibitem [{\citenamefont {{Bieber}}\ \emph {et~al.}(1996)\citenamefont
  {{Bieber}}, \citenamefont {{Wanner}},\ and\ \citenamefont
  {{Matthaeus}}}]{bieber1996dominant}%
  \BibitemOpen
  \bibfield  {author} {\bibinfo {author} {\bibfnamefont {J.~W.}\ \bibnamefont
  {{Bieber}}}, \bibinfo {author} {\bibfnamefont {W.}~\bibnamefont {{Wanner}}},\
  and\ \bibinfo {author} {\bibfnamefont {W.~H.}\ \bibnamefont {{Matthaeus}}},\
  }\href {https://doi.org/10.1029/95JA02588} {\bibfield  {journal} {\bibinfo
  {journal} {\jgr}\ }\textbf {\bibinfo {volume} {101}},\ \bibinfo {pages}
  {2511} (\bibinfo {year} {1996})}\BibitemShut {NoStop}%
\bibitem [{\citenamefont {{Hossain}}\ \emph {et~al.}(1995)\citenamefont
  {{Hossain}}, \citenamefont {{Gray}}, \citenamefont {{Pontius}}, \citenamefont
  {{Matthaeus}},\ and\ \citenamefont {{Oughton}}}]{hossain1995phenomenology}%
  \BibitemOpen
  \bibfield  {author} {\bibinfo {author} {\bibfnamefont {M.}~\bibnamefont
  {{Hossain}}}, \bibinfo {author} {\bibfnamefont {P.~C.}\ \bibnamefont
  {{Gray}}}, \bibinfo {author} {\bibfnamefont {J.}~\bibnamefont {{Pontius}},
  \bibfnamefont {Duane~H.}}, \bibinfo {author} {\bibfnamefont {W.~H.}\
  \bibnamefont {{Matthaeus}}},\ and\ \bibinfo {author} {\bibfnamefont
  {S.}~\bibnamefont {{Oughton}}},\ }\href {https://doi.org/10.1063/1.868665}
  {\bibfield  {journal} {\bibinfo  {journal} {Physics of Fluids}\ }\textbf
  {\bibinfo {volume} {7}},\ \bibinfo {pages} {2886} (\bibinfo {year}
  {1995})}\BibitemShut {NoStop}%
\bibitem [{\citenamefont {{Shalchi}}\ \emph {et~al.}(2004)\citenamefont
  {{Shalchi}}, \citenamefont {{Bieber}}, \citenamefont {{Matthaeus}},\ and\
  \citenamefont {{Qin}}}]{shalchi2004nonlinear}%
  \BibitemOpen
  \bibfield  {author} {\bibinfo {author} {\bibfnamefont {A.}~\bibnamefont
  {{Shalchi}}}, \bibinfo {author} {\bibfnamefont {J.~W.}\ \bibnamefont
  {{Bieber}}}, \bibinfo {author} {\bibfnamefont {W.~H.}\ \bibnamefont
  {{Matthaeus}}},\ and\ \bibinfo {author} {\bibfnamefont {G.}~\bibnamefont
  {{Qin}}},\ }\href {https://doi.org/10.1086/424839} {\bibfield  {journal}
  {\bibinfo  {journal} {\apj}\ }\textbf {\bibinfo {volume} {616}},\ \bibinfo
  {pages} {617} (\bibinfo {year} {2004})}\BibitemShut {NoStop}%
\bibitem [{\citenamefont {{Alves Batista}}\ \emph {et~al.}(2016)\citenamefont
  {{Alves Batista}}, \citenamefont {{Dundovic}}, \citenamefont {{Erdmann}},
  \citenamefont {{Kampert}}, \citenamefont {{Kuempel}}, \citenamefont
  {{M{\"u}ller}}, \citenamefont {{Sigl}}, \citenamefont {{van Vliet}},
  \citenamefont {{Walz}},\ and\ \citenamefont
  {{Winchen}}}]{alvesbatista2016CRPROPA}%
  \BibitemOpen
  \bibfield  {author} {\bibinfo {author} {\bibfnamefont {R.}~\bibnamefont
  {{Alves Batista}}}, \bibinfo {author} {\bibfnamefont {A.}~\bibnamefont
  {{Dundovic}}}, \bibinfo {author} {\bibfnamefont {M.}~\bibnamefont
  {{Erdmann}}}, \bibinfo {author} {\bibfnamefont {K.-H.}\ \bibnamefont
  {{Kampert}}}, \bibinfo {author} {\bibfnamefont {D.}~\bibnamefont
  {{Kuempel}}}, \bibinfo {author} {\bibfnamefont {G.}~\bibnamefont
  {{M{\"u}ller}}}, \bibinfo {author} {\bibfnamefont {G.}~\bibnamefont
  {{Sigl}}}, \bibinfo {author} {\bibfnamefont {A.}~\bibnamefont {{van Vliet}}},
  \bibinfo {author} {\bibfnamefont {D.}~\bibnamefont {{Walz}}},\ and\ \bibinfo
  {author} {\bibfnamefont {T.}~\bibnamefont {{Winchen}}},\ }\href
  {https://doi.org/10.1088/1475-7516/2016/05/038} {\bibfield  {journal}
  {\bibinfo  {journal} {\jcap}\ }\textbf {\bibinfo {volume} {2016}},\ \bibinfo
  {eid} {038} (\bibinfo {year} {2016})},\ \Eprint
  {https://arxiv.org/abs/1603.07142} {arXiv:1603.07142 [astro-ph.IM]}
  \BibitemShut {NoStop}%
\bibitem [{\citenamefont {Frigo}\ and\ \citenamefont {Johnson}(2005)}]{FFTW05}%
  \BibitemOpen
  \bibfield  {author} {\bibinfo {author} {\bibfnamefont {M.}~\bibnamefont
  {Frigo}}\ and\ \bibinfo {author} {\bibfnamefont {S.~G.}\ \bibnamefont
  {Johnson}},\ }\href@noop {} {\bibfield  {journal} {\bibinfo  {journal}
  {Proceedings of the IEEE}\ }\textbf {\bibinfo {volume} {93}},\ \bibinfo
  {pages} {216} (\bibinfo {year} {2005})},\ \bibinfo {note} {special issue on
  ``Program Generation, Optimization, and Platform Adaptation''}\BibitemShut
  {NoStop}%
\bibitem [{\citenamefont {Press}\ \emph {et~al.}(1992)\citenamefont {Press},
  \citenamefont {Teukolsky}, \citenamefont {Flannery},\ and\ \citenamefont
  {Vetterling}}]{press1992numerical}%
  \BibitemOpen
  \bibfield  {author} {\bibinfo {author} {\bibfnamefont {W.~H.}\ \bibnamefont
  {Press}}, \bibinfo {author} {\bibfnamefont {S.~A.}\ \bibnamefont
  {Teukolsky}}, \bibinfo {author} {\bibfnamefont {B.~P.}\ \bibnamefont
  {Flannery}},\ and\ \bibinfo {author} {\bibfnamefont {W.~T.}\ \bibnamefont
  {Vetterling}},\ }\href@noop {} {\emph {\bibinfo {title} {Numerical recipes in
  Fortran 77: volume 1, volume 1 of Fortran numerical recipes: the art of
  scientific computing}}}\ (\bibinfo  {publisher} {Cambridge university
  press},\ \bibinfo {year} {1992})\BibitemShut {NoStop}%
\bibitem [{\citenamefont {{Birdsall}}\ and\ \citenamefont
  {{Langdon}}(1991)}]{birdsall1991}%
  \BibitemOpen
  \bibfield  {author} {\bibinfo {author} {\bibfnamefont {C.~K.}\ \bibnamefont
  {{Birdsall}}}\ and\ \bibinfo {author} {\bibfnamefont {A.~B.}\ \bibnamefont
  {{Langdon}}},\ }\href@noop {} {\emph {\bibinfo {title} {{Plasma Physics via
  Computer Simulation}}}}\ (\bibinfo {year} {1991})\BibitemShut {NoStop}%
\bibitem [{\citenamefont {{Juneja}}\ \emph {et~al.}(1994)\citenamefont
  {{Juneja}}, \citenamefont {{Lathrop}}, \citenamefont {{Sreenivasan}},\ and\
  \citenamefont {{Stolovitzky}}}]{juneja1994synthetic}%
  \BibitemOpen
  \bibfield  {author} {\bibinfo {author} {\bibfnamefont {A.}~\bibnamefont
  {{Juneja}}}, \bibinfo {author} {\bibfnamefont {D.~P.}\ \bibnamefont
  {{Lathrop}}}, \bibinfo {author} {\bibfnamefont {K.~R.}\ \bibnamefont
  {{Sreenivasan}}},\ and\ \bibinfo {author} {\bibfnamefont {G.}~\bibnamefont
  {{Stolovitzky}}},\ }\href {https://doi.org/10.1103/PhysRevE.49.5179}
  {\bibfield  {journal} {\bibinfo  {journal} {\pre}\ }\textbf {\bibinfo
  {volume} {49}},\ \bibinfo {pages} {5179} (\bibinfo {year}
  {1994})}\BibitemShut {NoStop}%
\bibitem [{\citenamefont {{Giacalone}}\ and\ \citenamefont
  {{Jokipii}}(1999)}]{giacalone1999transport}%
  \BibitemOpen
  \bibfield  {author} {\bibinfo {author} {\bibfnamefont {J.}~\bibnamefont
  {{Giacalone}}}\ and\ \bibinfo {author} {\bibfnamefont {J.~R.}\ \bibnamefont
  {{Jokipii}}},\ }\href {https://doi.org/10.1086/307452} {\bibfield  {journal}
  {\bibinfo  {journal} {\apj}\ }\textbf {\bibinfo {volume} {520}},\ \bibinfo
  {pages} {204} (\bibinfo {year} {1999})}\BibitemShut {NoStop}%
\bibitem [{\citenamefont {{Malara}}\ \emph {et~al.}(2016)\citenamefont
  {{Malara}}, \citenamefont {{Di Mare}}, \citenamefont {{Nigro}},\ and\
  \citenamefont {{Sorriso-Valvo}}}]{malara2016fast}%
  \BibitemOpen
  \bibfield  {author} {\bibinfo {author} {\bibfnamefont {F.}~\bibnamefont
  {{Malara}}}, \bibinfo {author} {\bibfnamefont {F.}~\bibnamefont {{Di Mare}}},
  \bibinfo {author} {\bibfnamefont {G.}~\bibnamefont {{Nigro}}},\ and\ \bibinfo
  {author} {\bibfnamefont {L.}~\bibnamefont {{Sorriso-Valvo}}},\ }\href
  {https://doi.org/10.1103/PhysRevE.94.053109} {\bibfield  {journal} {\bibinfo
  {journal} {\pre}\ }\textbf {\bibinfo {volume} {94}},\ \bibinfo {eid} {053109}
  (\bibinfo {year} {2016})},\ \Eprint {https://arxiv.org/abs/1610.07333}
  {arXiv:1610.07333 [physics.flu-dyn]} \BibitemShut {NoStop}%
\bibitem [{\citenamefont {{Pucci}}\ \emph {et~al.}(2016)\citenamefont
  {{Pucci}}, \citenamefont {{Malara}}, \citenamefont {{Perri}}, \citenamefont
  {{Zimbardo}}, \citenamefont {{Sorriso-Valvo}},\ and\ \citenamefont
  {{Valentini}}}]{pucci2016energetic}%
  \BibitemOpen
  \bibfield  {author} {\bibinfo {author} {\bibfnamefont {F.}~\bibnamefont
  {{Pucci}}}, \bibinfo {author} {\bibfnamefont {F.}~\bibnamefont {{Malara}}},
  \bibinfo {author} {\bibfnamefont {S.}~\bibnamefont {{Perri}}}, \bibinfo
  {author} {\bibfnamefont {G.}~\bibnamefont {{Zimbardo}}}, \bibinfo {author}
  {\bibfnamefont {L.}~\bibnamefont {{Sorriso-Valvo}}},\ and\ \bibinfo {author}
  {\bibfnamefont {F.}~\bibnamefont {{Valentini}}},\ }\href
  {https://doi.org/10.1093/mnras/stw877} {\bibfield  {journal} {\bibinfo
  {journal} {\mnras}\ }\textbf {\bibinfo {volume} {459}},\ \bibinfo {pages}
  {3395} (\bibinfo {year} {2016})}\BibitemShut {NoStop}%
\bibitem [{\citenamefont {{Mace}}\ \emph {et~al.}(2012)\citenamefont {{Mace}},
  \citenamefont {{Dalena}},\ and\ \citenamefont
  {{Matthaeus}}}]{mace2012velocity}%
  \BibitemOpen
  \bibfield  {author} {\bibinfo {author} {\bibfnamefont {R.~L.}\ \bibnamefont
  {{Mace}}}, \bibinfo {author} {\bibfnamefont {S.}~\bibnamefont {{Dalena}}},\
  and\ \bibinfo {author} {\bibfnamefont {W.~H.}\ \bibnamefont {{Matthaeus}}},\
  }\href {https://doi.org/10.1063/1.3693379} {\bibfield  {journal} {\bibinfo
  {journal} {Physics of Plasmas}\ }\textbf {\bibinfo {volume} {19}},\ \bibinfo
  {eid} {032309} (\bibinfo {year} {2012})}\BibitemShut {NoStop}%
\bibitem [{\citenamefont {Hockney}\ and\ \citenamefont
  {Eastwood}(1988)}]{HockneyEastwood}%
  \BibitemOpen
  \bibfield  {author} {\bibinfo {author} {\bibfnamefont {R.~W.}\ \bibnamefont
  {Hockney}}\ and\ \bibinfo {author} {\bibfnamefont {J.~W.}\ \bibnamefont
  {Eastwood}},\ }\href@noop {} {\emph {\bibinfo {title} {Computer simulation
  using particles}}}\ (\bibinfo  {publisher} {crc Press},\ \bibinfo {year}
  {1988})\BibitemShut {NoStop}%
\bibitem [{\citenamefont {{Qin}}\ \emph {et~al.}(2013)\citenamefont {{Qin}},
  \citenamefont {{Zhang}}, \citenamefont {{Xiao}}, \citenamefont {{Liu}},
  \citenamefont {{Sun}},\ and\ \citenamefont {{Tang}}}]{QinEA13-boris}%
  \BibitemOpen
  \bibfield  {author} {\bibinfo {author} {\bibfnamefont {H.}~\bibnamefont
  {{Qin}}}, \bibinfo {author} {\bibfnamefont {S.}~\bibnamefont {{Zhang}}},
  \bibinfo {author} {\bibfnamefont {J.}~\bibnamefont {{Xiao}}}, \bibinfo
  {author} {\bibfnamefont {J.}~\bibnamefont {{Liu}}}, \bibinfo {author}
  {\bibfnamefont {Y.}~\bibnamefont {{Sun}}},\ and\ \bibinfo {author}
  {\bibfnamefont {W.~M.}\ \bibnamefont {{Tang}}},\ }\href
  {https://doi.org/10.1063/1.4818428} {\bibfield  {journal} {\bibinfo
  {journal} {Physics of Plasmas}\ }\textbf {\bibinfo {volume} {20}},\ \bibinfo
  {eid} {084503} (\bibinfo {year} {2013})}\BibitemShut {NoStop}%
\bibitem [{\citenamefont {Webb}(2014)}]{webb2014symplectic}%
  \BibitemOpen
  \bibfield  {author} {\bibinfo {author} {\bibfnamefont {S.~D.}\ \bibnamefont
  {Webb}},\ }\href {https://doi.org/https://doi.org/10.1016/j.jcp.2014.03.049}
  {\bibfield  {journal} {\bibinfo  {journal} {Journal of Computational
  Physics}\ }\textbf {\bibinfo {volume} {270}},\ \bibinfo {pages} {570 }
  (\bibinfo {year} {2014})}\BibitemShut {NoStop}%
\bibitem [{\citenamefont {{Ripperda}}\ \emph {et~al.}(2018)\citenamefont
  {{Ripperda}}, \citenamefont {{Bacchini}}, \citenamefont {{Teunissen}},
  \citenamefont {{Xia}}, \citenamefont {{Porth}}, \citenamefont {{Sironi}},
  \citenamefont {{Lapenta}},\ and\ \citenamefont
  {{Keppens}}}]{ripperda2018comprehensive}%
  \BibitemOpen
  \bibfield  {author} {\bibinfo {author} {\bibfnamefont {B.}~\bibnamefont
  {{Ripperda}}}, \bibinfo {author} {\bibfnamefont {F.}~\bibnamefont
  {{Bacchini}}}, \bibinfo {author} {\bibfnamefont {J.}~\bibnamefont
  {{Teunissen}}}, \bibinfo {author} {\bibfnamefont {C.}~\bibnamefont {{Xia}}},
  \bibinfo {author} {\bibfnamefont {O.}~\bibnamefont {{Porth}}}, \bibinfo
  {author} {\bibfnamefont {L.}~\bibnamefont {{Sironi}}}, \bibinfo {author}
  {\bibfnamefont {G.}~\bibnamefont {{Lapenta}}},\ and\ \bibinfo {author}
  {\bibfnamefont {R.}~\bibnamefont {{Keppens}}},\ }\href
  {https://doi.org/10.3847/1538-4365/aab114} {\bibfield  {journal} {\bibinfo
  {journal} {\apjs}\ }\textbf {\bibinfo {volume} {235}},\ \bibinfo {eid} {21}
  (\bibinfo {year} {2018})},\ \Eprint {https://arxiv.org/abs/1710.09164}
  {arXiv:1710.09164 [astro-ph.IM]} \BibitemShut {NoStop}%
\bibitem [{\citenamefont {{Hasselmann}}\ and\ \citenamefont
  {{Wibberenz}}(1970)}]{HasselmannWibberenz70}%
  \BibitemOpen
  \bibfield  {author} {\bibinfo {author} {\bibfnamefont {K.}~\bibnamefont
  {{Hasselmann}}}\ and\ \bibinfo {author} {\bibfnamefont {G.}~\bibnamefont
  {{Wibberenz}}},\ }\href {https://doi.org/10.1086/150736} {\bibfield
  {journal} {\bibinfo  {journal} {\apj}\ }\textbf {\bibinfo {volume} {162}},\
  \bibinfo {pages} {1049} (\bibinfo {year} {1970})}\BibitemShut {NoStop}%
\bibitem [{\citenamefont {{Urch}}(1977)}]{Urch77}%
  \BibitemOpen
  \bibfield  {author} {\bibinfo {author} {\bibfnamefont {I.~H.}\ \bibnamefont
  {{Urch}}},\ }\href {https://doi.org/10.1007/BF00644386} {\bibfield  {journal}
  {\bibinfo  {journal} {\apss}\ }\textbf {\bibinfo {volume} {46}},\ \bibinfo
  {pages} {389} (\bibinfo {year} {1977})}\BibitemShut {NoStop}%
\bibitem [{\citenamefont {{Qin}}\ \emph {et~al.}(2002)\citenamefont {{Qin}},
  \citenamefont {{Matthaeus}},\ and\ \citenamefont {{Bieber}}}]{QinEA02-apj}%
  \BibitemOpen
  \bibfield  {author} {\bibinfo {author} {\bibfnamefont {G.}~\bibnamefont
  {{Qin}}}, \bibinfo {author} {\bibfnamefont {W.~H.}\ \bibnamefont
  {{Matthaeus}}},\ and\ \bibinfo {author} {\bibfnamefont {J.~W.}\ \bibnamefont
  {{Bieber}}},\ }\href {https://doi.org/10.1086/344687} {\bibfield  {journal}
  {\bibinfo  {journal} {\apjl}\ }\textbf {\bibinfo {volume} {578}},\ \bibinfo
  {pages} {L117} (\bibinfo {year} {2002})}\BibitemShut {NoStop}%
\bibitem [{\citenamefont {{Matthaeus}}\ \emph {et~al.}(2003)\citenamefont
  {{Matthaeus}}, \citenamefont {{Qin}}, \citenamefont {{Bieber}},\ and\
  \citenamefont {{Zank}}}]{matthaeus2003nonlinear}%
  \BibitemOpen
  \bibfield  {author} {\bibinfo {author} {\bibfnamefont {W.~H.}\ \bibnamefont
  {{Matthaeus}}}, \bibinfo {author} {\bibfnamefont {G.}~\bibnamefont {{Qin}}},
  \bibinfo {author} {\bibfnamefont {J.~W.}\ \bibnamefont {{Bieber}}},\ and\
  \bibinfo {author} {\bibfnamefont {G.~P.}\ \bibnamefont {{Zank}}},\ }\href
  {https://doi.org/10.1086/376613} {\bibfield  {journal} {\bibinfo  {journal}
  {\apjl}\ }\textbf {\bibinfo {volume} {590}},\ \bibinfo {pages} {L53}
  (\bibinfo {year} {2003})}\BibitemShut {NoStop}%
\bibitem [{\citenamefont {{Kubo}}(1957)}]{kubo1957statistical}%
  \BibitemOpen
  \bibfield  {author} {\bibinfo {author} {\bibfnamefont {R.}~\bibnamefont
  {{Kubo}}},\ }\href {https://doi.org/10.1143/JPSJ.12.570} {\bibfield
  {journal} {\bibinfo  {journal} {Journal of the Physical Society of Japan}\
  }\textbf {\bibinfo {volume} {12}},\ \bibinfo {pages} {570} (\bibinfo {year}
  {1957})}\BibitemShut {NoStop}%
\bibitem [{\citenamefont {{Corrsin}}(1959)}]{corrsin1959progress}%
  \BibitemOpen
  \bibfield  {author} {\bibinfo {author} {\bibfnamefont {S.}~\bibnamefont
  {{Corrsin}}},\ }\href {https://doi.org/10.1016/S0065-2687(08)60102-8}
  {\bibfield  {journal} {\bibinfo  {journal} {Advances in Geophysics}\ }\textbf
  {\bibinfo {volume} {6}},\ \bibinfo {pages} {161} (\bibinfo {year}
  {1959})}\BibitemShut {NoStop}%
\bibitem [{\citenamefont {{Shalchi}}(2010)}]{shalchi2010unified}%
  \BibitemOpen
  \bibfield  {author} {\bibinfo {author} {\bibfnamefont {A.}~\bibnamefont
  {{Shalchi}}},\ }\href {https://doi.org/10.1088/2041-8205/720/2/L127}
  {\bibfield  {journal} {\bibinfo  {journal} {\apjl}\ }\textbf {\bibinfo
  {volume} {720}},\ \bibinfo {pages} {L127} (\bibinfo {year}
  {2010})}\BibitemShut {NoStop}%
\bibitem [{\citenamefont {{Aloisio}}\ and\ \citenamefont
  {{Berezinsky}}(2004)}]{aloisio2004diffusive}%
  \BibitemOpen
  \bibfield  {author} {\bibinfo {author} {\bibfnamefont {R.}~\bibnamefont
  {{Aloisio}}}\ and\ \bibinfo {author} {\bibfnamefont {V.}~\bibnamefont
  {{Berezinsky}}},\ }\href {https://doi.org/10.1086/421869} {\bibfield
  {journal} {\bibinfo  {journal} {\apj}\ }\textbf {\bibinfo {volume} {612}},\
  \bibinfo {pages} {900} (\bibinfo {year} {2004})},\ \Eprint
  {https://arxiv.org/abs/astro-ph/0403095} {arXiv:astro-ph/0403095 [astro-ph]}
  \BibitemShut {NoStop}%
\bibitem [{\citenamefont {{Caprioli}}\ and\ \citenamefont
  {{Spitkovsky}}(2013)}]{caprioli2013cosmicray}%
  \BibitemOpen
  \bibfield  {author} {\bibinfo {author} {\bibfnamefont {D.}~\bibnamefont
  {{Caprioli}}}\ and\ \bibinfo {author} {\bibfnamefont {A.}~\bibnamefont
  {{Spitkovsky}}},\ }\href {https://doi.org/10.1088/2041-8205/765/1/L20}
  {\bibfield  {journal} {\bibinfo  {journal} {\apjl}\ }\textbf {\bibinfo
  {volume} {765}},\ \bibinfo {eid} {L20} (\bibinfo {year} {2013})},\ \Eprint
  {https://arxiv.org/abs/1211.6765} {arXiv:1211.6765 [astro-ph.HE]}
  \BibitemShut {NoStop}%
\bibitem [{\citenamefont {{Shalchi}}(2014)}]{shalchi2014universality}%
  \BibitemOpen
  \bibfield  {author} {\bibinfo {author} {\bibfnamefont {A.}~\bibnamefont
  {{Shalchi}}},\ }\href {https://doi.org/10.1016/j.asr.2014.01.006} {\bibfield
  {journal} {\bibinfo  {journal} {Advances in Space Research}\ }\textbf
  {\bibinfo {volume} {53}},\ \bibinfo {pages} {1024} (\bibinfo {year}
  {2014})}\BibitemShut {NoStop}%
\bibitem [{\citenamefont {{Shalchi}}(2015)}]{shalchi2015perpendicular}%
  \BibitemOpen
  \bibfield  {author} {\bibinfo {author} {\bibfnamefont {A.}~\bibnamefont
  {{Shalchi}}},\ }\href {https://doi.org/10.1063/1.4906359} {\bibfield
  {journal} {\bibinfo  {journal} {Physics of Plasmas}\ }\textbf {\bibinfo
  {volume} {22}},\ \bibinfo {eid} {010704} (\bibinfo {year} {2015})},\ \Eprint
  {https://arxiv.org/abs/1501.06482} {arXiv:1501.06482 [astro-ph.SR]}
  \BibitemShut {NoStop}%
\bibitem [{\citenamefont {{Minnie}}\ \emph {et~al.}(2009)\citenamefont
  {{Minnie}}, \citenamefont {{Matthaeus}}, \citenamefont {{Bieber}},
  \citenamefont {{Ruffolo}},\ and\ \citenamefont {{Burger}}}]{minnie2009when}%
  \BibitemOpen
  \bibfield  {author} {\bibinfo {author} {\bibfnamefont {J.}~\bibnamefont
  {{Minnie}}}, \bibinfo {author} {\bibfnamefont {W.~H.}\ \bibnamefont
  {{Matthaeus}}}, \bibinfo {author} {\bibfnamefont {J.~W.}\ \bibnamefont
  {{Bieber}}}, \bibinfo {author} {\bibfnamefont {D.}~\bibnamefont
  {{Ruffolo}}},\ and\ \bibinfo {author} {\bibfnamefont {R.~A.}\ \bibnamefont
  {{Burger}}},\ }\href {https://doi.org/10.1029/2008JA013349} {\bibfield
  {journal} {\bibinfo  {journal} {Journal of Geophysical Research (Space
  Physics)}\ }\textbf {\bibinfo {volume} {114}},\ \bibinfo {eid} {A01102}
  (\bibinfo {year} {2009})}\BibitemShut {NoStop}%
\bibitem [{\citenamefont {{Shalchi}}(2019)}]{shalchi2019heuristic}%
  \BibitemOpen
  \bibfield  {author} {\bibinfo {author} {\bibfnamefont {A.}~\bibnamefont
  {{Shalchi}}},\ }\href {https://doi.org/10.3847/2041-8213/ab379d} {\bibfield
  {journal} {\bibinfo  {journal} {\apjl}\ }\textbf {\bibinfo {volume} {881}},\
  \bibinfo {eid} {L27} (\bibinfo {year} {2019})},\ \Eprint
  {https://arxiv.org/abs/1908.00694} {arXiv:1908.00694 [astro-ph.SR]}
  \BibitemShut {NoStop}%
\bibitem [{\citenamefont {{Arendt}}\ and\ \citenamefont
  {{Shalchi}}(2020)}]{arendt2020detailed}%
  \BibitemOpen
  \bibfield  {author} {\bibinfo {author} {\bibfnamefont {V.}~\bibnamefont
  {{Arendt}}}\ and\ \bibinfo {author} {\bibfnamefont {A.}~\bibnamefont
  {{Shalchi}}},\ }\href {https://doi.org/10.1016/j.asr.2020.07.024} {\bibfield
  {journal} {\bibinfo  {journal} {Advances in Space Research}\ }\textbf
  {\bibinfo {volume} {66}},\ \bibinfo {pages} {2001} (\bibinfo {year}
  {2020})}\BibitemShut {NoStop}%
\bibitem [{\citenamefont {{Reichherzer}}\ \emph {et~al.}(2020)\citenamefont
  {{Reichherzer}}, \citenamefont {{Becker Tjus}}, \citenamefont {{Zweibel}},
  \citenamefont {{Merten}},\ and\ \citenamefont
  {{Pueschel}}}]{2020MNRAS.498.5051R}%
  \BibitemOpen
  \bibfield  {author} {\bibinfo {author} {\bibfnamefont {P.}~\bibnamefont
  {{Reichherzer}}}, \bibinfo {author} {\bibfnamefont {J.}~\bibnamefont {{Becker
  Tjus}}}, \bibinfo {author} {\bibfnamefont {E.~G.}\ \bibnamefont {{Zweibel}}},
  \bibinfo {author} {\bibfnamefont {L.}~\bibnamefont {{Merten}}},\ and\
  \bibinfo {author} {\bibfnamefont {M.~J.}\ \bibnamefont {{Pueschel}}},\ }\href
  {https://doi.org/10.1093/mnras/staa2533} {\bibfield  {journal} {\bibinfo
  {journal} {\mnras}\ }\textbf {\bibinfo {volume} {498}},\ \bibinfo {pages}
  {5051} (\bibinfo {year} {2020})},\ \Eprint {https://arxiv.org/abs/1910.07528}
  {arXiv:1910.07528 [astro-ph.HE]} \BibitemShut {NoStop}%
\end{thebibliography}%

\end{document}